\documentclass[11pt,tightenlines,showpacs,nofootinbib,twocolumn]{revtex4-1}
\usepackage[dvips]{graphicx}
\graphicspath{{figures/}}

\usepackage[bbgreekl]{mathbbol}
\usepackage{tabularx}
\usepackage[caption=false]{subfig}
\usepackage{amsmath}
\usepackage{amssymb}
\usepackage{verbatim}
\usepackage[usenames,dvipsnames]{color} 
\usepackage[colorlinks=true,linkcolor=blue]{hyperref}
\usepackage{cleveref} 
\crefname{figure}{Fig.}{Figs.}
\Crefname{figure}{Figure}{Figures}

\crefname{equation}{Eq.}{Eqs.}
\Crefname{equation}{Equation}{Equations}

\crefname{table}{Table}{Tables}
\Crefname{table}{Table}{Tables}

\crefname{section}{Sec.}{Secs.}
\Crefname{section}{Section}{Sections}

\usepackage{makecell}
\usepackage{pbox}

\usepackage{bbm}

\usepackage[american]{babel}

\DeclareMathOperator*{\argmin}{arg\,min}
\DeclareMathOperator*{\argmax}{arg\,max}
\DeclareMathOperator{\Tr}{Tr}

\usepackage{xpatch}
\xpatchcmd\bibsection{\begingroup}{\vskip 40pt\begingroup}{}{}

\usepackage{etoolbox}

\makeatletter
\appto{\appendix}{%
  \@ifstar{\def\theequation@prefix{A.}}%
          {}%
}
\makeatother

\begin{document}

\title{Engineering single-polymer micelle shape using non-uniform spontaneous surface curvature} 
\date{\today} 
\author{Brian Moths and Thomas Witten}

\begin{abstract}
Conventional micelles, composed of simple amphiphiles, exhibit only a few standard morphologies, each characterized by its mean surface curvature set by the amphiphiles.
Here we demonstrate a rational design scheme to construct micelles of more general shape from polymeric amphiphiles.
We replace the many amphiphiles of a conventional micelle by a single flexible, linear, block copolymer chain containing two incompatible species arranged in multiple alternating segments.
With suitable segment lengths, the chain exhibits a condensed spherical configuration in solution, similar to conventional micelles.
Our design scheme posits that further shapes are attained by altering the segment lengths.
To assess the power of this scheme, we exhibit stable micelles of horseshoe form using conventional bead-spring simulations in two dimensions.
Modest changes in the segment lengths produce smooth changes in the micelle's shape and stability.
\end{abstract}

\maketitle 

\section{Introduction}
\label{sec:introduction}

\begin{figure} 
 \centering \includegraphics[width=0.7\linewidth]{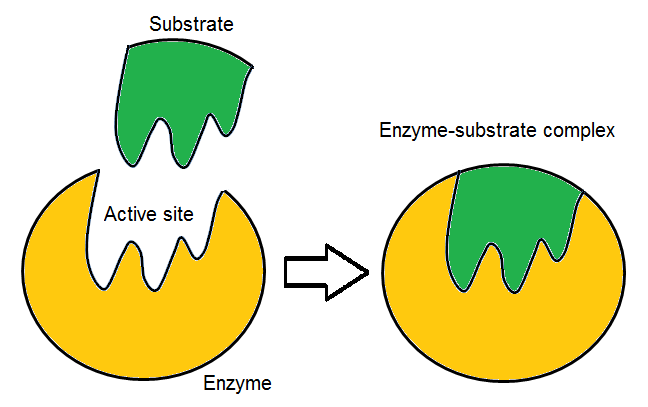}
 \caption{(Color online) Illustration of lock and key mechanism \cite{lockAndKeyWiki}.
  The enzyme, in yellow, is meant to interact specifically with a substrate, shown in green.
  To avoid unwanted interactions, the enzyme has a specific shape to which only a substrate of complementary shape may bind.
  More generally, the term ``lock and key mechanism" may refer to any interaction controlled by shape.
 } 
 \label{fig:lockandkey}
\end{figure}

Amphiphilic molecules have self-organizing behavior, which makes them useful for a variety of applications. 
One application that has received much attention is drug delivery using micelle carriers~\cite{Cho2015,KATAOKA2001113,mixedMicelle2011}.
Among other things, it is found that the shape of the micelles affect their drug delivery performance, for example by altering how much drug can be loaded into the micelle, where in the body the drug accumulates, or how long the drug remains in the body~\cite{Cai2007,VENKATARAMAN2011}.
More generally, shape can be used to facilitate or inhibit interactions, as in the well-known ``lock and key'' mechanism~\cite{fisherLockAndKey}, shown in \cref{fig:lockandkey}.
In fact, the lock and key mechanism has already been used with dimpled sphere-shaped colloids to create self-assembled structures~\cite{Sacanna2010Lock}, and it would seem straightforward to apply this same concept to micelles. 
This paper presents a strategy for influencing the natural shape of a micelle by controlling the way it is constructed. 
Specifically, we demonstrate, through simulation, the ability to design the shape of a micelle constructed from a linear multiblock copolymer by choosing the lengths of its constituent blocks.
Our strategy is motivated by recent advances in polymer synthesis allowing for realization of linear multiblock copolymers with individually controlled block lengths~\cite{Matyjaszewski12}.

Much work has been done to study the factors influencing micelle shape.
One line of investigation is to assume a continuum energy model for the micelle surface, and study the resulting ground states and fluctuations~\cite{BERGSTROM200815,Khokhlov2005,LIPOWSKY201414,Watson2011}.
To make contact between the continuum parameters of these models and the physics of the micelle on the scale of a single amphiphile, simulations of suitable structures (bilayers, tethers, etc.) made out of the micelle's constituent amphiphiles may be conducted to determine the continuum parameters inherent to the amphiphiles (such as the amphiphile surface density, bending modulus, etc)~\cite{BRANDT20112104,Goetz1999,Harmandaris2006,Hu2012,Hu2015,Laradji2000,Rekvig2004,Rozycki2015,Tarazona2013,Venable201560}.
Instead of investigating the continuum properties of the micelle shape to infer geometrical features,  micelles of interest may be directly simulated with a particle-based model (with either atomic or coarse-grained resolution)~\cite{Guo2010,Loverde2010,Poorgholami-Bejarpasi2010,Sheng2013,Srinivas2004Selfassembly,Velinova2011}.
Additionally, micelle shape can be studied experimentally~\cite{JELONEK2015357,Lof2009,Rikken2016,Zhao2010}.  
In the context of the approaches described in the previous paragraph, our work falls into the category of directly simulating the micelle using a particle-based model.
We choose this approach over a continuum representation for two reasons.
First, we are interested in micelles whose size is on the order of the amphiphile length, where the scale of surface fluctuations can be roughly ten percent of the micelle size \cite{ljunggren89}.
Second,  we attempt to resolve average positions of individual amphiphile junction points.
However, unlike the micelle simulations of~\cite{Guo2010,Loverde2010,Poorgholami-Bejarpasi2010,Sheng2013,Srinivas2004Selfassembly,Velinova2011}, which study only the topology or rough shape features such as size or aspect ratio, this work seeks to obtain precise control of the micelle shape.
Fine shape control is desirable both because it is a requirement of the lock and key interactions referenced in \cite{Sacanna2010Lock}, and it can also be used to achieve full optimization of a micelle's drug delivery properties, as will be discussed further in \cref{sec:conclusion}.

\subsection{Shape-design rationale}
\label{subsec:shapeDesign}

\begin{figure}
\centering
\includegraphics[width=\linewidth]{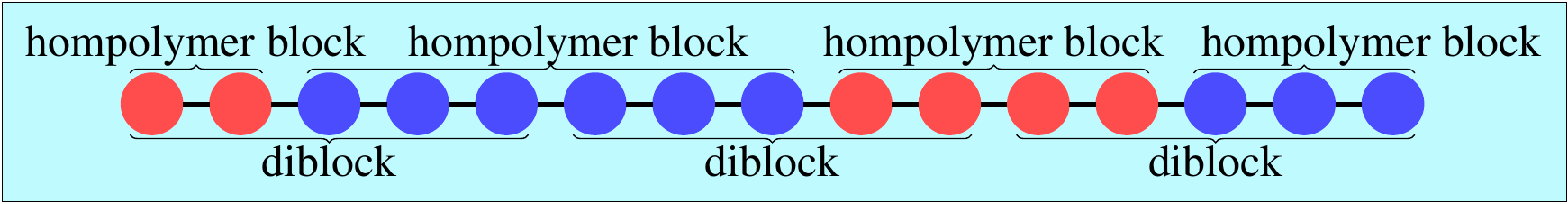}
\caption{(Color online) 
Two views of a model multiblock copolymer. 
The multiblock contains two species of monomer beads, shown in red and blue, connected by bonds shown in black. 
One view of the multiblock is as a collection of homopolymer blocks.
The other view represents the multiblock as a collection of diblocks joined end to end.
}
\label{fig:multiblockdiblock}
\end{figure}

To obtain this fine shape control we construct the micelle from one linear multiblock copolymer containing two species of monomer, a solvophobic species that is immiscible with the solvent and a solvophilic species that dissolves well in the solvent, arranged into segments of judiciously chosen length made purely of one species or the other.
To better explain how the choice of segment length affects the micelle shape, we view the multiblock copolymer not as a sequence of chemically pure homopolymer segments joined together, but rather as a sequence of diblocks joined end to end so that the chemically similar ends of sequential diblocks are joined.
In this view, the ends of each homopolymer segment correspond to diblock junction points, and a bond joining sequential diblocks occurs in the middle of a homopolymer segment.
\Cref{fig:multiblockdiblock} illustrates the two ways of viewing the polymer chain.

\begin{figure}
 \centering \includegraphics[width=0.7\linewidth]{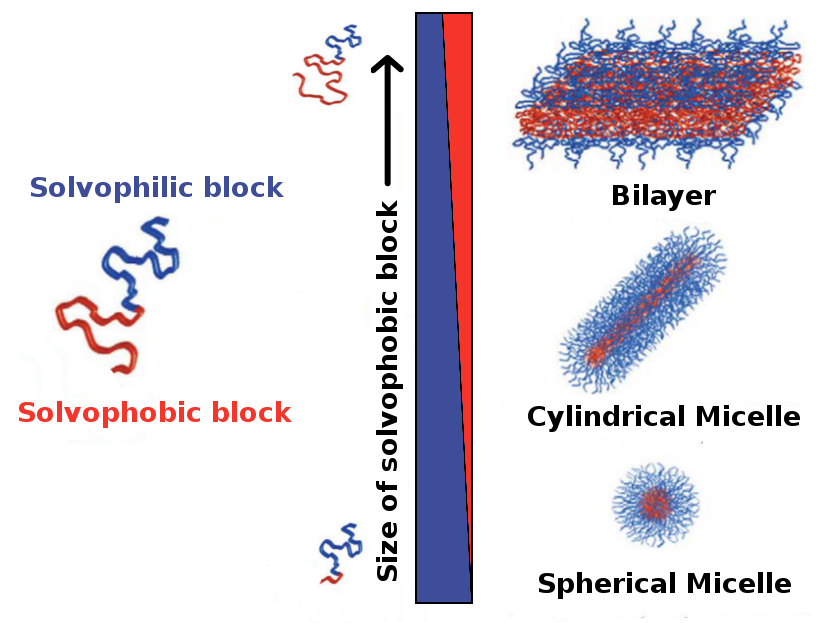}
 \caption{(Color online) Illustration of micelle shape dependence on diblock composition.
  Diblocks with a very small solvophobic block tend to form highly curved spherical micelles.
  Diblocks with a more symmetric composition form flat bilayers.
  The case of cylindrical micelles is intermediate to these two. 
  Figure adapted from \cite{Zhang98}.
 }
 \label{fig:micelleShapesMassRatio}
\end{figure}

There are two reasons it is advantageous to view the micelle as a collection of diblocks.
The first is that the applicability to drug delivery applications cited above, in which the drug carriers are micelles formed from (disconnected) diblock copolymers, becomes more apparent.
The second reason is that viewing the micelle as a collection of diblocks having adjustable block lengths provides a straightforward, theoretically informed strategy for choosing the multiblock's segment lengths.
The relationship between the diblocks' block lengths and micelle shapes is determined by the requirement that diblocks pack efficiently on the micelle surface.
There are two factors that affect surface packing: the first is the energetic interactions between the monomers, and the second is the diblock chain stretching entropy.
Intuition may be gained by considering the limit of long blocks, where scaling arguments yield analytic results, as reported in~\cite{Wang91}.
For example, one particularly relevant result of \cite{Wang91} is an analytical expression for the dependence of the preferred mean curvature of an interface containing a monolayer of diblocks on the block lengths.
Even though a continuous range of preferred mean curvature can be achieved by adjusting the relative block lengths of a diblock, it is known that only three shapes can be achieved by micelles composed of a single species of diblock: spherical, cylindrical, and bilayer \cite{Israelachvili2010Intermolecular} (see \cref{fig:micelleShapesMassRatio}).

By contrast, we expect micelles containing several species of diblock (or, in the alternate view, multiblocks containing homopolymer segments of varying lengths) to exhibit a much larger variety of shapes.
The design method for achieving a desired shape presented in this paper is to choose the block lengths of the constituent diblocks so that their associated preferred curvatures matches the curvature of the desired shape.
In practice, it is not sufficient to simply choose a set of block lengths; additionally, the diblock positions must be controlled so that the desired curvature is imprinted at the desired location on the surface.
It is exactly for this reason that the diblocks are joined together into one linear multiblock copolymer---the added bonds between the diblocks hinder unwanted movement across the micelle surface.
An illustration of this shape-design mechanism is given in \cref{fig:interfaceCurvatureWithMicelle}.

\subsection{Motivation}
\label{subsec:motivation}

The micelle shape-design method described above is one means of creating self-assembled globular objects of controlled, macromolecular size.
It is useful to contrast this method with other means of making globules of regulated form.
The first of these is perhaps the most familiar: crystal growth.
Like the micelles we propose in the this paper, crystals have well-defined geometrical characteristics (e.g., lattice planes) that emerge from the local interactions between their constituents.
Another similarity with our micelles is that crystals result from non-specific interactions resulting from only a few chemical species.
Due to the non-specificity of the interactions, a crystal may deform (e.g., by dislocation glide), without losing its natural geometric characteristics, as each atom or molecule is left in an identical environment after the deformation.
However, there are many differences between our micelles and crystals.
The most important difference for our purposes is that crystals do not naturally form well-defined finite shapes; instead, the size of the self-assembled structure is determined only by the amount of constituents present.
Additionally, the shapes formed by crystals can be categorized into only a few classes, further limiting the shape control that can be achieved through selecting the constituents.
Another important difference is that crystals are solid, and therefore do not have fluctuations.

The second naturally occurring system, perhaps more similar to our micelles, is a globular protein.
It could be said that globular proteins are more similar to our micelles because they both have a well-defined shape and size determined by their composition.
However, unlike the micelles we propose, the shape of a globular protein is determined by specific, high-energy, localized interactions between its constituent amino acids.
This leads to the ``protein folding problem": the folded shape of the protein is difficult to predict from the sequence of amino acids.
If the amino acid sequence is even slightly altered, the shape is often completely destroyed.
Also owing to the specific nature of the interactions, if a protein's shape is significantly deformed, many atoms' environments become completely different, so that the shape is irreversibly lost.
By contrast, the non-specific interactions responsible for amphiphile aggregation allow for a smooth dependence of energy on the micelle configuration, so a perturbed micelle returns to its equilibrium shape.
Also, the simplicity of the non-specific interaction allows for the straightforward design strategy described in \cref{subsec:shapeDesign}; there should be no analogy to the ``protein folding problem" for the micelles we consider.
Additionally, the tight nature of the bonds in proteins gives a solid-like character leading to low shape fluctuations, while micelles may have large fluctuations, which may be used, e.g., to regulate drug delivery or reduce the shape specificity of lock-and-key interactions.

\subsection{Scope of this paper}
\label{subsec:scope}
In \cref{subsec:shapeDesign}, we described our shape design strategy as judiciously selecting block lengths for diblock copolymers to achieve a desired preferred curvature profile, and then joining these diblocks into one multiblock copolymer to constrain their positions on the micelle surface.
To organize the following discussion, we distinguish two challenges associated with this shape-design mechanism.
The first challenge is to determine which block lengths should be selected for each diblock on the micelle surface to produce the desired shape.
The second challenge is to constrain the diblocks so they keep their intended positioning on the micelle surface, which we attempt to do by joining the diblocks together.

Were the diblocks not joined together, there would be a number of ways the second challenge could fail to be met.
For example, the diblocks might diffuse on the surface of the micelle, washing out the intended curvature profile and leaving only a uniform spontaneous curvature profile in its place.
A more extreme example is for the micelle to divide into two disconnected pieces.
We use the term ``malformed" to refer to any such micelle where the diblocks do not have their intended relative positioning.
Conversely, if the diblocks do have their intended positions, we call the shape ``well-formed".

It is clear that the relative positioning of diblocks can be enforced, and therefore malformed shapes prevented, by adding sufficiently many bonds between diblocks that are intended to be near each other.
There are many ways one could imagine introducing bonds besides joining the diblocks end to end, as described above.
For example, one could bond one end of each diblock to a common linear backbone polymer producing a comb polymer.
One may ask what scheme of bonding diblocks is optimal, or if other methods of preventing malformed shapes are possible.
However, in this paper, we do not address this question; we focus instead on the first challenge of selecting diblock compositions to produce a desired shape, given that the micelle shape remains well-formed.
Still, some bond scheme is necessary to address this second challenge.
This is why we join the diblocks end to end to form a linear multiblock copolymer.
(While a multiblock copolymer is more difficult to experimentally realize a collection of diblocks, techniques have been developed to synthesize multiblock copolymers having a specified sequence of block lengths~\cite{golas2009}.)
We find that this scheme causes most simulated micelles to be well-formed.
Any simulations that result in malformed micelles are simply discarded since we focus exclusively on the first challenge, which has only to do with well-formed micelles.

\begin{figure}
\centering
\includegraphics[width=0.7\linewidth]{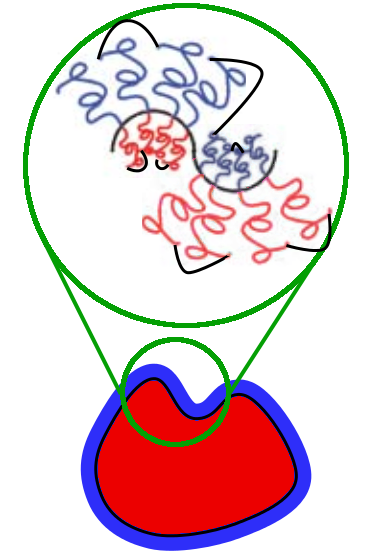}
\caption{(Color online) Schematic of proposed mechanism for making micelles of designed shape. 
A shape-designed micelle, shown in two dimensions for simplicity, has a solvophobic interior (shown in red). 
At the surface, there are diblocks containing both solvophobic and solvophilic (shown in blue) blocks. 
The interface between the solvophobic and solvophilic regions is shown in black, and it has a concave dimple. 
The inset shows how our shape-design mechanism gives rise to the designed shape: regions of the micelle surface where a convex curvature is desired are populated with diblocks having a larger solvophilic block and consequently preferring a convex curvature, while regions to be made concave are populated with diblocks containing larger solvophobic blocks, thereby preferring more concave curvature. 
Bonds, indicated in black, connect the diblocks end to end forming a multiblock copolymer in order to fix diblocks in their intended positions.
}
\label{fig:interfaceCurvatureWithMicelle}
\end{figure}
The purpose of the paper, then, is to address the first challenge: to show that the shape features can be controlled by selecting the species of constituent diblocks at each point on the micelle surface.
For simplicity, we consider a two-dimensional system, and we restrict our attention to a case study of a shape with a single concave dimple similar to the one in \cref{fig:interfaceCurvatureWithMicelle}.
We choose this shape because it is a minimal example requiring our shape-design mechanism: while it is simple, it does not arise as an equilibrium shape of a micelle composed of diblocks of a single species.

The remainder of this paper is organized as follows: In \cref{sec:method}, we describe our polymer model and how this model is simulated.
Then, we verify that our model and simulation method give physically reasonable results.
Lastly, we describe how the polymeric micelles are represented in the simulation. 
In \cref{sec:analysis}, we describe how micelle shape properties are extracted from the simulation results and which specific shape features we study.
In \cref{sec:results}, we apply the analysis methods of \cref{sec:analysis} to micelle simulations, showing that these methods give self-consistent results and demonstrating the extent to which our shape-design mechanism affects features of the micelle shape.
In \cref{sec:discussion}, we discuss future work suggested by this research and the implications our results have for the applications mentioned above.

\section{Method}
\label{sec:method}

In this section, we describe how our micelles are simulated.
A simulation requires an underlying physical model specifying, for example, the degrees of freedom used to represent the system and the interactions governing the system.
With the model specified, it is necessary to choose a method to simulate the model system.
Because the applications we consider in \cref{sec:introduction} concern micelles in thermodynamic equilibrium at some finite temperature, the goal of the simulation is to produce a Boltzmann-distributed ensemble of micelle configurations.
After describing our model and simulation method, we validate the method by applying it to a testbed homopolymer system and verifying that the resulting configuration ensembles have the expected statistics.
After demonstrating the validity of our simulation method, we describe how we apply it to demonstrating the utility of our shape-design mechanism.
\subsection{Model}
\label{subsec:model}
Since we expect our shape-design mechanism ought to apply very generally without regard to the specific features of a particular chemical structure, we choose a simple model having the minimum content necessary to exhibit our shape-designed mechanism. 
Specifically, we use a coarse-grained two dimensional bead-spring model with implicit solvent, similar to the models used in~\cite{detcheverry2009,Binder11,Dimitrakopoulos04,Hsieh06}.
In this model, a polymer is represented as a sequence of beads, each described by only a position and a common diameter.
By ``coarse-grained", we mean a simulation bead does not represent just a single atom or even a single monomer, but rather several chemical repeat units.
Any sequential pair of beads in the polymer is connected by a bond represented by a harmonic pair potential.
In addition to these bond potentials, the beads also interact through a short-range pair potential.
Since the solvent is treated implicitly, the nature of the short-range interaction between two beads depends not only on the material composing the two beads, but also on the solvent.
For example, beads representing the same non-polar hydrocarbon would have a more attractive pair potential when immersed in a polar solvent than they would when immersed in a non-polar solvent.
To represent diblocks, our model has two species of bead: one solvophobic and one solvophilic.

To completely specify the model, we now give a precise description of the pair potentials governing the beads.
First we establish a system of units.
Since we are ultimately interested in finding the thermal equilibrium properties of the micelles, a natural unit of energy is the thermal energy $k_B T$, where $k_B$ is Boltzmann's constant and $T$ is the temperature of the system being simulated.
As we are interested in simulating polymers, we choose the unit of length to be the root-mean-square thermal length $L_{\textrm{therm}}$ of the harmonic spring connecting two adjacent beads (ignoring the close-range potential).
In two dimensions, the system has two internal degrees of freedom, so we find by the equipartition theorem that the unit of length $L_{\textrm{therm}}$ is given by 
\begin{equation}
\label{eq:thermalLength}
k_B T = \dfrac{1}{2} k L_{\textrm{therm}}^2,
\end{equation}
where $k$ is the spring constant of the harmonic potential.
To complete our system of units, we may take the our unit of mass to be the bead mass.
For the remainder of this paper, we nondimensionalize all physical quantities using this system of units.
For example, the bond interaction $U_{\textrm{bond}}(d)$ felt by two adjacent beads displaced by a distance $d$, is given by
\begin{equation}
U_{\textrm{bond}}(d) = d^2.
\label{eq:uBond}
\end{equation}

\begin{figure}
\subfloat[][]{ %
\label{fig:hardPlot} %
\centering %
\includegraphics[width=\columnwidth]{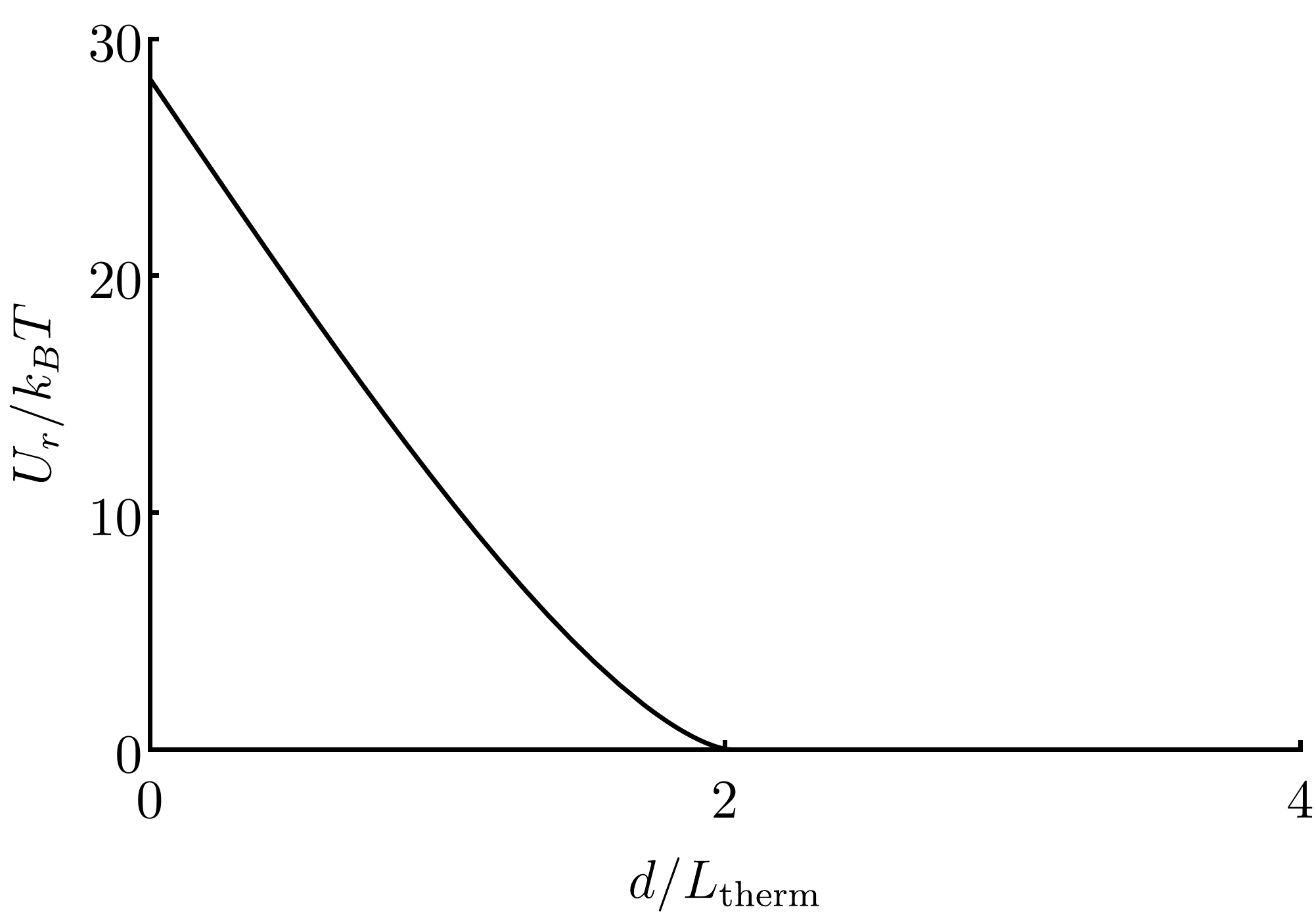} %
}

\subfloat[][]{ %
\label{fig:bluePlot} %
\centering %
\includegraphics[width=\columnwidth]{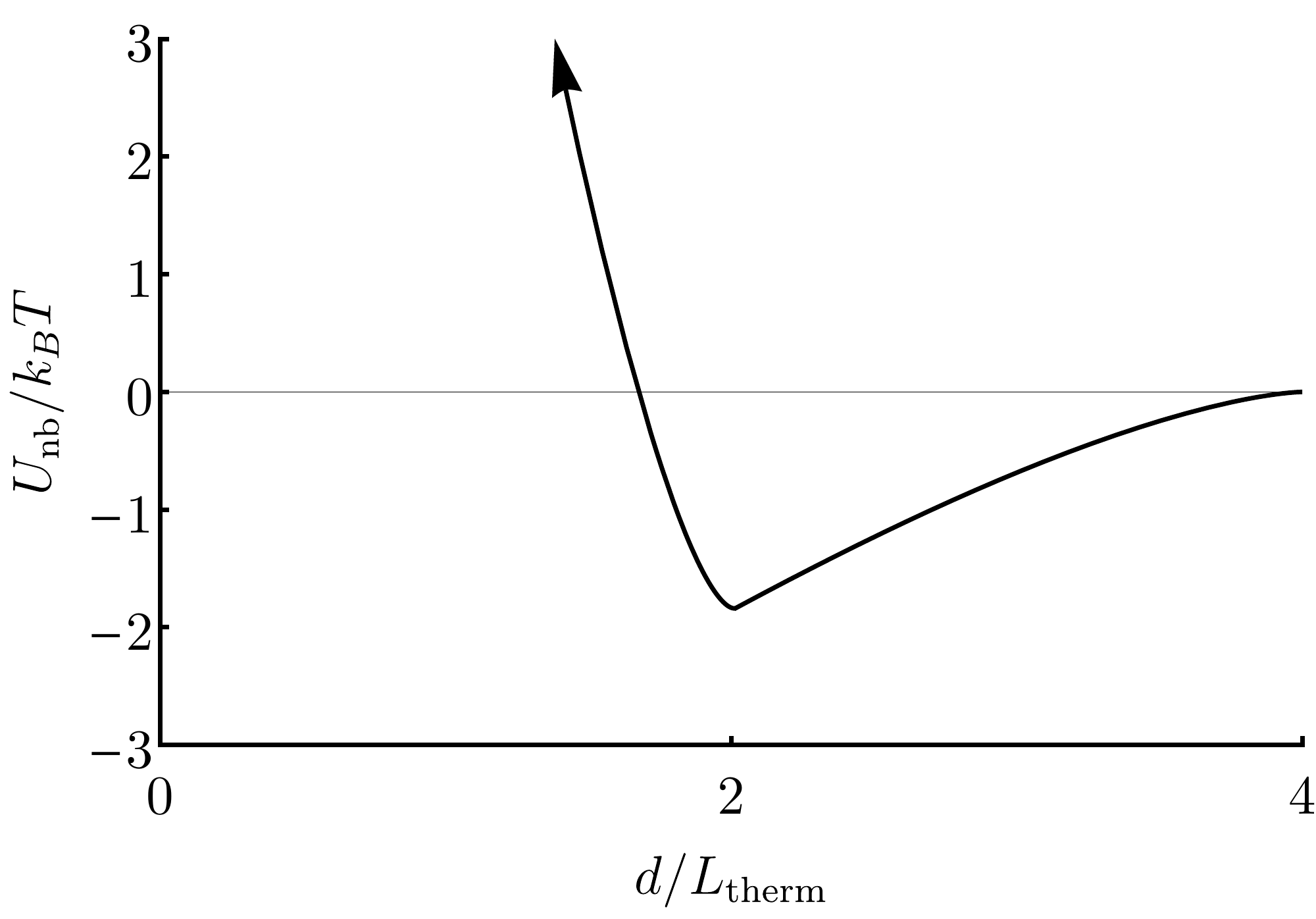} %
}

\caption{Plots of non-bonded interaction potential. 
Energies and lengths have been nondimensionalized using $k_B T$ and $L_{\textrm{therm}}$ respectively, as described in the second paragraph of \cref{subsec:model}.
In \protect\subref{fig:hardPlot}, the purely repulsive interaction potential $U_r$ of \cref{eq:coreEnergy} for a bead pair containing a solvophilic bead is plotted. 
In \protect\subref{fig:bluePlot}, the interaction potential for two solvophobic beads is plotted. 
This interaction potential contains an attractive term in addition to the repulsive potential $U_r$ plotted in \protect\subref{fig:hardPlot}.
}
\label{fig:potentialPlots}
\end{figure}

The form of the non-bonded interaction $U_{\textrm{nb}}$ is more complicated.
This interaction is the sum of two terms.
One term is a stiff repulsion $U_r$ enforcing that no two beads have the same position.
Following~\cite{detcheverry2009}, we take an interaction whose strength is proportional to the size of the overlap region of the beads. 
Therefore the stiff repulsion is given by
\begin{equation}
\frac{U_r(d)}{P_r D_r^2/2} =  \cos^{-1}\left(\frac{d}{D_r}\right) - \frac{d}{D_r} \sqrt{1 - \frac{d^2}{D_r^2}},
\label{eq:coreEnergy}
\end{equation}
where $D_r$ is the maximum range of the repulsive interaction, $P_r$ is a constant setting the strength of the interaction, and again $d$ is the distance between the bead centers.
For simplicity, we simply take the same value of $P_r$ and $D_r$ to govern all pairs of beads.
In addition to this stiff repulsion, there is an attraction $U_a(d)$ between solvophobic beads of the same form as \cref{eq:coreEnergy}, but with a longer interaction range $D_a$, and a negative strength parameter $P_a$.
The values of the constants are $D_r = 2.015873$, $P_r=8.870637$, $D_a = 4$, and $P_a=-0.378$.
These values were chosen to produce a homopolymer with physically reasonable properties, as will be seen in \cref{subsec:validation}.
Plots of the pair potentials are shown in \cref{fig:potentialPlots}.

\subsection{Simulation method}
\label{subsec:simulationMethod}
To determine the average micelle shape resulting from this model, we perform a constant temperature molecular dynamics simulation using LAMMPS~\cite{lammps95}.
Although LAMMPS is a molecular dynamics simulator, meaning it essentially works by calculating forces from the potentials and then using Newton's second law to get the accelerations from these forces, it provides for running simulations of a fixed number of particles ``\verb|n|" at constant temperature ``\verb|t|" and either constant volume ``\verb|v|" or pressure ``\verb|p|", through its \verb|fix nvt| and \verb|fix npt| commands, respectively.
Since we are interested in obtaining a thermal ensemble of configurations, these are exactly the commands we used to time-evolve the system.
There are two parameters for the \verb|fix nvt| command: a timestep and a time constant \verb|Tdamp| setting how quickly the simulation thermalizes the system. 
Additionally, for constant pressure simulations, there is an additional time constant \verb|Pdamp| determining how quickly the volume in a constant pressure simulation responds to an unbalanced pressure.
Choosing a large value for the timestep and small values for \verb|Tdamp| and \verb|Pdamp| has the benefit of reducing the computational expense of the simulation, but, taken too far, may lead to instability in the simulation.
The instability associated with too large of a timestep occurs when the force acting on a particle at the beginning of the timestep differs significantly from the force at the end of the timestep.
This causes an instability because the molecular dynamics integrator uses a constant force over the course of the timestep, so that the particle may move very far during the timestep with no opposing force to stop it.
Often, as in our simulations, the particles are near the minimum of a potential well, so if a very large displacement does occur, then the particle will typically be subjected to a large restoring force in the following timestep.
This large restoring force leads to an even larger displacement in the following timestep, and in this way, the particle experiences unphysically large and high-frequency oscillations in position.
A similar thing happens if \verb|Tdamp| (or \verb|Pdamp|) is too small.
In this case, a deviation in the temperature (or pressure) from the set point is overcorrected each timestep, leading to increasingly large fluctuations.
After some experimentation, we found that setting the timestep to $0.003$ and setting \verb|Tdamp| and \verb|Pdamp| to $0.5$ allows for efficient simulation, and, as demonstrated in \cref{subsec:validation}, does not lead to instability.
Another detail of the simulation is the initialization.
The initial bead velocities were chosen from a thermal distribution at the temperature of the simulation thermostat.
The choice of initial configuration is discussed later in this paper in \cref{subsec:micelleDesign}.

\subsection{Validation}
\label{subsec:validation}
Before employing our model and simulation method to study our shape-design mechanism, we first verify that they produce physically reasonable results in cases where the expected behavior is known.
One such case is a homopolymer melt: a system consisting of a dense phase of polymers all made from the same single species of bead (in our case, solvophobic).
The melt ought to have a well-defined density, compressibility, and surface tension.
Additionally, the density and mean square end to end distance of the polymer ought to have an expected dependence on the number of monomers in the polymer.
In addition to verifying that the simulation is consistent with these expectations, we also test for quantitative agreement between the melt properties and those of real polymer; specifically, we compare with poly(dimethylsiloxane) (PDMS).
A comparison of our system with PDMS is apt because both systems are only weakly insoluble.

\begin{figure}
\centering
\includegraphics[width=\linewidth]{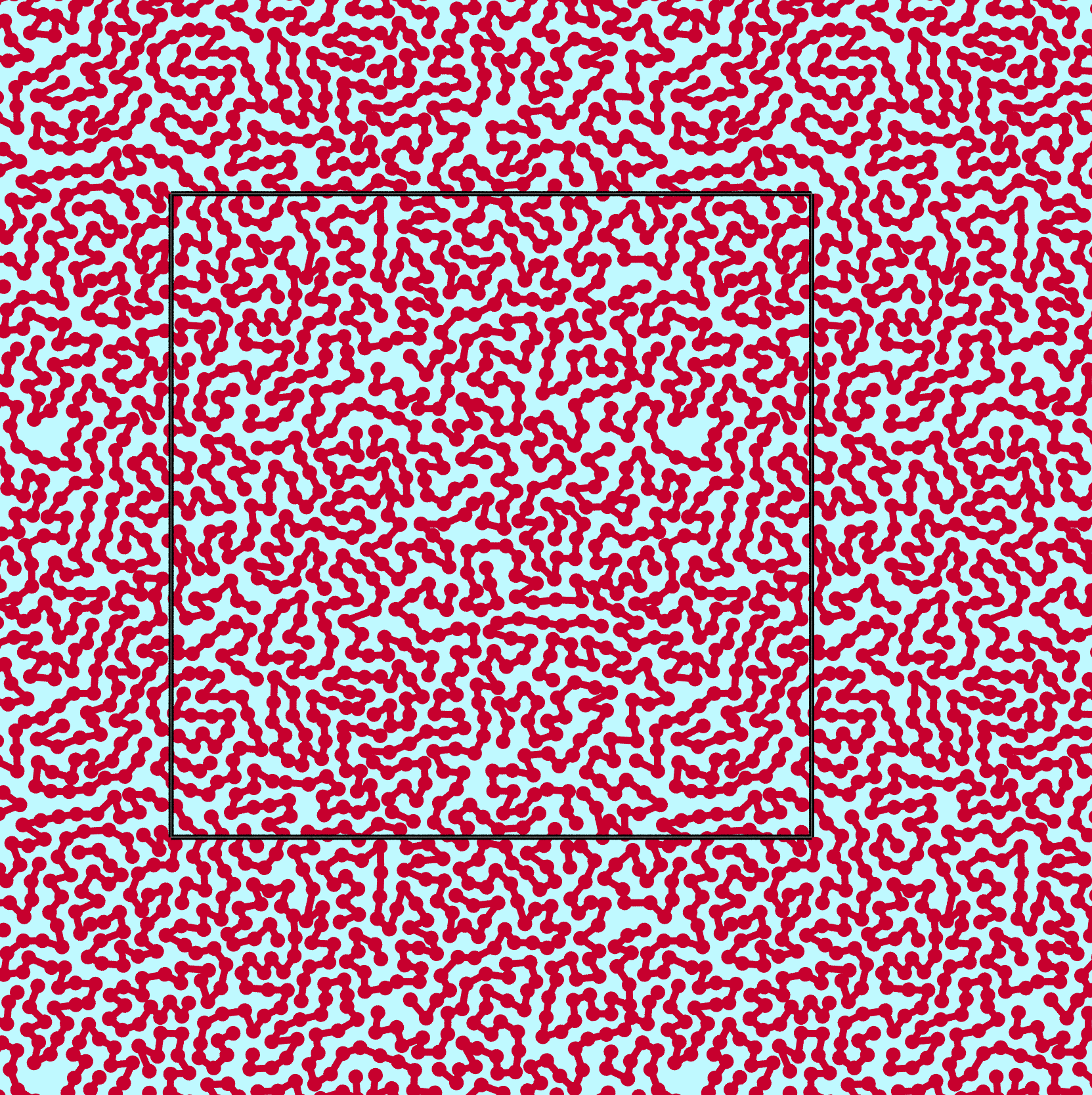}
\caption{(Color online) Visualization of simulated homopolymer melt.
The simulated system contains 30 solvophobic homopolymer chains, each having 35 beads, shown as red disks.
The springs connecting adjacent beads are shown as red segments.
The system is periodic, its boundary indicated by the black square.
}
\label{fig:bulksnapshot}
\end{figure}

\begin{figure}
\centering
\includegraphics[width=\linewidth]{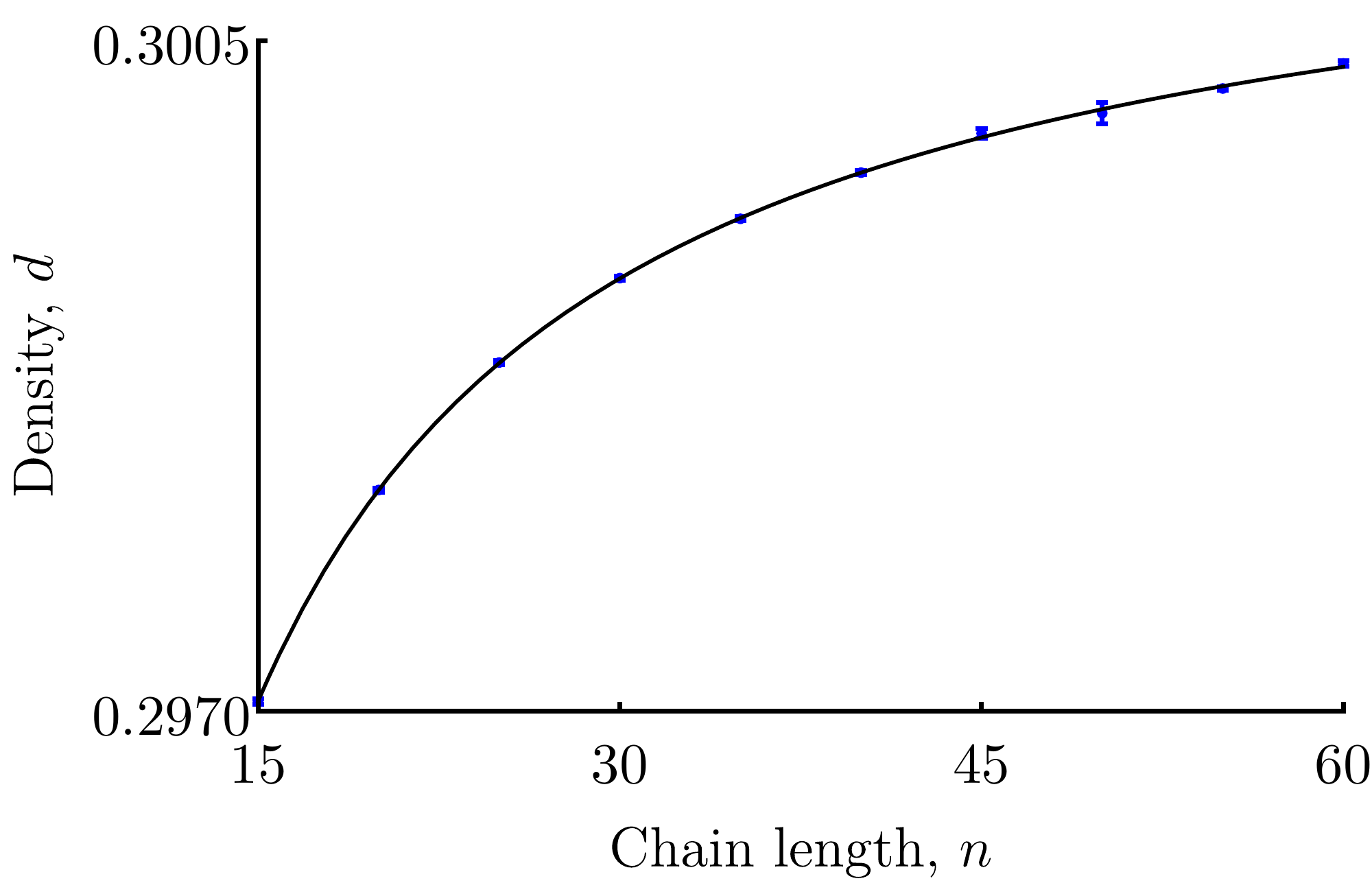}
\caption{(Color online) 
Solid curve: expected functional form form \cref{eq:densityExpectation}, fitted to the simulation values shown as points.
The resulting reduced chi-squared is $0.5$, indicating that this functional form is consistent with the data.
The best-fit value of $d_\infty$ is $0.30146\pm 0.00001$.
}
\label{fig:chainLengthDensity}
\end{figure}

First we consider the melt density.
We simulate a periodic homopolymer system, shown in \cref{fig:bulksnapshot}, at zero pressure.
In this case, the equilibrium density is the number of beads divided the average volume of the system.
Multiple independent simulations for different chain lengths $n$ were run in order to find the dependence of the simulated density $d$ on chain length and compare this dependence to theoretical expectations.
Theoretically, the density is expected to reach a finite value $d_\infty$ as the chain length goes to infinity.
For large values of the chain length, the dependence of the density on the chain length $d(n)$ can be expanded in the small parameter $1/n$:
\begin{equation}
\label{eq:densityExpectation}
d(n)=d_\infty + \frac{a}{n},
\end{equation}
where $a$ is a parameter representing how strongly the density depends on the chain length.
We find our data is indeed well-fitted by this functional form, as seen in \cref{fig:chainLengthDensity}.

\begin{figure}

\subfloat[][]{ %
\centering
\includegraphics[width=\columnwidth]{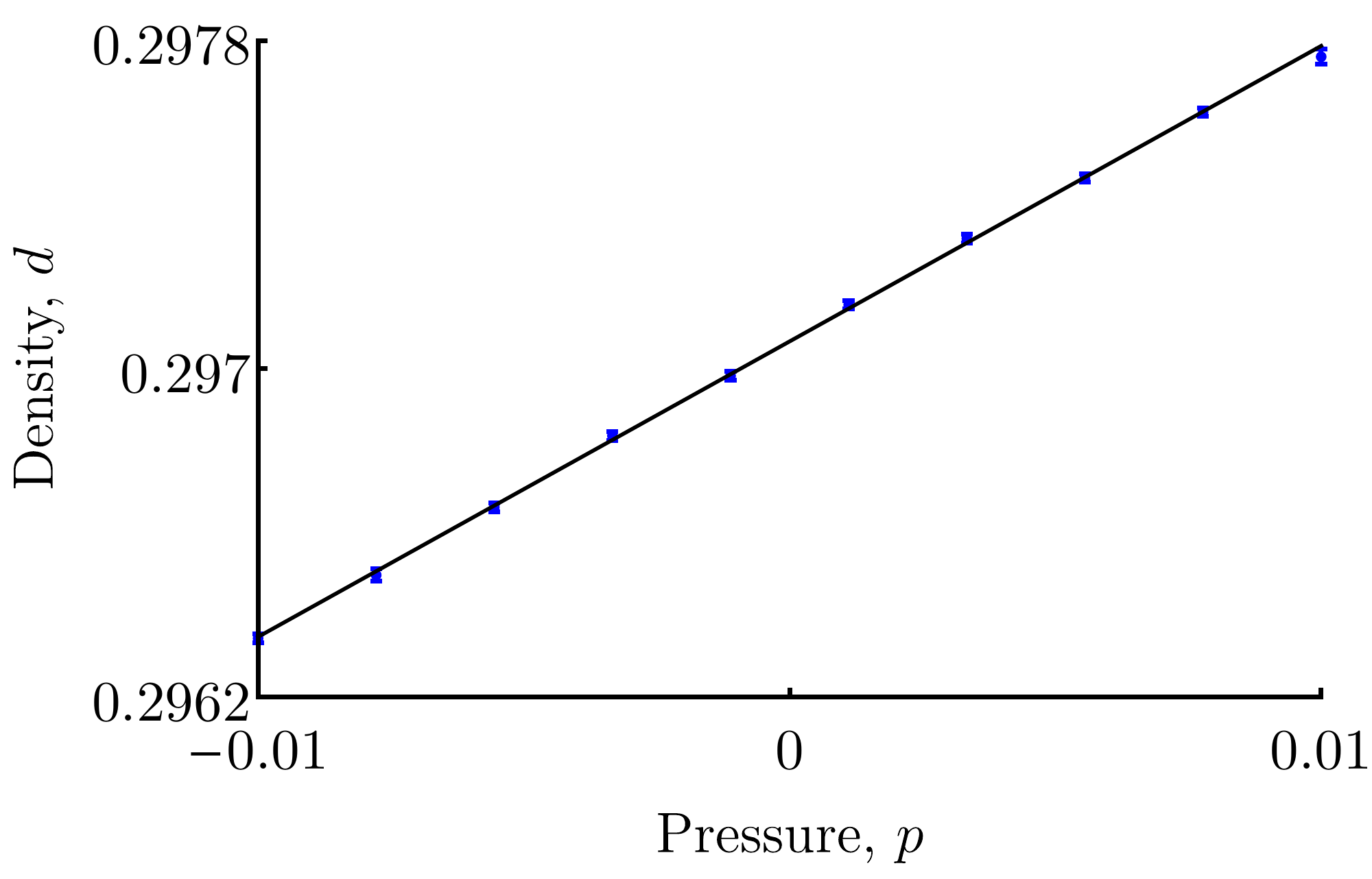}
\label{fig:pressureDensitySmallerRange}
}

\subfloat[][]{ %
\centering
\includegraphics[width=\columnwidth]{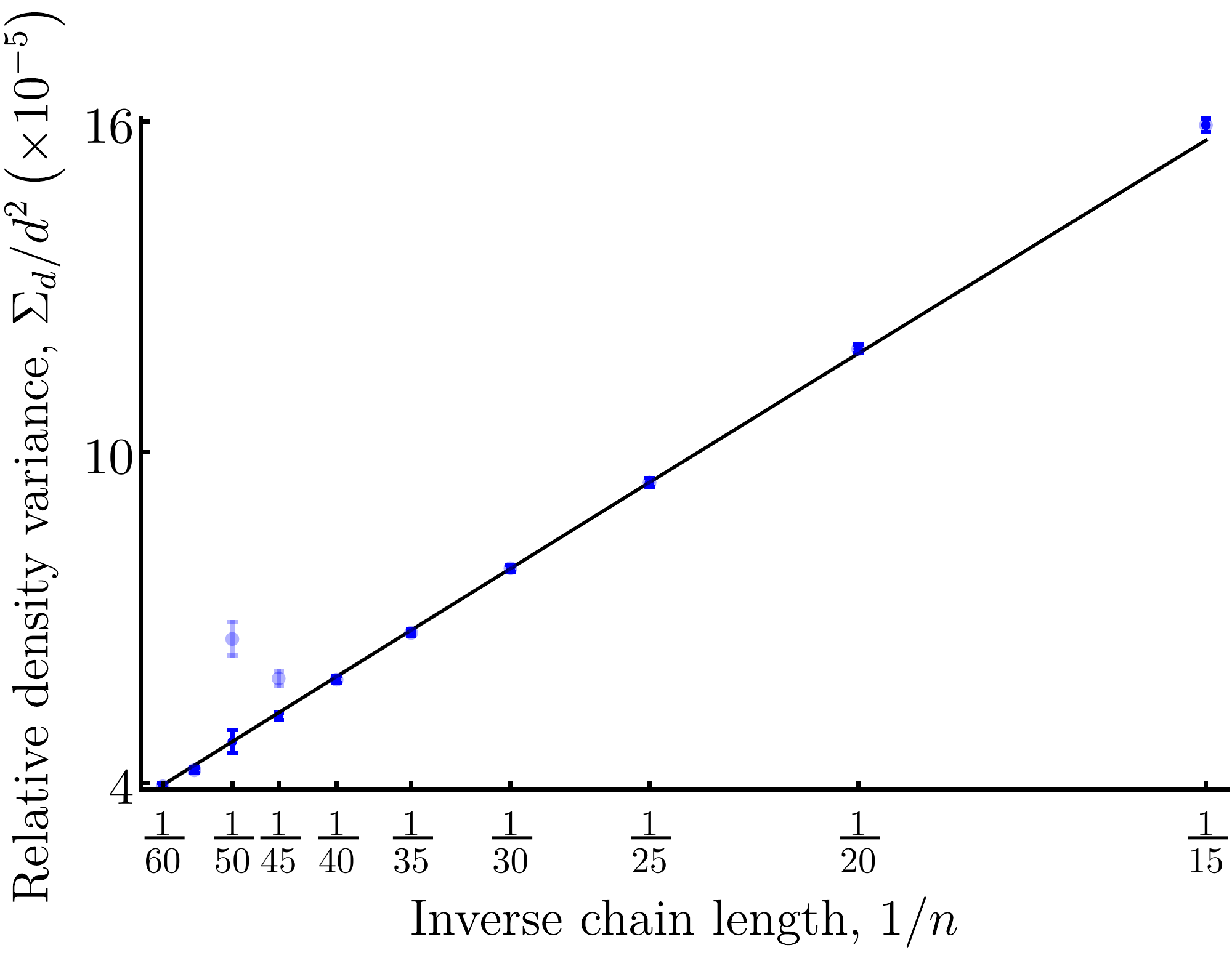}
\label{fig:chainLengthCompressibilityBlack}
}

\caption{(Color online) 
Two independent fits to determine the compressibility $\beta$. 
In \protect\subref{fig:pressureDensitySmallerRange}, the density $d$ of the homopolymer system of \cref{fig:bulksnapshot} with 30 polymer chains and 15 beads per chain is plotted against the applied pressure (data points).
The solid line shows a fit of the functional form $d(p)=d_0\left(1+\beta p\right)$ to the data, which yields a compressibility $\beta$ of $0.242 \pm 0.002$, and reduced chi-squared of $0.65$.
In \protect\subref{fig:chainLengthCompressibilityBlack}, data from the same simulations of \cref{fig:chainLengthDensity} are plotted.
The variance of the sampled densities, normalized by the square density $d(n)^2$ found from the fit of \cref{fig:chainLengthDensity}, are plotted against the reciprocal of the chain length. 
Two systems having $45$ and $50$ beads per chain produced anomalous results, which could not be reproduced after several attempts.
These anomalous results are shown as partially transparent data points, and the solid data points at the same chain lengths are results from representative repeats of the simulation.
A fit of the expected functional form $\beta d(n)/(N_c n)$ to the data excluding the outliers was performed, where
$N_c$ is the number of chains (namely, $30$) in the simulation.
The fit gives a best-fit compressibility of $0.2372 \pm 0.0007$, and the chi-squared of the fit is $1.4$.
}
\label{fig:compressibilityComparison}
\end{figure}

Having found the equilibrium density at zero pressure, we may also find the equilibrium density at finite pressure.
The response of the density to an applied pressure is characterized by the compressibility, denoted $\beta$.
Alternatively, $\beta$ may be determined by analyzing the density fluctuations at constant pressure.
If the simulation produces a proper Boltzmann ensemble, these two ways of determining the compressibility ought to agree.
Indeed, the compressibilities determined from these two methods do in fact agree, as can be seen in \cref{fig:compressibilityComparison}.

In \cref{fig:chainLengthCompressibilityBlack}, there are two outliers exhibiting anomalously large fluctuations, and therefore excluded from the compressibility analysis.
Since these outliers could not be reproduced, they were replaced by data from repeated simulations.
In any case, the outliers represent only small fluctuations in density (less than one percent).
Additionally, the average density observed from these two simulations is consistent with the trend exhibited by the rest of the simulations, as shown in \cref{fig:chainLengthDensity}.
This suggests that the samples are incompletely equilibrated for purposes of determining these small fluctuations, even though they are well equilibrated for determining the density.
Consequently, we conclude that care is needed when the magnitude of, and uncertainty in, a physical quantity's thermal fluctuations.
Accordingly, we provide validation of our estimates of micelle shape fluctuations in \cref{subsec:analysisValidation}.

\begin{figure}
\centering
\includegraphics[width=\linewidth]{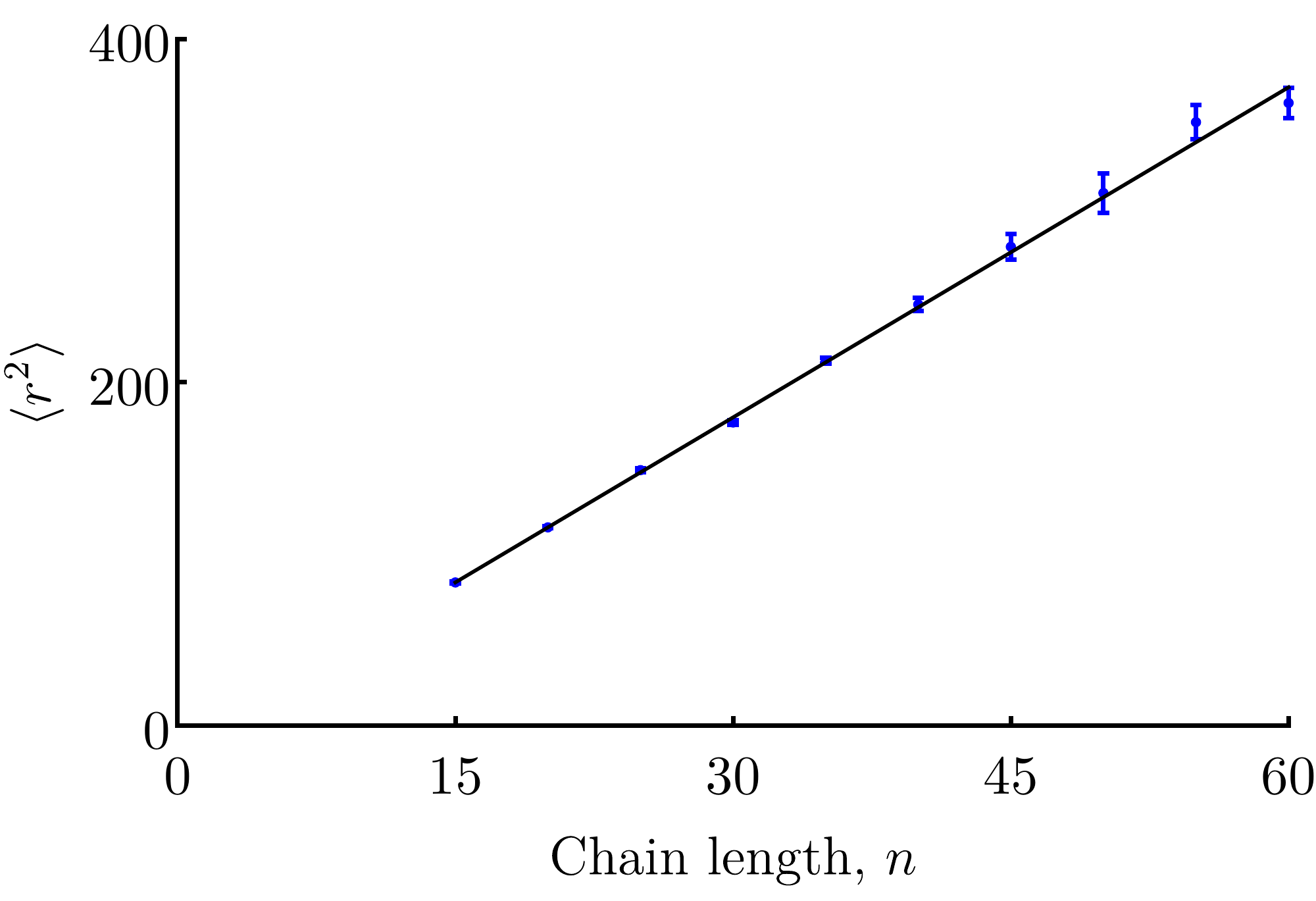}
\caption{(Color online) 
Plot of the mean-square end-to-end chain distance $\langle r^2 \rangle$ as a function of chain length for the same systems of \cref{fig:chainLengthDensity} and \ref{fig:chainLengthCompressibilityBlack}.
The data are fit to a the theoretical expectation given in \cref{eq:msdExpectation}, resulting in a best-fit value of $b^2$ equal to $6.41 \pm 0.05$ and a reduced chi-squared of $1.1$.}
\label{fig:chainLengthMSD}
\end{figure}

In addition to the density, the mean-square end-to-end chain distance $\langle r^2 \rangle$ also has an expected dependence on chain length \cite{Doi1986ChainLength}
\begin{equation}
\label{eq:msdExpectation}
\langle r^2 \rangle(n) = b^2 n + \langle r^2 \rangle_0,
\end{equation}
where $b^2$ is a parameter giving the size of each bead's contribution to the mean-square end-to-end distance, and $\langle r^2 \rangle_0$ is a subleading correction.
As can be seen in \cref{fig:chainLengthMSD}, our data does match this expectation well.

\begin{figure}
\centering
\includegraphics[width=\linewidth]{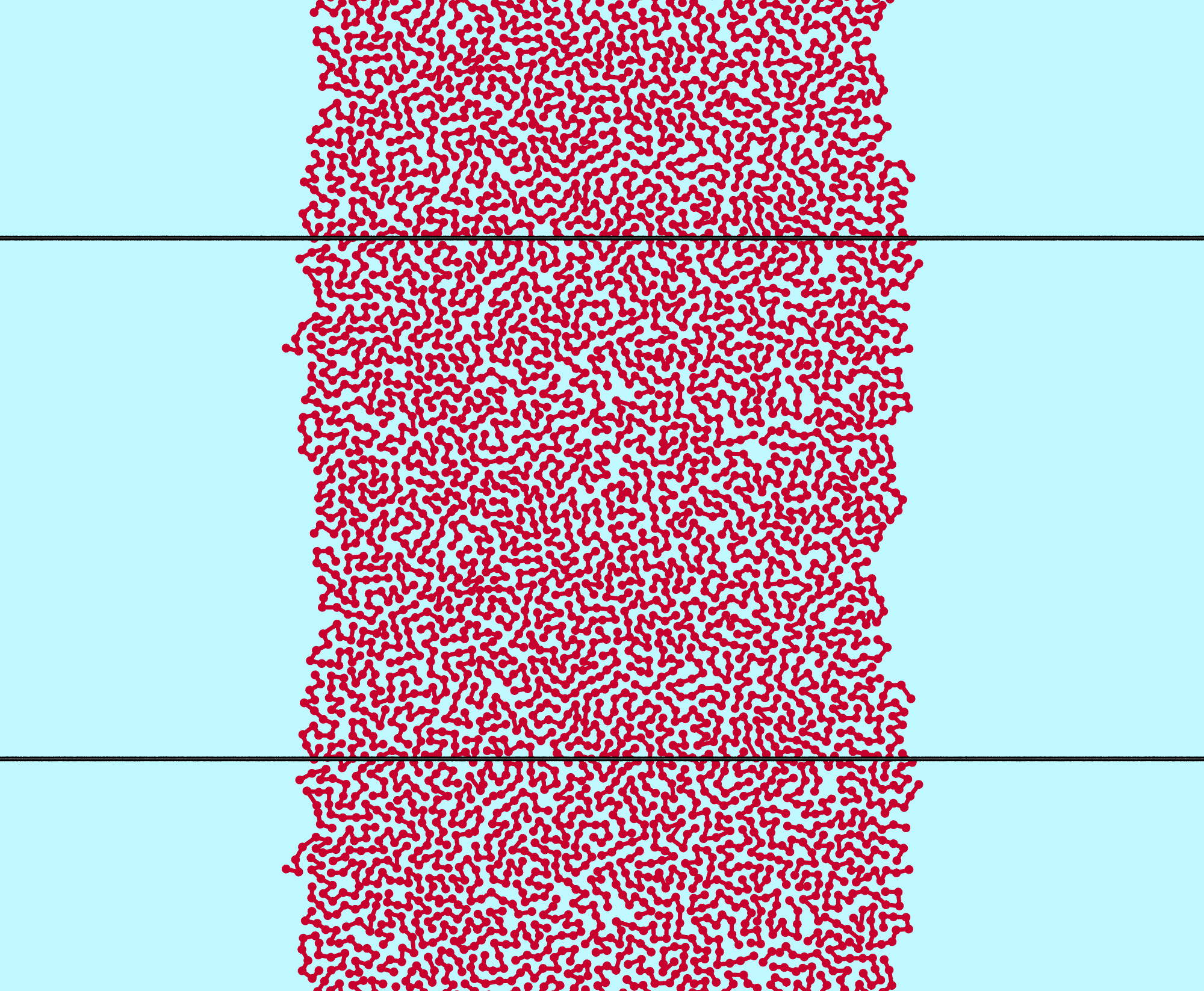}
\caption{(Color online) 
Typical configuration of a strip of homopolymer in (implicit) solvent.
The simulated system contains 143 solvophobic homopolymer chains, each having 15 beads, shown as red disks.
The springs connecting adjacent beads are shown as red segments.
The system is periodic in one direction, and the fixed periodic boundaries are shown as black lines.}
\label{fig:stripsnapshot}
\end{figure}

\begin{figure}
\centering
\includegraphics[width=\linewidth]{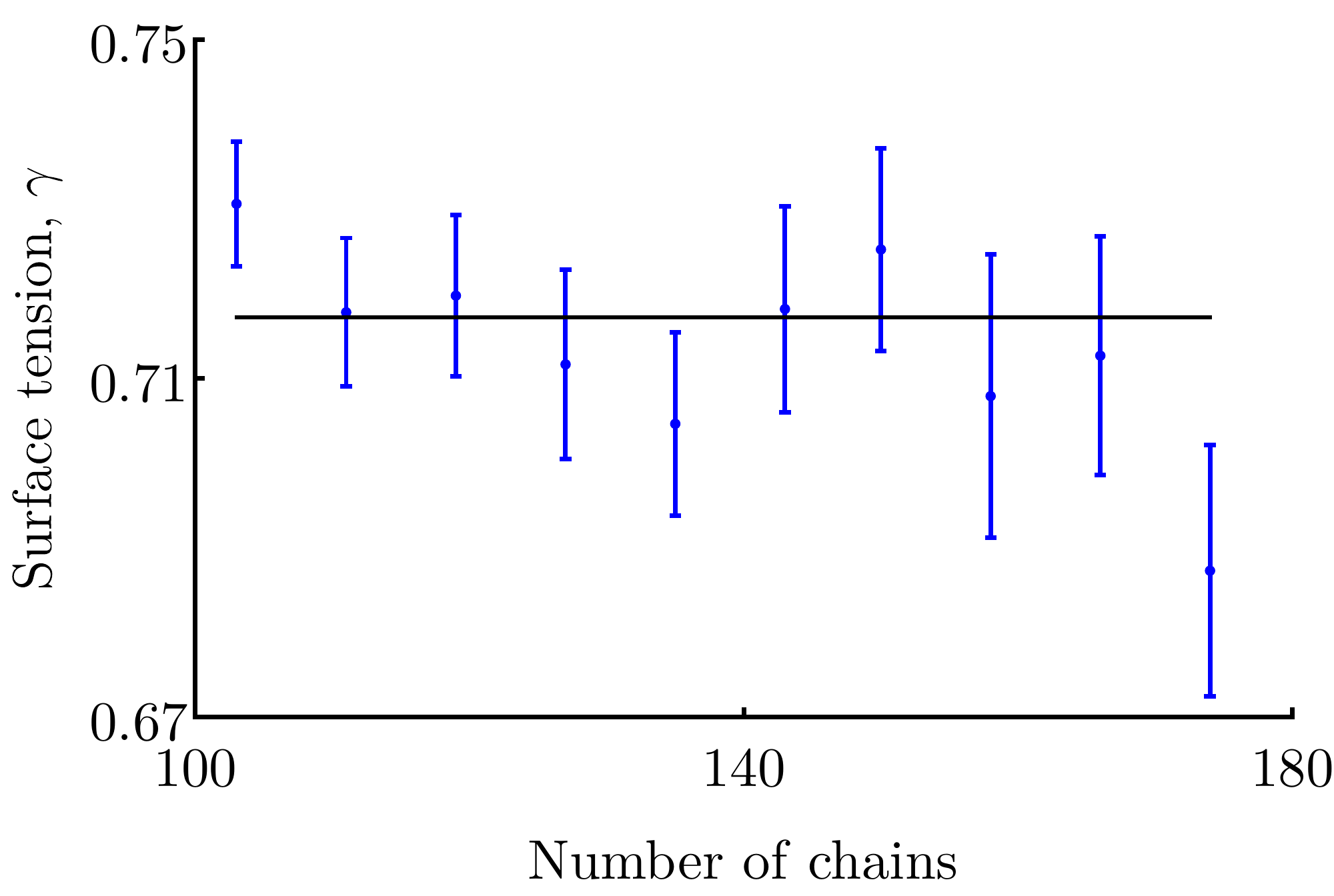}
\caption{(Color online) 
Plot of inferred surface tensions vs number of chains for system simulated in the geometry of \cref{fig:stripsnapshot}.
A best fit to a constant yields a surface tension of $0.717 \pm 0.003$.
The reduced chi-squared of the fit is 1.1, indicating that the data are consistent with the surface tension being independent of the number of chains in the system, as theoretically expected.} 
\label{fig:surfacetensionfit} 
\end{figure}

The final property of our system considered here is its interfacial tension with the solvent, which we determine using a simulation cell of fixed volume and two free surfaces, as shown in \cref{fig:stripsnapshot}.
Since the simulation has two surfaces, the surface tension is half of the force transmitted across the simulation cell.
Because we know the inter-bead forces as a function of bead position, calculating the transmitted force in the simulation is a simple matter.
We expect a well-defined surface tension independent of the number of chains in the simulation, and indeed this is what we find, as can be seen in \cref{fig:surfacetensionfit}.

\newcommand{\kuhnLength}{\ell_K}
Having verified that the homopolymer melt behavior matches theoretical expectations, we may ask if the actual values of the density, compressibility and mean-square end-to-end distance per monomer are similar to those of real polymer.
We compare to poly(dimethylsiloxane) at room temperature.
To form a basis for comparison, we must make contact between simulation units and the physical units in which PDMS is measured.
One correspondence of units may be set by identifying the temperature of the simulation with the room temperature.
Another correspondence can be made by identifying the Kuhn length \cite{Doi1986KuhnLength}, which we denote by $\kuhnLength$, of the model system with that of PDMS.
To compare the number density of the simulated system with PDMS, it is necessary to specify the number of dimethylsiloxane monomers corresponding to one simulated bead.
We make the choice that one Kuhn segment in the simulation ought to correspond to one Kuhn segment of PDMS.
The mean square bond length $l^2$ of our simulation is $3.2$, so the number of beads per Kuhn segment $b^2/l^2$ is equal to $2.0$, and the Kuhn length $b^2/l$ is $3.6$.
Using the correspondences we have just described, physical properties of our simulation and those of PDMS are tabulated in \cref{tab:peoComparison}.

From \cref{tab:peoComparison}, we see that the density, compressibilities, and surface tensions of our simulated system are all on the same order of magnitude of those of PDMS, validating that our simulated system has properties similar to real polymer.
Further, the nature of the difference of the two systems' properties can be partially explained:
our simulated system's compressibility is higher than that of PDMS, and its surface tension, lower.
This is explained by our tuning of the interaction parameters of our model to create a ``soft" system with a low energy barrier for bead rearrangements, leading to shorter simulation times.
With this in mind, we conclude that our model polymer is reasonably similar to PDMS, if a little softer.

\begin{table}
\begin{tabular}{lccc}
Quantity & Unit & \makecell{PDMS\\ value} & \makecell{Value in\\ simulation} \\
\hline
\makecell{Kuhn segment\\ density} & $\kuhnLength^{-d}$ & $2.5$ & $1.9$ \\[.3cm]
Compressibility & $\dfrac{\kuhnLength^d \times 10^{-3}}{ k_B T}$ & $4.0$ & $19 $\\[0.5cm]
\vspace{1em}
Surface tension & $\dfrac{k_B T}{\kuhnLength ^{d-1}}$ & $5.6$ & $2.6$

\end{tabular} 
\caption{Properties of our simulated homopolymer system compared with those of poly(dimethylsiloxane).
Since only comparison of nondimensional ratios are meaningful, we express is property in a system where the unit of length is the Kuhn length $\kuhnLength$, the unit of energy is thermal energy $k_B T$, and the amount of polymer is measured by the number of Kuhn segments.
Nondimensionalized this way, the values of the density are similar, but our simulated system has a lower surface tension and higher compressibility
The physical properties of PDMS needed to calculate the values in this table may be found in~\cite{crcHandbook,pdmsPolymerHandbook,physPropsOfPols}.
}
\label{tab:peoComparison}
\end{table}

\subsection{Micelle design}
\label{subsec:micelleDesign}

Having demonstrated the validity of our simulation method, we now describe how it is applied to demonstrate our shape-design mechanism.
Specifically, we discuss how the shape-designed micelle is represented in the simulation. 
Recall that the goal of this paper is to create a micelle with a concave dimple by constructing it from multiple diblock species. 
In \cref{subsec:scope}, we identified two challenges in achieving this goal.
The first challenge is selecting diblock species to produce the desired shape, and the second challenge is to ensure the diblocks have their intended positioning over the micelle surface.

First, we describe how the first challenge is addressed in our model.
For simplicity, we construct the micelle using just two diblock compositions.
\newcommand{\nin}{n_{\textrm{int}}}
\newcommand{\nex}{n_{\textrm{ext}}}
A diblock composition is characterized by the number of its solvophilic and solvophobic beads, denoted $\nex$ and $\nin$, respectively.
As explained earlier, we expect a diblock containing relatively more solvophilic beads to prefer a more convex curvature, and a diblock containing relatively more solvophobic beads to prefer a more concave curvature.
For a pragmatic measure of the relative prevalence of either species of bead in a diblock, we introduce the ``asymmetry ratio" $r$ of a diblock, given by:
\begin{equation}
\label{eq:asymmetryRatio}
r=\frac{\nex - \nin}{\nex+\nin}.
\end{equation}
The asymmetry ratio is zero for diblocks having an equal number of solvophobic and solvophilic beads, and it is $1$ or $-1$ for polymers made purely out of solvophilic or solvophobic beads respectively.
We expect a positive correlation between the asymmetry ratio of a diblock and its preferred curvature. 
Therefore to implement our shape-design mechanism, we should position higher asymmetry ratio diblocks (we call these diblocks ``solvophilic-rich") along the most of the surface of the micelle, and lower asymmetry ratio diblocks (we call these diblocks ``solvophobic-rich") where the dimple is intended to be.

\newcommand{\ncore}{n_{\textrm{core}}}

In addition to the diblocks, we add one more ingredient to the micelle: a solvophobic homopolymer chain, which we call a ``core chain".
The number of beads in the core chain gives an additional degree of control over the size of the micelle, which, as will be shown in \cref{subsec:shapeFeaturesResult}, affects other properties of interest.
In our simulations, we chose specific block lengths for the core chain and the diblocks; these are given in \cref{sec:results}.

\begin{figure}
\centering
\includegraphics[width=\linewidth]{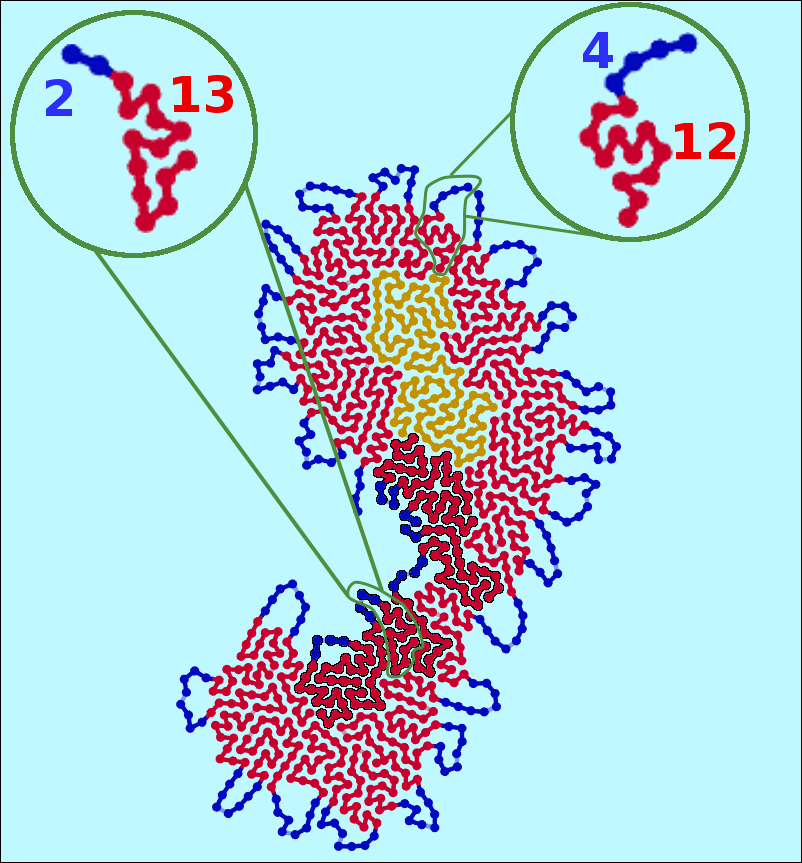}
\caption{(Color online) 
Illustration of how a micelle designed to have a dimple is constructed in our model.
The micelle contains both species of beads present in our model: solvophobic (shown in red or tan) and solvophilic (shown in blue).
The micelle consists of a long chain of solvophobic beads shown in tan, and a collection of diblocks. 
There are two species of diblocks: a ``solvophobic-rich" species with two solvophilic beads and thirteen solvophobic beads (outlined with black), and a ``solvophilic-rich" species with four solvophilic beads and twelve solvophobic beads.
These diblocks are joined end to end.
The dimple is intended to appear in the region occupied by the solvophobic-rich diblocks, as shown (indeed, the above configuration was taken from a simulation).
}
\label{fig:frameWithInsets}
\end{figure}

Next, we describe how the second challenge is addressed.
We constrain the diblock positions by introducing additional bonds joining the diblocks end to end so that they form a linear multiblock copolymer, as shown in \cref{fig:multiblockdiblock}.
It remains to specify the order of the core chain and the two species of diblocks within the multiblock copolymer.
The core chain appears on one end of the multiblock, followed first by all the solvophobic-rich diblocks and then the solvophilic-rich diblocks.
An example of a micelle formed by a multiblock copolymer so constructed is shown in \cref{fig:frameWithInsets}.

\subsection{Simulation initialization}
\label{subsec:initialization}

\begin{figure}
\centering
\includegraphics[width=\linewidth]{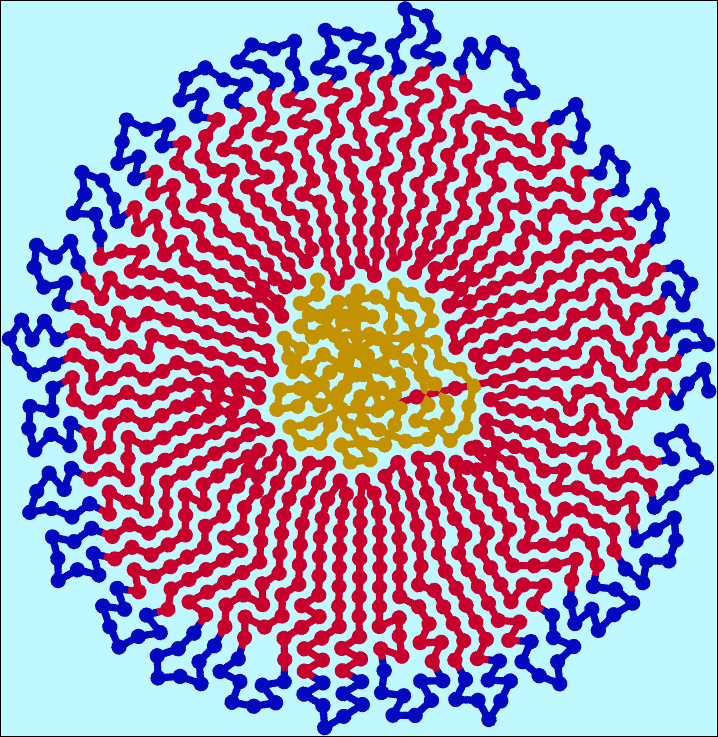}
\caption{(Color online) Initial relaxed configuration of simulated model shape-designed micelle, as described in the text.
Color coding is as in \cref{fig:frameWithInsets}.
The resulting micelle configuration has the intended topology and surface diblock ordering, but is not biased toward its intended dimpled shape.} 
\label{fig:initializationcropped}
\end{figure}

To begin a molecular dynamics simulation of a micelle, an initial configuration is required.
We generate initial micelle configurations (see \cref{fig:initializationcropped}) the following way: we begin by initializing the core chain as a random walk starting at the origin with step length similar to the average bond length.
Next, each diblock is initialized in a straight line pointing away from the origin.
The solvophobic ends of each diblock are evenly spaced on a circle centered at the origin, whose area is equal to the area occupied by the core chain as calculated from the equilibrium homopolymer density.
After this configuration is constructed, a conjugate gradient minimization of the micelle energy is performed to relax any extreme forces that may arise due to unnaturally large bond stretching or bead overlaps.
The relaxed configuration produced by the energy minimization determines the initial bead positions for the LAMMPS simulation.
Since we perform molecular dynamics, the initial velocities must be specified in addition to the initial positions.
These initial velocities are drawn from a Boltzmann distribution having the same temperature as the simulation thermostat.
In this initialization scheme, there are two sources of randomness which ultimately lead to variable results from the otherwise deterministic simulation procedure: the first is the random initial velocities just discussed; the second is the random initial configuration of the core chain.

This initialization was chosen because, while it does not bias the micelle shape towards forming a dimple, the initial configuration is well-formed, meaning the diblocks have the intended relative positioning on the micelle surface.
Since we are interested in studying only well-formed micelles, we might as well initialize the micelles in such a configuration.
Indeed, if our multiblock copolymer is initialized in a less favorable configuration, (say, a random walk or linear configuration), we find that it does not self-assemble into a well-formed configuration during the course of a simulation, consistent with the results of \cite{LEWANDOWSKI20091289}. 

\section{Analysis}
\label{sec:analysis}
As the simulation runs, it records information about the micelle configuration at regular intervals of simulation time.
In this section we discuss how this output of the molecular dynamics simulations is analyzed.
Since the goal of this work is to create single-polymer micelles of a designed shape, one goal of the analysis to determine the average micelle shape from the simulation output.
As with any simulation, it is only possible to determine properties of the system, in our case the average shape, to a finite precision; accordingly, we also make an estimate of the uncertainty in the micelle shape.
In addition to the average micelle shape, we explained that the fluctuations in the shape are of interest, and so another goal of the analysis is to determine these fluctuations.

In \cref{subsec:average}, we formalize what we mean by micelle ``shape", and we describe how the average shape is determined.
In \cref{subsec:shapeVariance}, we discuss how we determine both the shape fluctuations and our uncertainty in the average shape.
Next, in \cref{subsec:misshapenMicelles} and \cref{subsec:combiningRuns}, we say how we combined multiple independent simulations to get a best estimate of the quantities of interest.
In \cref{subsec:consistency}, we describe how the results of these independent simulations can be compared with each other to test for consistency.
Finally, in \cref{subsec:shapeFeatures}, we define two scalar properties---one characterizing the average shape and one, the shape fluctuation---in order to distill the geometric features we are most interested in.

\subsection{Average shape from simulation run}
\label{subsec:average}
\newcommand{\br}{\mathbf{r}}
\newcommand{\bbr}{\mathbbm{r}}
\newcommand{\bbbr}{\bar{\bbr}}
\newcommand{\cbbbr}{\check{\bbbr}}

Since we are only interested in the shape of the micelle, we do not record the position of every bead.
Instead, we record only the location of the junction points: the midpoint of two beads that are adjacent along the polymer backbone but have opposite solvophobicity.
A micelle has only as many junction points as diblocks it was constructed from.
For the remainder of this paper, we define the term ``micelle shape" to mean the ordered sequence of junction points.
As such, a micelle shape $\bbr$ has the mathematical form
\begin{equation}
\label{eq:shapeDefinition}
\bbr=\left( \br_1,\br_2,\dots,\br_n,\dots,\br_{N_j} \right),
\end{equation}
where $N_j$ is the number of junction points in the micelle, and each $\br_n$ is itself a two-dimensional spatial vector representing the $n$th junction point's position: 
\begin{equation}
\label{eq:spatialVector}
\br_n=\left( r_{n1},r_{n2}\right).
\end{equation}

Since we are interested only in the relative positions of the junction points, and not overall translations of the shape, we assume without loss of generality that each shape $\bbr$ is geometrically centered at the origin so that
\begin{equation}
\label{eq:geometricallyCentered}
\sum_{n=1}^{N_j}\br_n=\mathbf{0}.
\end{equation}

A micelle simulation outputs a time series of micelle shapes; however, the statistical information of interest can be summarized by just one average micelle shape and the fluctuations about this average.
The process of obtaining this average contains some subtlety.
For example, a simple arithmetic average will not suffice because the arithmetic average of a micelle shape and the same shape rotated by $180^\circ$ is a micelle shape with every junction point at the origin.
Instead, since the shapes are merely different by rotations, these two shapes should be considered the same, so that the average would somehow give the same shape back again.

We define an appropriate average in two steps: first we introduce a metric giving the true difference between two shapes, then we define the average of a set of shapes to be the shape that minimizes the sum of these differences.
\newcommand{\dex}{\Delta_{\textrm{ext}}}
\newcommand{\din}{\Delta_{\textrm{int}}}
\newcommand{\bba}{\mathbbm{a}}
\newcommand{\bbb}{\mathbbm{b}}
\newcommand{\bbc}{\mathbbm{c}}
To define the metric, we first define an ``extrinsic" difference metric $\dex(\bba,\bbb)$ between two shapes $\bba$ and $\bbb$ as the sum of square differences of corresponding position components. Introducing some notation, this difference may be written as follows
\begin{equation}
\label{eq:dex}
\dex(\bba,\bbb) = \left(\bba -\bbb\right)^2,
\end{equation}
where the square $\bbr^2$ of a shape $\bbr$ is given by the dot product $\bbr \cdot \bbr$ of the shape with itself, and the dot product of two shapes is defined as the sum of dot products of corresponding junction points:
\begin{equation}
\label{eq:shapeDotProduct}
\bba \cdot \bbb = \sum_{n=1}^{N_j} \mathbf{a}_n \cdot \mathbf{b}_n.
\end{equation}
An illustration of this distance metric is given in \cref{fig:dextPicture}.

\begin{figure}

\subfloat[][]{ %
\centering
\includegraphics[width=\columnwidth]{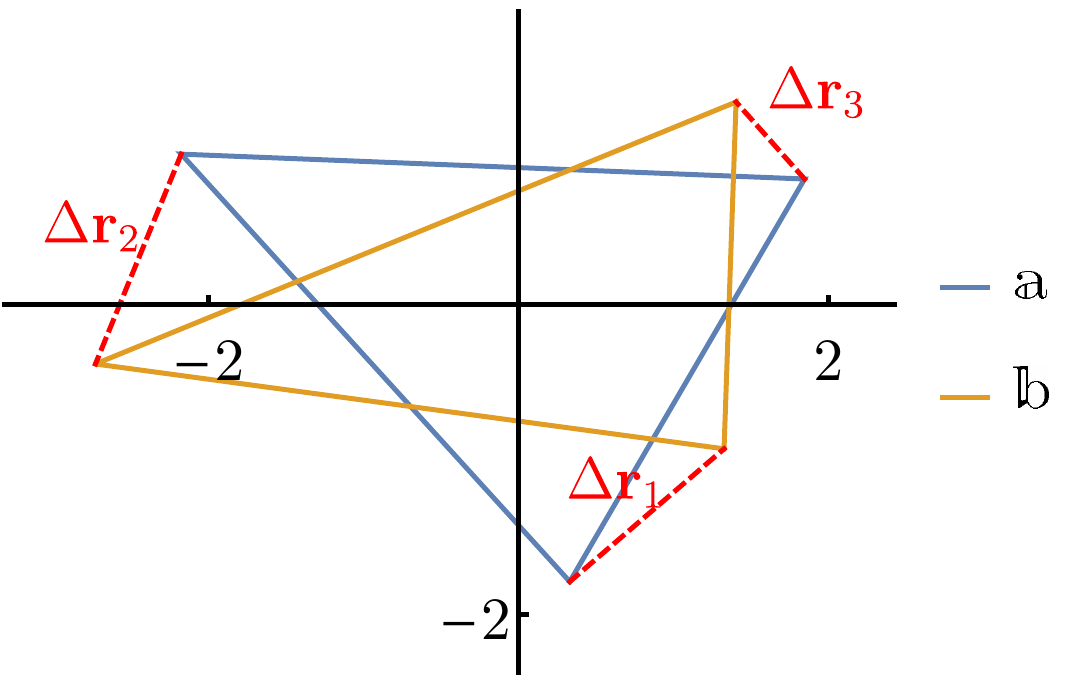}
\label{fig:dextPicture}
}

\subfloat[][]{ %
\centering
\includegraphics[width=\columnwidth]{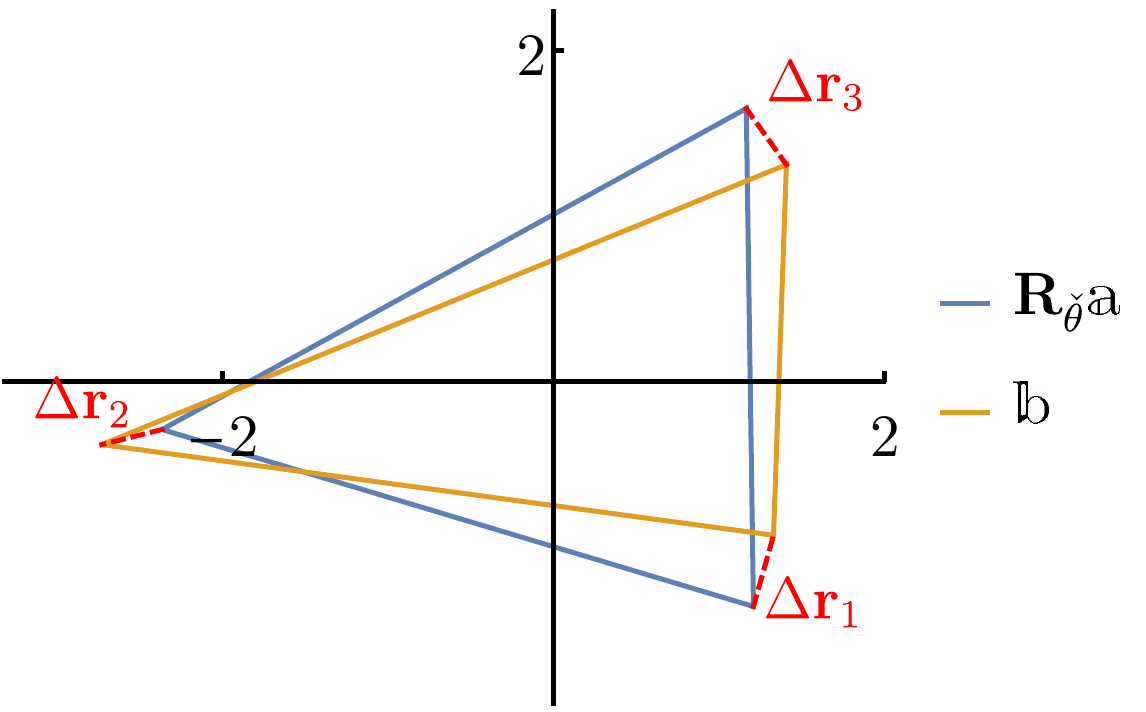}
\label{fig:dintPicture}
}

\caption{(Color online)
Illustration of the difference metrics $\dex$, defined in \cref{eq:dex} and $\din$ defined in \cref{eq:dinDefinition}.
In \protect\subref{fig:dextPicture}, two shapes, $\bba$ and $\bbb$ are shown, together with the displacement vectors $\Delta \mathbf{r}_1$, $\Delta \mathbf{r}_2$, and $\Delta \mathbf{r}_3$ connecting corresponding vertices of the two shapes.
As specified by \cref{eq:geometricallyCentered}, both shapes are geometrically centered on the origin.
The difference $\dex$ is given by the sum of square lengths of these vectors: $\dex(\bba,\bbb) = \sum_{i=1}^{3} \Delta \mathbf{r}_i^2$.
In \protect\subref{fig:dintPicture} the same two shapes are shown, except $\bba$ is rotated about the origin by the angle $\check{\theta}$ that minimizes $\dex$.
Consequently, the displacement vectors $\Delta \mathbf{r}_i$ are clearly smaller here than in \protect\subref{fig:dextPicture}.
The intrinsic difference $\din$ is defined as this minimum value of $\dex$.
}
\label{fig:dextdintPicture}
\end{figure}

The difference metric $\dex$ has a shortcoming: it has a nonzero value when evaluated on two shapes differing only by a rotation.
Since there is no natural frame in which the shapes are defined, we desire a metric which is insensitive to rotation of either of its arguments.
To this end, we define an ``intrinsic" difference metric $\din$ defined as the minimum of $\dex$ with respect to rotations of one of its arguments.
\newcommand{\bR}{\mathbf{R}}
\newcommand{\bRt}{\bR_\theta}
The action of a spatial rotation 
\begin{equation}
\label{eq:rotationMatrix}
\bRt = \begin{pmatrix}
\cos \theta & -\sin \theta \\
\sin \theta & \cos \theta
\end{pmatrix}
\end{equation}
on a shape $\bbr$ is defined junction point by junction point:
\begin{equation}
\label{eq:rotationAction}
\left( \bRt \bbr \right)_n = \bRt \left(\br_n\right).
\end{equation}
Then the intrinsic difference $\din$ has been defined by 
\begin{equation}
\label{eq:dinDefinition}
\din(\bba,\bbb) = \min_\theta \dex\left(\bRt \bba, \bbb\right).
\end{equation}
An illustration of the intrinsic difference $\din$ is given in \cref{fig:dintPicture}.

It can be shown\footnote{The proof is simple.
$\Delta_{\textrm{ext}}$ in \cref{eq:dinDefinition} is minimized when $\left(\bRt \bba\right) \cdot \bbb $ is maximized.
Now $\bRt \bba$ may be written as $ \bba \cos \theta + \wedge \bba  \sin \theta$.
Using this representation, the product $\left(\bRt \bba\right) \cdot \bbb $ can be transformed into a two dimensional dot product $\left(\cos \theta, \sin \theta \right) \cdot \left(\bba \cdot \bbb, \bba \wedge \bbb \right)$.
This product is maximized when the unit vector $\left(\cos \theta, \sin \theta \right)$ points in the direction of $\left(\bba \cdot \bbb, \bba \wedge \bbb \right)$, hence \cref{eq:vectorForMinTheta}.
With no more effort, we see that the maximum value of the dot product is the length of the same vector, leading to \cref{eq:dinformula}. }
that the angle $\check{\theta}$ minimizing \cref{eq:dinDefinition} is the signed angle that the two dimensional vector 
\begin{equation}
\label{eq:vectorForMinTheta}
\left(\bba \cdot \bbb, \bba \wedge \bbb \right)
\end{equation}
makes with the vector $\left(1,0\right)$, 
where the wedge product of two shapes is defined by
\begin{equation}
\label{eq:shapeWedgeProduct}
\bba \wedge \bbb = \left(\wedge \bba \right) \cdot \bbb,
\end{equation}
and where the wedge $\wedge \bba $ of a shape $\bba$ is simply the result of rotating it by $\pi/2$:
\begin{equation}
\label{eq:wedgeDefinition}
\wedge \bba =\bR_{\pi/2} \bba.
\end{equation}
Thus e.g., if $\bbb = \bba$, then $\bba \cdot \bbb$ is positive and $\bba \wedge \bbb=0$, so that $\check{\theta}=0$.
Indeed of $\bba$ and $\bbb$ are any two aligned shapes, so that $\check{\theta}=0$, then $\bba \cdot \bbb$ must be positive and $\bba \wedge \bbb$ must be zero.
If instead $\bbb=\wedge \bba$, then $\bba \cdot \bbb=0$ and $\bba \wedge \bbb$ is positive, so that $\check{\theta}=+\pi/2$.
\newcommand{\cbba}{\check{\bba}}
It can be shown that the intrinsic difference between two shapes can be calculated explicitly as
\begin{equation}
\label{eq:dinformula}
\din(\bba,\bbb) = \bba^2+\bbb^2-2\sqrt{\left(\bba\cdot\bbb\right)^2+\left(\bba\wedge\bbb\right)^2}.
\end{equation}

If the shape $\bba$ is then rotated by the minimizing angle $\check{\theta}$, then we say $\bba$ has been aligned with $\bbb$.
(In general we will use a ``\,\,$\check{ }$\,\," to indicate that a quantity has been somehow aligned.)
We note that, somewhat counter-intuitively, the relationship of being aligned is not transitive: if $\bba$ is aligned with $\bbb$, and $\bbb$ is aligned with some third shape $\mathbb{c}$, $\bba$ is typically not aligned with $\mathbb{c}$.

\newcommand{\cbbr}{\check{\bbr}}

\begin{figure}
\centering
\includegraphics[width=\linewidth]{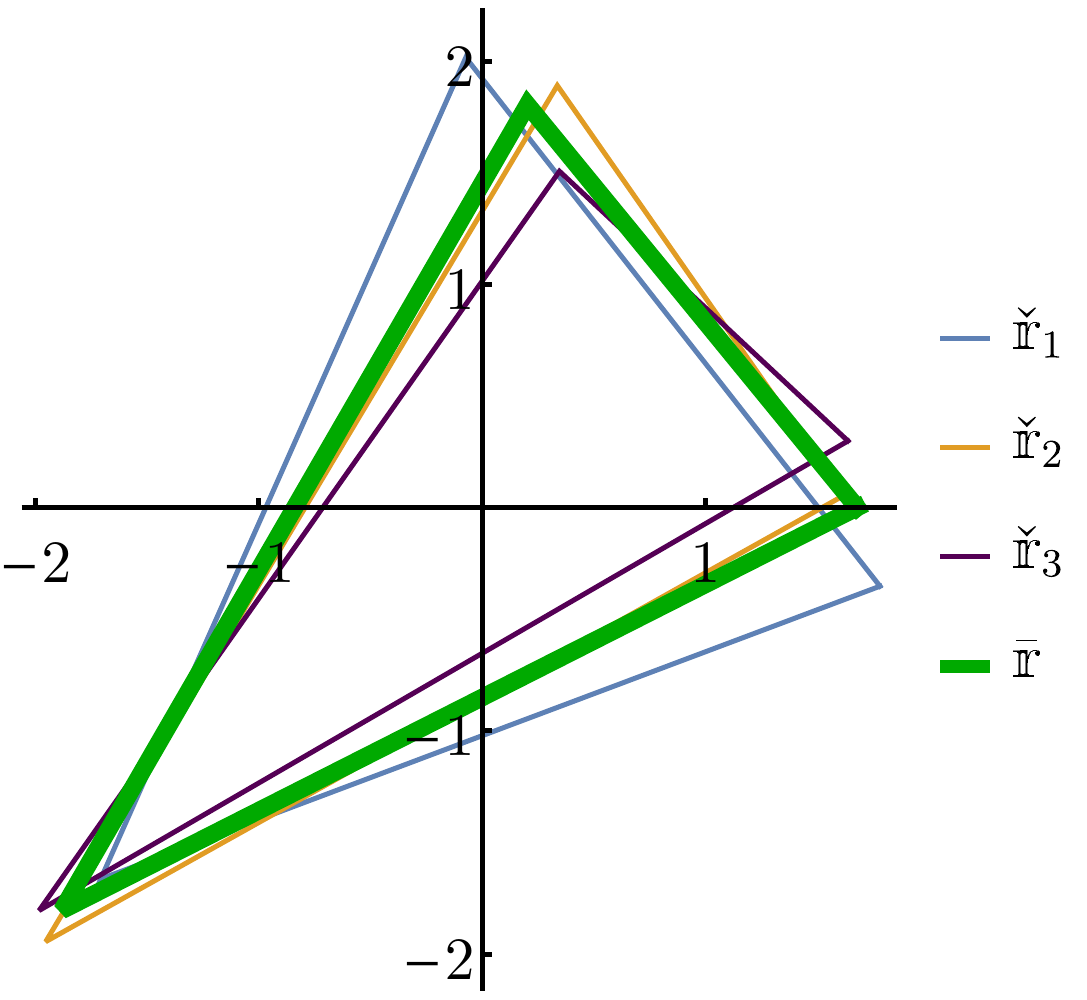}
\caption{(Color online) Illustration of the average $\bbbr$ of three shapes $\bbr_1$, $\bbr_2$, and $\bbr_3$.
The original $\bbr_i$ are not shown; instead, the result $\cbbr_i$ of aligning $\bbr_i$ with $\bbbr$ is shown. For definiteness, we have fixed the otherwise arbitrary orientation of $\bbbr$ by positioning its first junction point on the $x$ axis.
}
\label{fig:alignedAverageExamplePicture}
\end{figure}

Having defined the appropriate notion of shape difference, we now define the average. 
Owing to the nontransitivity of alignment,the definition is more involved than might be expected: it is not possible to simply align all the shapes with each other and then do a simple average, because, as stated above, it is typically impossible for even three shapes to be pairwise aligned with each other.
Instead, given a time series of $N_s$ simulated shape samples $\bbr_\alpha$, $\alpha = 1,2,\dots, N_s$, we define their average $\bbbr$ as the shape that minimizes the sum of $\din$ with the samples\footnote{
We use $\argmin$  to represent the operation of finding the argument which minimizes a function. That is, if $x^*=\argmin\limits_x f(x)$, then $f(x^*)$ is the minimum value of $f$.
}:
\begin{equation}
\label{eq:averageDefinition}
\bbbr = \argmin_{\bba} \sum_{\alpha=1}^{N_s} \din\left(\bba,\bbr_\alpha\right).
\end{equation}
Because $\din$ is defined only up to rotations, this definition of $\bbbr$ is defined only up to rotation.
The rotational degree of freedom can be fixed e.g. by making the first junction point $\bar{\mathbf{r}}_1$ lie on the positive $x$ axis.
Once a choice of orientation of $\bbbr$ has been made, the orientation of each sample $\bbr_\alpha$ can be fixed by aligning it with $\bbbr$; we denote this aligned shape sample $\cbbr_\alpha$.
This method of averaging is illustrated in \cref{fig:alignedAverageExamplePicture}.
To justify our choice of this method of averaging, we show in the \cref{sec:appendix}that it is equivalent to another natural method of averaging.
\newcommand{\ct}{\check{\theta}}

\subsection{Shape variance from simulation run}
\label{subsec:shapeVariance}
In addition to the average micelle shape, we have stated that the shape fluctuations can be used to control the micelle's interactions.
The fluctuations are characterized by a $2N_j \times 2N_j$ variance matrix $\mathbb{\Sigma}$, defined by
\newcommand{\bbs}{\mathbb{\Sigma}}
\begin{equation}
\label{eq:shapeVarianceDefinition}
\mathbb{\Sigma} = \frac{1}{N_s -1} \sum_{\alpha=1}^{N_s} \left(\cbbr_\alpha - \bar{\bbr}\right) \otimes \left(\cbbr_\alpha - \bar{\bbr}\right),
\end{equation}
where the tensor product $\bba\otimes\bbb$ gives the $2N_j \times 2N_j$ second rank tensor that acts on a shape $\bbc$ according to the rule ${(\bba \otimes \bbb) \cdot \bbc = \bba \left( \bbb \cdot \bbc \right)}$.

\newcommand{\bbbs}{\bar{\bbs}}

These shape fluctuations limit the precision to which the mean shape $\bbbr$ is determined.
Naturally, one wants to be able to quantify this precision.
Indeed, to convincingly show that the micelle's shape depends on its composition, as we indeed set out to do, we must show that the variability in the mean shapes cannot be explained only by the uncertainty caused by the limited precision of the simulation technique.
Consequently, it is necessary to estimate this uncertainty in the mean shape.
A naive estimate for the variance matrix $\bbbs$ representing the degree of uncertainty in $\bbbr$ is $\bbs/N_s$.
However, this would underestimate the uncertainty since the shape samples are correlated.
If the system was described by a single correlation time of $\tau$ sampling intervals, then the correlations could be accounted for by using the estimate $\bbs \tau /N_s$.
However, to make matters more complicated, there is no single correlation time describing the correlations in the sample.
Typically we find large scale cooperative fluctuations have longer correlation times than high wavenumber fluctuations.

\newcommand{\ml}{\mathbb{m}_\ell}

To address this complication, we find the autocorrelation time of each fluctuation mode independently, as we now describe in more detail.
A mode $\ml$ is an eigenvector of the variance matrix $\bbs$.
This eigenvector has a corresponding eigenvalue $\Sigma_\ell$, so that 
\begin{equation}
\label{eq:sigmaDecomposition}
\bbs = \sum_{\ell=1}^{2 N_j} \Sigma_\ell \mathbb{m}_\ell \otimes \mathbb{m}_\ell.
\end{equation}
(To fix the ordering of the $\mathbb{m}_\ell$ with respect to $\ell$, we arrange them in decreasing order of $\Sigma_\ell$.)
For each $\ell$, we define time series of mode amplitudes $A_{\ell \alpha}$ given by
\begin{equation}
\label{eq:modeAmplitudes}
A_{\ell \alpha} = \left(\check{\bbr}_\alpha - \bbbr\right) \cdot \mathbb{m}_\ell.
\end{equation}

\begin{figure}
\centering
\includegraphics[width=\linewidth]{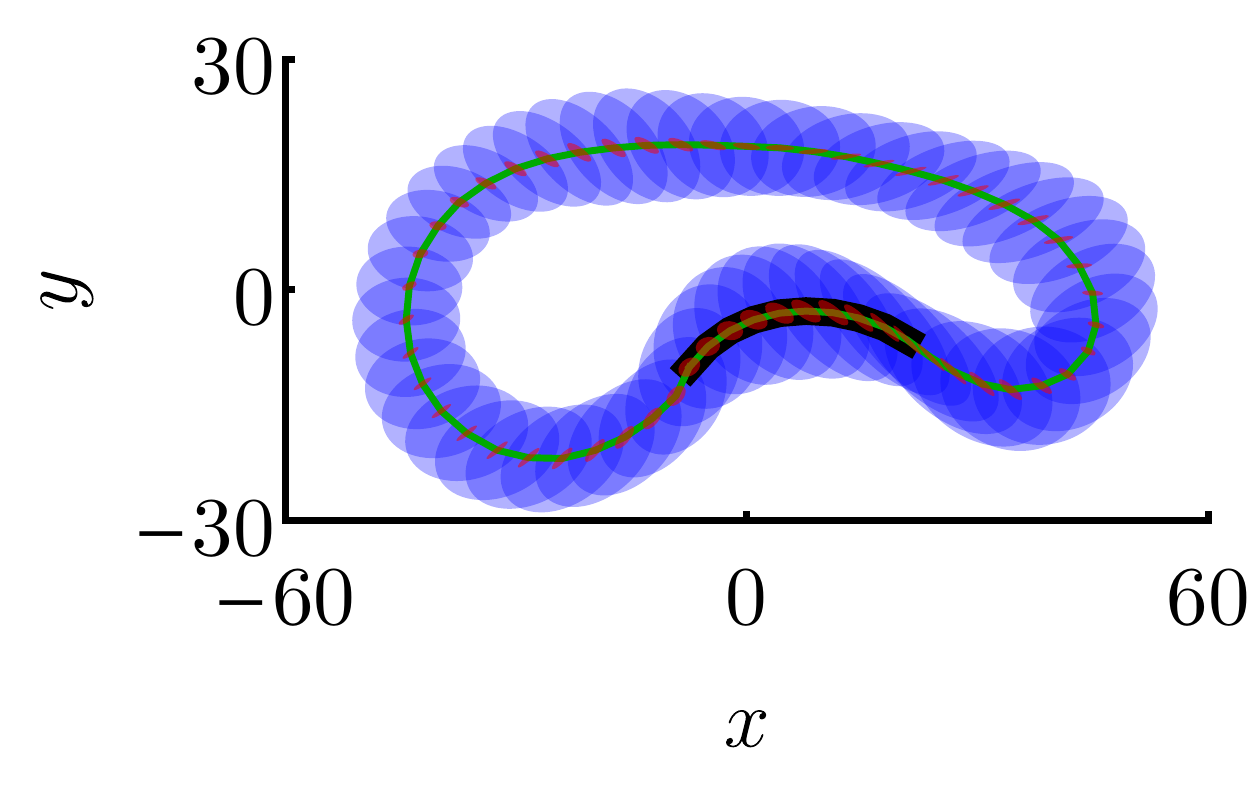}
\caption{(Color online) Graphical representation of an average micelle shape and its fluctuations and uncertainty. 
The green curve linearly interpolates the average position of each diblock junction point. 
The region of the surface occupied by solvophobic-rich diblocks is outlined in black. 
For each junction point, a blue ellipse and a red ellipse is drawn. 
A blue ellipse represents the $40\%$ confidence region, corresponding to one standard deviation from the mean, for a junction point assuming a Gaussian distribution with variance given by the shape variance $\bbs$ of \cref{eq:shapeVarianceDefinition}.
In the same way, a red ellipse represents the variance in the mean shape $\bbbs$ of \cref{eq:varianceInMean}.
}
\label{fig:plotofaverage} 
\end{figure}

Under sufficient simplifying assumptions, the mode amplitudes $A_{\ell\alpha}$ may be used to estimate the uncertainty in $\bbbr$, following canonical methods~\cite{dataReduction}.
Specifically, this estimate of uncertainty assumes that the mode amplitudes $A_{\ell\alpha}$ are sufficiently small that shape fluctuates as a harmonic system in thermal equilibrium and consequently the normal mode amplitudes fluctuate independently.
In such a harmonic system each mode $\ml$ has a characteristic autocorrelation time $\tau_\ell$ (in units of the sampling interval), which we estimate using the initial convex sequence estimator of \cite{geyer2011}.
Samples of the amplitude separated by times longer than $\tau_\ell$ sampling intervals may be viewed as statistically independent, so that the number of independent samples of the $\ell$th mode amplitude is $N_s/\tau_\ell$.
Accordingly, the squared uncertainty in the mean along the $\ml$ direction is estimated to be the variance $\Sigma_\ell$ of the shape distribution along this direction divided by the number of independent samples $N_s/\tau_\ell$.
We therefore estimate the variance matrix $\bbbs$ given the uncertainty in the mean by
\begin{equation}
\label{eq:varianceInMean}
\bbbs = \sum_{\ell=1}^{2 N_j} \frac{\Sigma_\ell}{N_s/ \tau_\ell} \mathbb{m}_\ell \otimes \mathbb{m}_\ell.
\end{equation}
A graphical representation of an average micelle shape together with its variance and variance in mean is shown in \cref{fig:plotofaverage}.

Since this uncertainty estimation ignores uncertainties in the mode eigenvectors $\ml$, and moreover our system's fluctuations may be too large to permit a harmonic approximation, the estimate of \cref{eq:varianceInMean} may be inaccurate.
However in \cref{subsec:consistency}, we will describe a way of validating \cref{eq:varianceInMean} by checking whether the expected uncertainty in $\bbbr$ within a simulation run is consistent with the repeatability of $\bbbr$ over several simulation runs. 
Then in \cref{subsec:analysisValidation}, we will use this validation to show our results are largely consistent with this harmonic scheme.

\subsection{Rejecting malformed micelles}
\label{subsec:misshapenMicelles}
\begin{figure}

\subfloat[][]{ %
\centering
\includegraphics[width=\columnwidth]{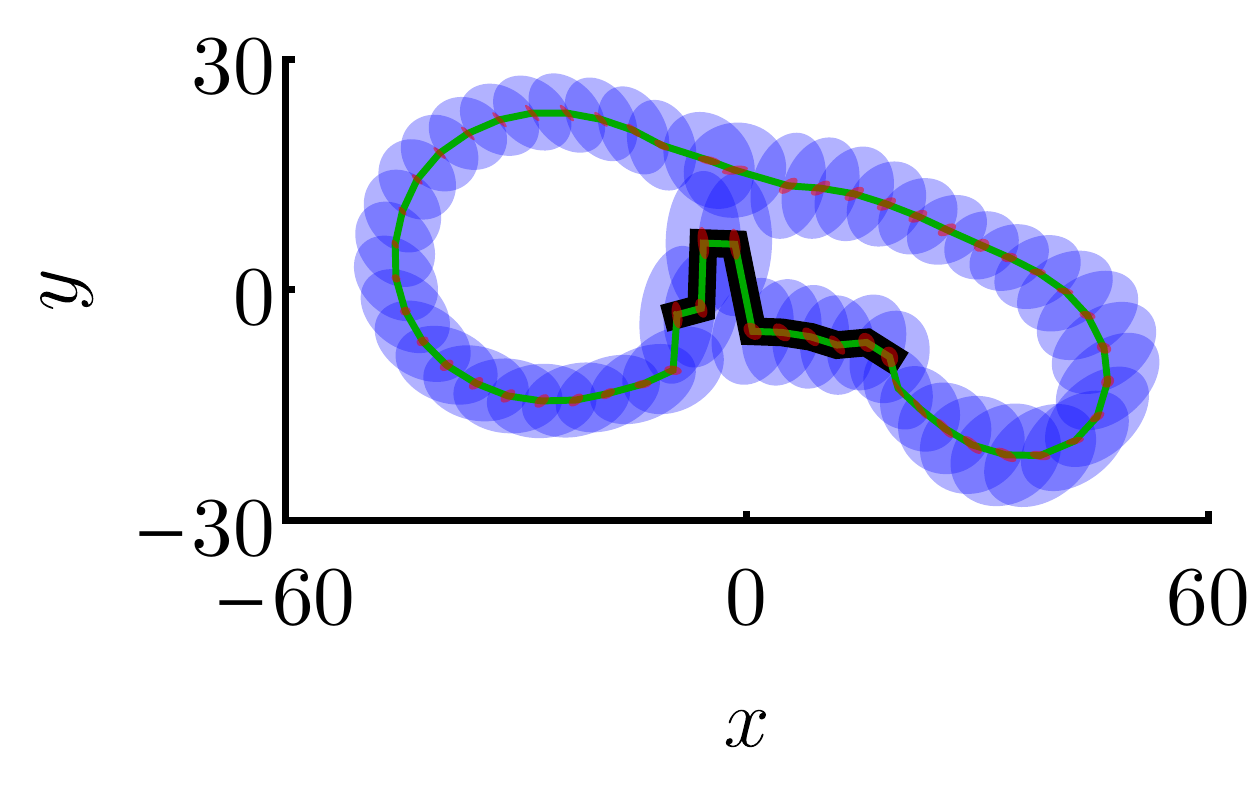}
\label{fig:plotofaveragewithjump}
}

\subfloat[][]{ %
\centering
\includegraphics[width=\columnwidth]{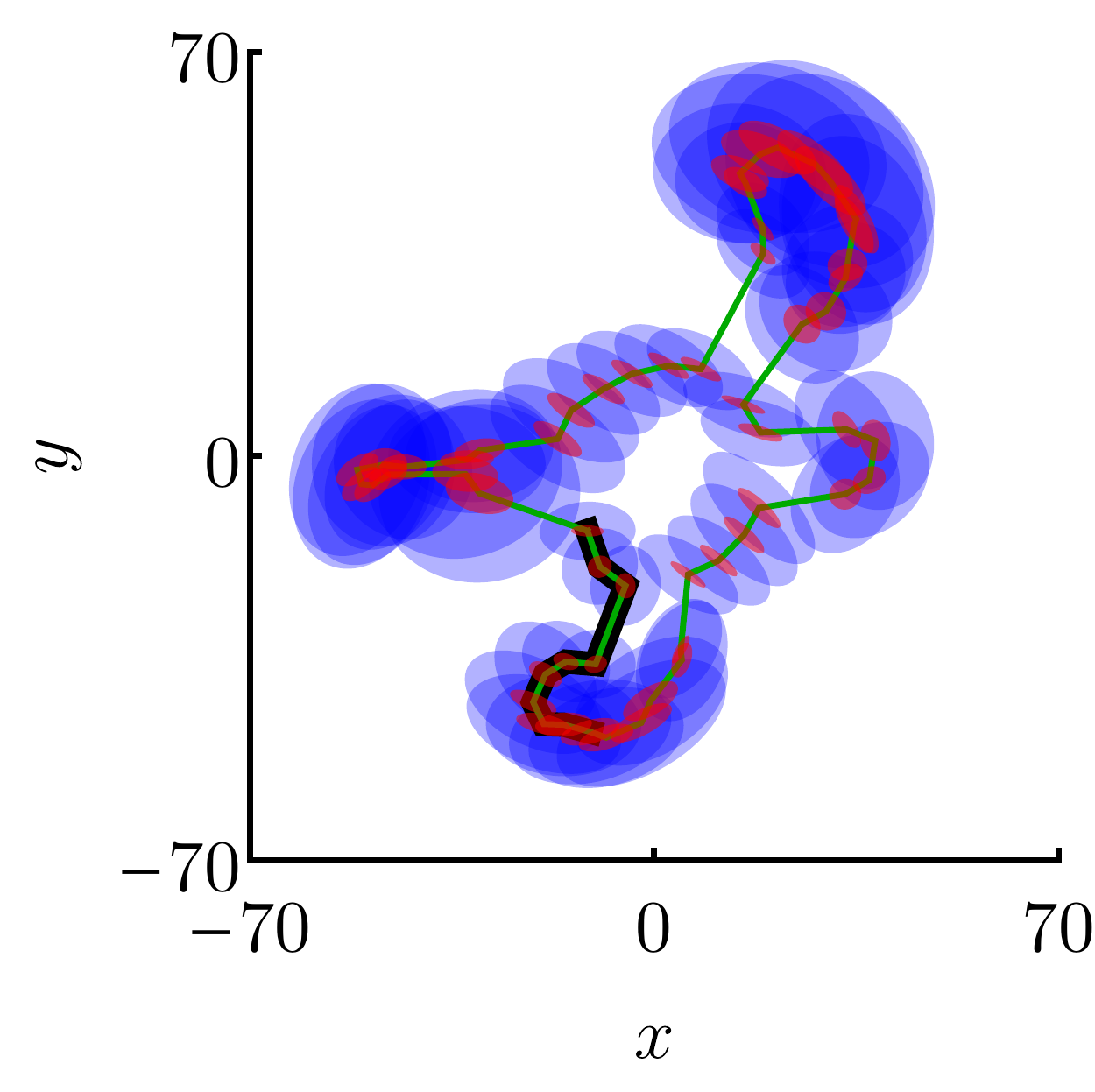}
\label{fig:plotOfBrokenUpAverage}
}

\caption{(Color online) 
Two different types of malformed micelles.
The curves and ellipses have the same meaning as in \cref{fig:plotofaverage}.
In \protect\subref{fig:plotofaveragewithjump}, the average micelle shape contains two junction points in the concave region which approach the opposite surface of the micelle.
This occurs because the junction points crossed to the other side during the simulations.
In \protect\subref{fig:plotOfBrokenUpAverage}, the micelle is broken up into multiple disconnected regions outlined by the junction points.
Such malformed micelles are not expected to take the designed shape, and so they are excluded from our analysis.}
\label{fig:badShapes} 
\end{figure}

In the case shown in \cref{fig:plotofaverage}, the diblock junction points make a smooth curve along the surface of the micelle, indicating that the bonds joining the diblocks were sufficient to make the micelle well-formed. 
However, this is not the case with every simulation.
\cref{fig:plotofaveragewithjump} shows a case where multiple diblock junction points crossed from one side of the micelle to the other.
In addition to this example, we have observed cases where the solvophobic region of the micelle splits into two or more disconnected pieces, as shown in \cref{fig:plotOfBrokenUpAverage}.
Since this work is concerned only with the behavior of well-formed micelles, we simply discard any such results where the average micelle shape is malformed.

If data is to be discarded in a consistent manner, a precise definition of ``well-formed" is needed.
We considered a shape to be well-formed if it satisfies two criteria: an ordering criterion and a smoothness criterion.
The ordering criterion is satisfied if the shortest closed path visiting each junction exactly once visits the junction points in the intended order.
This criterion detects whether diblocks cross from one side of the micelle surface to another, and it also detects smaller defects such as a transposition of two diblocks.
The smoothness criterion is satisfied if the maximum distance between any two sequential junction points exceeds the median distance by less than forty percent.
The smoothness criterion detects whether diblocks have broken off from the main surface either to form a small aggregate of diblocks outside the micelle, or form a cluster of solvophilic beads in the interior of the micelle.
We found that about half of the average micelle shapes resulting from our simulations satisfied both criteria for being well-formed.

\subsection{Combining simulation runs}
\label{subsec:combiningRuns}
So far, we have discussed how to determine an average micelle shape and its fluctuation from a single simulation run.
However, we performed multiple simulation runs of each micelle with different random initial velocities (see \cref{subsec:initialization}) to both generate more statistical data, and more crucially to confirm that the uncertainties in each simulation run's mean shape are well estimated.
Therefore, for each micelle composition there is not one, but $N_a$ average shapes, denoted $\bbbr_\xi$, $\xi = 1,2,\dots,N_a$, and each of these has a corresponding variance $\bbs_\xi$ and variance in mean $\bbbs_\xi$.
In this section, we describe how these quantities are combined to produce a best estimate of the shape, its fluctuations, and its error.
To represent these best estimates, we will use the corresponding symbols, but without the $\xi$ subscript.
More explicitly, $\bbbr$ denotes the combined average; $\bbs$, the combined variance; and $\bbbs$, the error in the combined mean.

It would be possible to combine the means $\bbbr_\xi$ via the simple minimization given in \cref{eq:averageDefinition}.
However this formula would ignore the uncertainties $\bbbs_\xi$ in the means.
These uncertainties ought to be taken into account, because mean shapes with lower uncertainty should be given more weight in determining the combined mean.
Indeed, as might be predicted from the outliers of \cref{fig:chainLengthCompressibilityBlack}, we do find that there is a significant variability in simulation runs' uncertainties in the mean.
To account for the uncertainties, we may define the combined mean using a maximum likelihood estimate~\cite{dataReduction}.
Under the canonical assumptions of a harmonic system in thermal equilibrium, we expect the probability distribution of $\bbbr_\xi$ to be Gaussian, so the log-likelihood $L(\bbbr,\bbbr_\xi,\bbs_\xi)$ of a true average shape $\bbbr$ given an estimated average $\bbbr_\xi$ and its error $\bbs_\xi$ is (up to an unimportant constant offset) proportional to following quadratic form:
\begin{equation}
\label{eq:errorUsingUncertainty}
L\left(\bbbr,\bbbr_\xi,\bbbs_\xi\right)=-\left(\check{\bbbr}-\bbbr_\xi\right)\cdot\bbbs_\xi^{-1}\cdot\left(\check{\bbbr}-\bbbr_\xi\right),
\end{equation}
where $\check{\bbbr}$ is the result of aligning $\bbbr$ with $\bbbr_\xi$.
Since this log-likelihood depends only on $\check{\bbbr}$, it is independent of rotations of $\bbbr$, as desired.
The product in \cref{eq:errorUsingUncertainty} involving a matrix inversion is indeed well-defined since $\bbbs_\xi$ does have full rank when viewed as an operator on the space of shapes aligned with $\bbbr_\xi$, and the shapes $\check{\bbbr}-\bbbr_\xi$ indeed lie in this space.

Having defined the log-likelihood of a mean given a single simulation, we may now define the same for multiple simulation runs.
This is facilitated by the fact that the runs are statistically independent from one another, so that the joint probability of measuring each $\bbbr_\xi$ with the uncertainty $\bbbs_\xi$ given the true mean shape $\bbbr$ is the product of the individual probabilities, and therefore the log-likelihood of the true mean shape $\bbbr$ given the $\bbbr_\xi$ and $\bbbs_\xi$ is simply the sum of the individual log-likelihoods.
Therefore the maximum likelihood estimate for the true mean shape $\bbbr$ is given by maximizing the sum of log-likelihoods:
\begin{equation}
\label{eq:combindAverage}
\bbbr= \argmax_\bba \sum_{\xi=1}^{N_a}L\left(\bba,\bbbr_\xi,\bbbs_\xi\right).
\end{equation}

\newcommand{\cbbs}{\check{\bbs}}

Once the average $\bbbr$ is found, one can ask what is the best estimate for the fluctuations that can be made from each run's mean $\bbbr_\xi$ and variance $\bbs_\xi$.
We make this best estimate by finding the variance matrix which best represents the fluctuations aggregated over all simulation runs.
The variance $\bbs_\xi$ represents a set of samples $\bbr_{\xi \alpha}$ aligned with $\bbbr_{\xi}$.
The first step in estimating the variance from the shape samples via \cref{eq:shapeVarianceDefinition} is to align the samples with the mean (in the present case, $\bbbr$).

This alignment can be decomposed into two steps.
First we apply to each sample the rotation $\mathbf{R}_\xi$ which aligns $\bbbr_\xi$ with $\bbbr$, producing the aligned mean shape $\cbbbr_\xi$ given by $\mathbf{R}_\xi \bbbr_\xi$ and the rotated samples $\cbbr_{\xi\alpha}$, given by $\mathbf{R}_\xi \bbr_{\xi\alpha}$.
However, the alignment of the sample $\bbr_{\xi \alpha}$ with the combined mean $\bbbr$ is not yet complete: although $\cbbr_{\xi \alpha}$ is aligned with $\cbbbr_\xi$, and $\cbbbr_\xi$ is aligned with $\bbbr$, nontransitivity implies that $\cbbr_{\xi \alpha}$ is typically not aligned with $\bbbr$.

An additional, albeit small, rotation $\bR_{\xi \alpha}$ specific to each sample is necessary to complete the alignment.
If the rotation is to align the sample with $\bbbr$, then, by the reasoning found below \cref{eq:wedgeDefinition}, we must have $\wedge \bbbr \cdot\bR_{\xi \alpha} \cbbr_{\xi \alpha} =0$, where $\wedge \bbbr$ is defined in \cref{eq:wedgeDefinition} as a $90^\circ$ counter-clockwise rotation of $\bbbr$.

To make progress, we observe that since the misalignment caused by nontransitivity is typically small, and so the required rotation angle $\theta_{\xi \alpha}$ is small, we may make the Taylor expansion approximation $\bR_{\xi \alpha} \cbbr_{\xi \alpha} \approx \cbbr_{\xi \alpha} + \theta_{\xi \alpha} \left(\wedge \cbbr_{\xi \alpha}\right)$.
Since each sample is near its corresponding mean, we further make the approximation $\wedge \cbbr_{\xi \alpha} \approx \wedge \cbbbr_\xi$.
Combining these two approximations, we find that the action of $\bR_{\xi \alpha}$ on the sample $\cbbr_{\xi \alpha}$ is given by
\begin{equation}
\bR_{\xi \alpha} \cbbr_{\xi \alpha} \approx \cbbr_{\xi \alpha} + \theta_{\xi \alpha} \left(\wedge \cbbbr_\xi\right).
\label{eq:approximateRotationRule}
\end{equation}

Inserting the approximation into the condition for alignment, we obtain $\wedge \bbbr \cdot \left(\cbbr_{\xi \alpha} + \theta_{\xi \alpha} \left( \wedge \cbbbr_\xi\right) \right)=0$, so that $\theta_{\xi \alpha} = -\frac{\left(\wedge\bbbr\right) \cdot \cbbr_{\xi \alpha}}{\bbbr \cdot \cbbbr_{\xi }}$.
Inserting this $\theta_{\xi \alpha}$ into \cref{eq:approximateRotationRule}, we find the action of the second rotation on a sample to be 
\begin{equation}
\bR_{\xi \alpha}\cbbr_{\xi \alpha} \approx  \left(\mathbb{1} - \frac{\wedge\cbbbr_\xi \otimes \wedge{\bbbr} }{\cbbbr_\xi \cdot \bbbr}\right)  \cbbr_{\xi \alpha} \equiv  \mathbb{\Pi}_\xi \cbbr_{\xi \alpha},
\label{eq:projectionDefinition}
\end{equation} 
where we have defined the projection operator $\mathbb{\Pi}_\xi$ to be the tensor in parentheses above.
We see that the combined effect of the two rotations on a sampled fluctuation is given by the linear operator $  \mathbb{\Pi}_\xi\mathbf{R}_\xi $.
Consequently, the variance matrix for the aligned samples is $ \mathbb{\Pi}_\xi\mathbf{R}_\xi \bbs_\xi  \mathbf{R}_\xi^T \mathbb{\Pi}_\xi^T$, which we denote by $\cbbs_\xi$.

If the aligned fluctuations from each simulation run were aggregated, the resulting variance matrix would be the weighted average of the individual $\cbbs_\xi$ of each run, weighted by the number of samples.
However, since each run had approximately the same length, they generate approximately the same number of samples, and so we make the approximation that the best estimate for the combined variance is a simple average of the $\cbbs_\xi$:
\begin{equation}
\label{eq:varianceAverageOfRuns}
\bbs=\frac{1}{N_a} \sum_{\xi=1}^{N_a} \check{\bbs}_\xi.
\end{equation}

\newcommand{\cbbbs}{\check{\bbbs}}
To estimate the variance in the mean $\bbbs$ giving the uncertainty in $\bbbr$, we also perform an average.
Just as in \cref{eq:varianceAverageOfRuns}, we must revise the original $\bbbs_\xi$ to account for the fact that the best estimate of the mean is now $\bbbr$, and to align $\bbbs_\xi$ with this $\bbbr$.
Accordingly we perform the same alignment transformation on each $\bbbs_\xi$ that was done for the sample variances $\bbs_\xi$ above.
The result is denoted $\cbbbs_\xi$.
We then average these $\cbbbs_\xi$ values as in \cref{eq:varianceAverageOfRuns}.
This average doesn't change systematically as $N_a$ increases.
However the overall variance in the mean of $N_a$ independent samples is $1/N_a$ times this average.
Thus 
\begin{equation}
\label{eq:varianceInMeanAverageOfRuns}
\bbbs=\frac{1}{N_a^2} \sum_{\xi=1}^{N_a} \check{\bbbs}_\xi.
\end{equation}

\subsection{Testing consistency between simulation runs}
\label{subsec:consistency}
At the end of \cref{subsec:shapeVariance}, we stated that the uncertainties defined by \cref{eq:varianceInMean} could be tested by comparing them to the observed variability in the mean shape.
Having defined the average $\bbbr$, we may use $L\left(\bbbr,\bbbr_\xi,\bbbs_\xi\right)$ defined in \cref{eq:errorUsingUncertainty} to measure the differences between the individual run averages $\bbbr_\xi$ and the combined mean $\bbbr$.
Since $L\left(\bbbr,\bbbr_\xi,\bbbs_\xi\right)$ is the negative of the chi-squared statistic and each mean has $2 N_j$ degrees of freedom, we expect that this $L$ should be of order $-2N_j$. 
We may therefore define the reduced chi-squared $\chi^2_\nu$ of a combined average $\bbbr$ by
\begin{equation}
\label{eq:reducedChiSquared}
\chi^2_\nu=\frac{-1}{2N_j N_a}\sum_{\xi=1}^{N_a}L\left(\bbbr,\bbbr_\xi,\bbbs_\xi\right).
\end{equation}
We expect this $\chi^2_\nu$ to be near unity; however, if the simulation runs were too short so that the full range of thermal shapes is not explored in a single simulation, then the difference in the means $\bbbr_\xi$ would be larger than the errors in the means $\bbbs_\xi$ would suggest, and $\chi^2_\nu$ would be much larger than unity.
Thus $\chi^2_\nu$ is a statistic that tests how well-estimated are the uncertainties $\bbbs_\xi$ from each simulation run of a given micelle composition.
Since the $\bbbs_\xi$ are defined using $\bbs_\xi$ (see \cref{eq:varianceInMean}), the $\chi^2_\nu$ provide an indirect test of the $\bbs_\xi$ as well.

In fact, it is possible to validate the uncertainties $\bbbs_\xi$ in more detail.
From \cref{eq:varianceInMean}, we expect that the uncertainty of the mean $\bbbs_\xi$ in the direction of its $\ell$th mode $\mathbb{m}_{\xi\ell}$ is given by $\frac{\Sigma_{\xi\ell} \tau_{\xi\ell}}{N_{s,\xi}}$.
To verify this expectation, we can define the chi-squared of the $\ell$th mode by
\begin{equation}
\label{eq:chiell}
\chi^2_{\xi\ell} = \frac{\left(\mathbb{m}_{\xi\ell} \cdot \left(\bbbr_\xi - \bbbr\right)\right)^2}{\Sigma_{\xi\ell} \tau_{\xi\ell}/N_{s,\xi}}.
\end{equation}
For each $\ell$, we expect that $\chi^2_{\xi\ell}$ should be near unity.
Testing this expectation gives a more thorough validation of \cref{eq:varianceInMean}, and in particular that finding the correlation time $\tau_{\xi\ell}$ of each mode is sufficient to characterize the full correlations of the shape fluctuations.
The reduced chi-squared $\chi^2_\nu$ for each micelle composition and a representative set of chi-squareds $\chi^2_{\xi\ell}$ for the modes of one simulation run are presented in \cref{subsec:analysisValidation}.

In addition to testing the formulas for the simulation runs $\bbbr_\xi$ and their uncertainties $\bbbs_\xi$ defined in \cref{subsec:average} and \cref{subsec:shapeVariance}, we would like to validate the combined average $\bbbr$ and its uncertainty $\bbbs$.
One method of validation is to verify that the result of combining the $\bbbr_\xi$ and the $\bbbs_\xi$ is consistent with the result of first concatenating the list of samples $\bbr_{\xi \alpha}$, and then finding a mean from these samples, using \cref{eq:averageDefinition} as if the samples were generated from a single simulation.
We chose to compare with the average of the combined samples, denoted by $\tilde{\bbbr}$, since the procedure for averaging a time series of shapes can itself be validated using $\chi^2_\nu$ and $\chi^2_{\xi\ell}$.
Once the average of the combined samples has been found, a $\chi^2_\ell$ statistic similar to the one in \cref{eq:chiell} can be computed.
More precisely, we define $\chi^2_\ell$ by 
\begin{equation}
\chi^2_\ell = \frac{(\ml \cdot (\tilde{\bbbr} -\bbbr ))^2}{\bbbs_\ell}
\label{eq:combinedChiEll}
\end{equation}
where $\ml$ is the $\ell$th eigenmode of $\bbbs$ and $\bbbs_\ell$ is the associated eigenvalue.
An example calculation of this statistic will be presented in \cref{subsec:analysisValidation}.

\subsection{Shape features}
\label{subsec:shapeFeatures}
\begin{figure}
\centering
\includegraphics[width=0.7\linewidth]{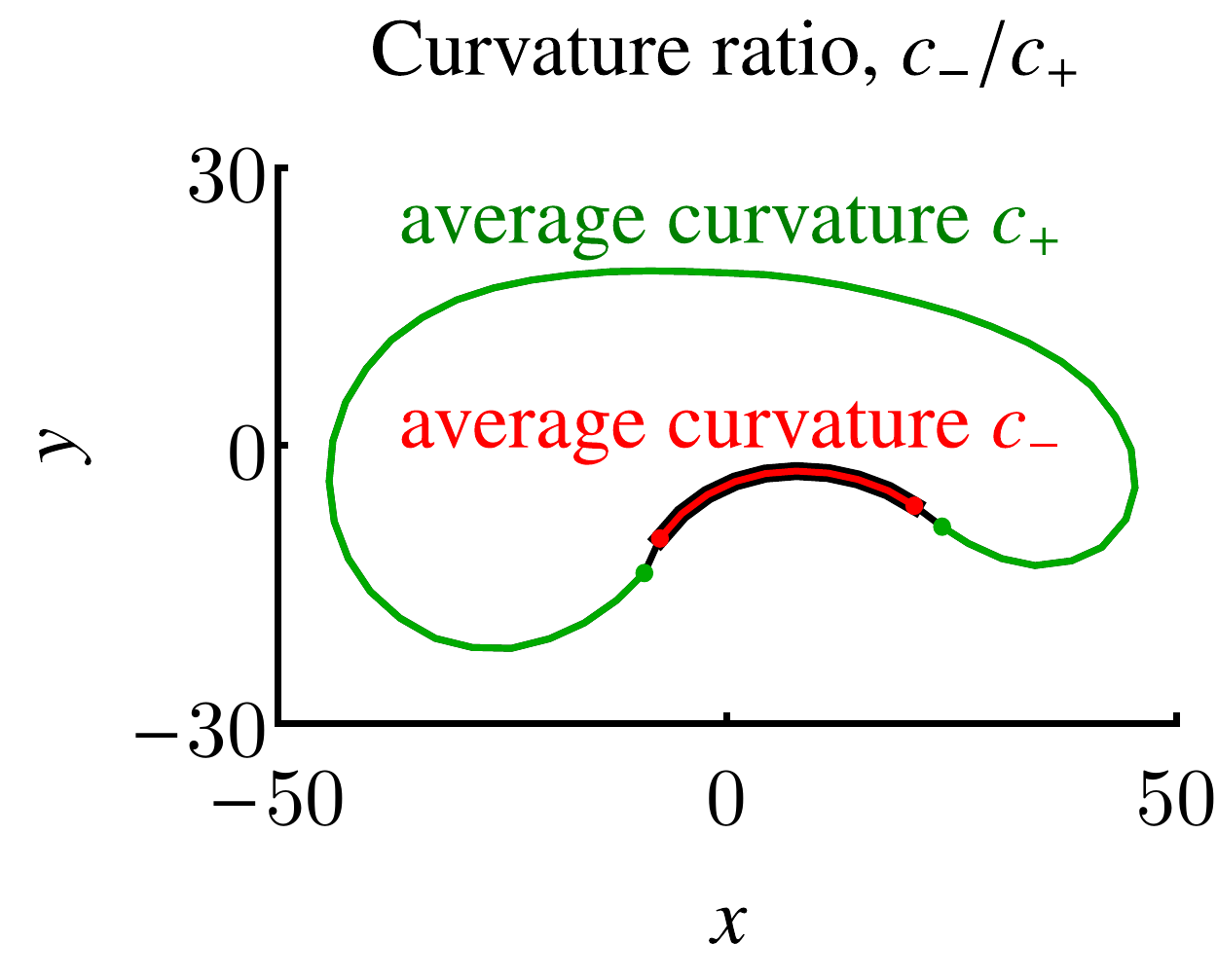}
\caption{(Color online) Illustration of curvature ratio definition. 
The curvature ratio is defined as the ratio of the average curvature $c_-$ in the region occupied by the solvophobic-rich diblocks divided by the average curvature $c_+$ in the region occupied by the solvophilic blocks. 
The curvature ratio is $1$ for a circle, and is negative for a shape with a dimple such as the one shown in the figure, becoming more negative as the dimple grows more pronounced.}
\label{fig:curvatureRatio}
\end{figure}

We have now given a complete description about how to extract micelle shape information from the simulation.
However, in this paper we will pay special attention to two features of the micelle shape: the ``curvature ratio" and the ``normalized fluctuation".
The curvature ratio $c_-/c_+$ is defined as the ratio of average signed curvature of the micelle surface region occupied by the solvophobic-rich diblocks divided by that of the solvophilic-rich diblocks as illustrated in \cref{fig:curvatureRatio}.
We are interested in this quantity because our intent is to use our shape-design mechanism to produce a micelle of unusual shape, specifically one with a dimple.
The curvature ratio quantifies the strength of this dimple, and therefore can help detect the conditions under which our shape-design mechanism works best.

In addition to the average shape, we are interested in the shape fluctuations, as noted previously. 
We have introduced the variance matrix $\bbs$ to characterize the shape fluctuations.
We summarize the size of the fluctuations represented by this $2 N_j \times 2 N_j$ matrix with a single scalar, the normalized fluctuation $\delta$, to measure the amount of fluctuation in the micelle shape, normalized so as not to scale with the number of junction points or the size of the micelle.
The normalized fluctuation is defined by
\begin{equation}
\label{eq:normalizedFluctuation}
\delta = \sqrt{\frac{\Tr \bbs}{\bbbr^2}}.
\end{equation}
It is enlightening to notice a connection between the normalized fluctuation $\delta$ and the distance metric $\din$: if $\bbbr$ and $\bbs$ are the mean and variance of a single simulation run, then by \cref{eq:shapeVarianceDefinition} and \cref{eq:dinDefinition}, the $\delta$ for this simulation run is given by
\begin{equation}
\sqrt{\frac{\frac{1}{N_s-1}\sum_{\alpha=1}^{N_s}\din(\bbbr,\bbr_\alpha)}{\bbbr^2}}.
\label{eq:normalizedFluctuationAlternate}
\end{equation}

Since we will be analyzing these two features of micelle shape, we must make an estimate of their uncertainty.
The uncertainty of the curvature ratio can be straightforwardly derived from the full variance matrix $\bbbs$ giving the variance in the mean.

Estimating the uncertainty in the fluctuation $\delta$ is more subtle because $\delta$ is defined in terms of the variance $\bbs$, and so the error in delta represent the error in the fluctuations of a quantity rather than the error in the quantity itself.
Consequently, we attempt only a rough estimate to the error in $\delta$.
The dominant source of uncertainty in $\delta$ comes from $\Tr \bbs$.
We find the uncertainty in $\Tr \bbs$ by recognizing that the trace is the sum of eigenvalues: 
\begin{equation}
\label{eq:traceEigenvalues}
\Tr \bbs= \sum_\ell \Sigma_{\ell}.
\end{equation}
Evidently, it sufficient to estimate the uncertainty in each variance.
To perform this estimate, we resort to assuming that each $\Sigma_{\ell}$ represents the variance of a Gaussian distribution.
Given $n$ samples of a univariate Gaussian random variable with variance $\Sigma$, a formula for the uncertainty $\sigma_\Sigma$ of the estimate of the distribution's variance is given by (see~\cite{Rao73}) 
\begin{equation}
\label{eq:standardDeviationOfVariance}
\sigma_\Sigma=\Sigma \sqrt{\frac{2}{n-1}}.
\end{equation}

To use this formula to estimate the uncertainty in $\Sigma_\ell$, we must choose a value for the number of independent samples $n_\ell$ contributing to the estimation of $\Sigma_\ell$. 
We estimate this number of samples for each mode $\mathbb{m}_\ell$ of $\bbs$, by taking the ratio of the fluctuation in the mode amplitude $\Sigma_\ell$ with the uncertainty in the mean along the $\mathbb{m}_\ell$ direction:
\begin{equation}
\label{eq:independentSamplesForVariance}
n_\ell=\frac{\mathbb{m}_\ell^T \bbs \mathbb{m}_\ell}{\mathbb{m}_\ell^T \bbbs \mathbb{m}_\ell}.
\end{equation}

As this estimate of the normalized fluctuation uncertainty requires approximation in the form of \cref{eq:standardDeviationOfVariance,eq:independentSamplesForVariance}, some degree of validation is in order.
To this end, we calculate the normalized fluctuation from each simulation run (i.e., substitute $\bbbr_\xi$ and $\bbs_\xi$ in \cref{eq:normalizedFluctuation}), and compute the standard error in the resulting normalized fluctuations.
While this second method may seem reasonable, we prefer the uncertainty estimate described in the preceding paragraphs to the second estimate for two reasons: first, the second estimate cannot be used if there is only one simulation run producing a well-formed micelle; and second, the first estimate is less likely to underestimate the uncertainty because it takes into account all modes while the second estimate can result in an underestimate if the normalized fluctuations from individual happen to be similar.
The results of using these two methods is compared in \cref{subsec:analysisValidation}.

\section{Results}
\label{sec:results}
In this section, we present the results of simulating micelles of several compositions.
The range of micelle compositions was not chosen to be exhaustive but only to demonstrate a significant degree of control over the micelle shape.
To this end, we varied two aspects of the micelle composition: the length of the core chain and the composition of the solvophobic-rich diblocks.
The length of the core chain ranged from 600 beads to 1000 beads.
Two solvophobic-rich diblock compositions were studied, the first being $30$ solvophobic beads and $2$ solvophilic beads and the second being  $27$ solvophobic beads and $4$ solvophilic beads.
Since the first composition has a larger asymmetry, and therefore the micelles containing these diblocks have a larger asymmetry contrast between their solvophobic-rich and solvophilic-rich diblocks, we refer to these micelles as ``high contrast".
Conversely, we refer to the other micelles, whose solvophobic-rich diblocks contain $27$ solvophobic beads and $4$ solvophilic beads, as ``low contrast".
Other aspects of the micelle composition were held constant: each micelle had $12$ solvophobic-rich diblocks and $55$ solvophilic-rich diblocks, and the solvophilic-rich diblocks each had $24$ solvophobic beads and $7$ solvophilic beads.
To obtain sufficient statistics, each simulation was run in parallel on nine cores for $70$ hours, during which time LAMMPS completed about one billion timesteps.
The mean shapes, fluctuations, and uncertainties resulting from these simulations are plotted in \cref{tab:resultShapeTable}.

In \cref{subsec:analysisValidation}, we discuss the results of validation tests discussed in \cref{subsec:consistency} and \cref{subsec:shapeFeatures}.
In \cref{subsec:shapeFeaturesResult}, we show the average shapes of the micelles, and plot the shape features introduced in \cref{subsec:shapeFeatures} as a function of the size of the core and the asymmetry ratio of the solvophobic-rich diblocks.

\newcommand{\figureSize}{.22}
\begin{table*}
\begin{tabular}{|c|cccc|}
\hline 
& 600 Core & 700 Core & 848 Core & 1000 Core\\
\hline 
\makecell{high\\ contrast} &
\raisebox{-.5\height}{\includegraphics[width=\figureSize \linewidth]{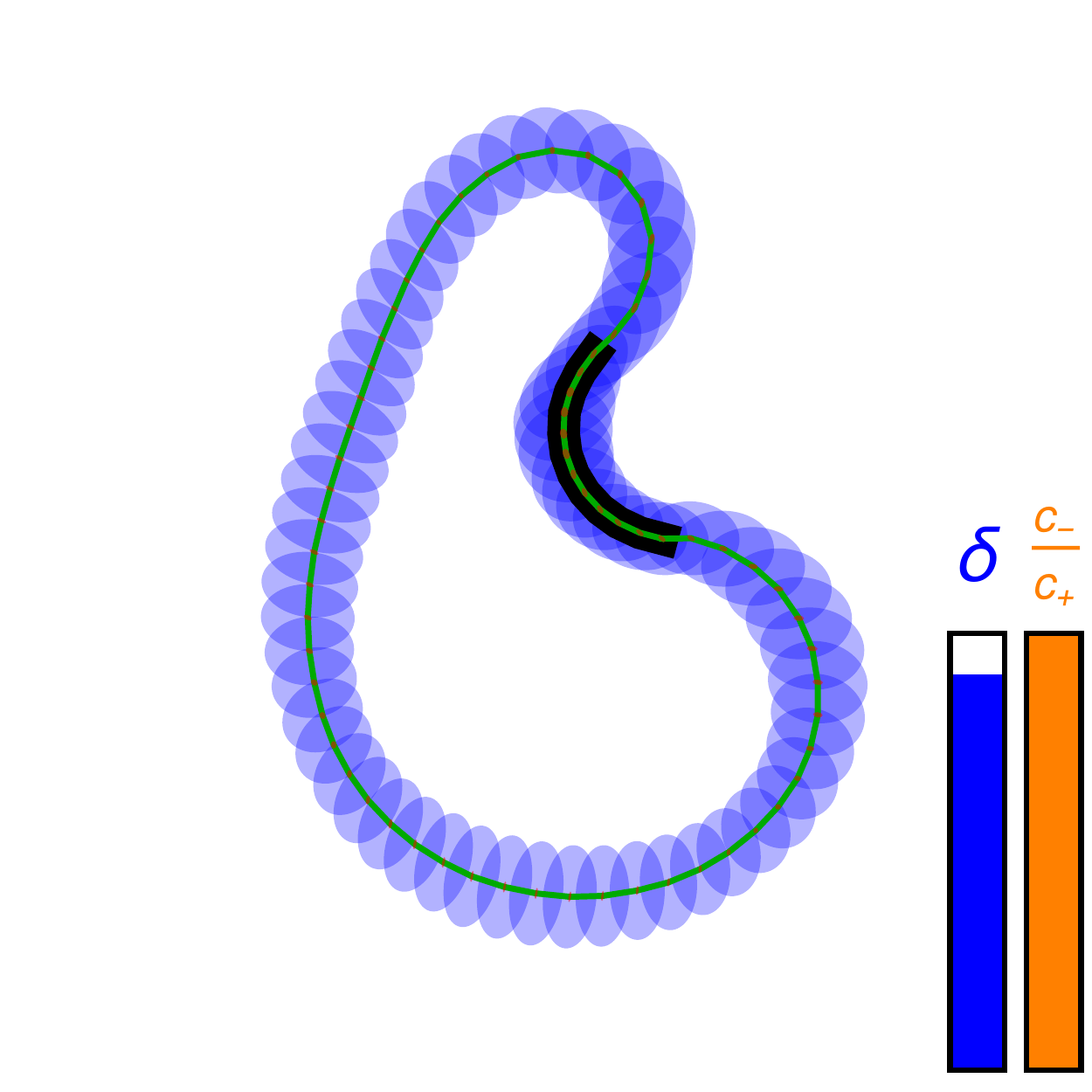}}
&
\raisebox{-.5\height}{\includegraphics[width=\figureSize \linewidth]{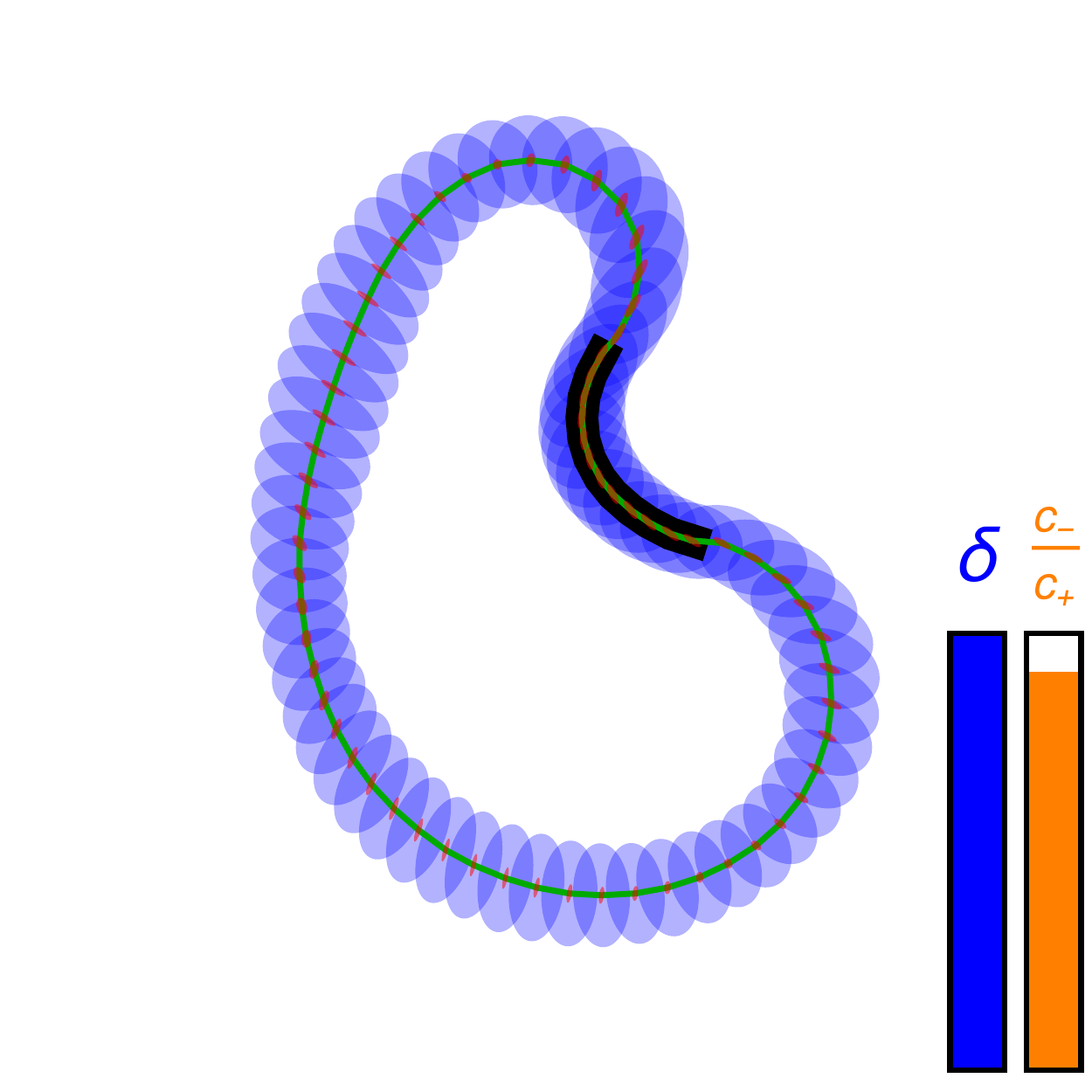}}
&
\raisebox{-.5\height}{\includegraphics[width=\figureSize \linewidth]{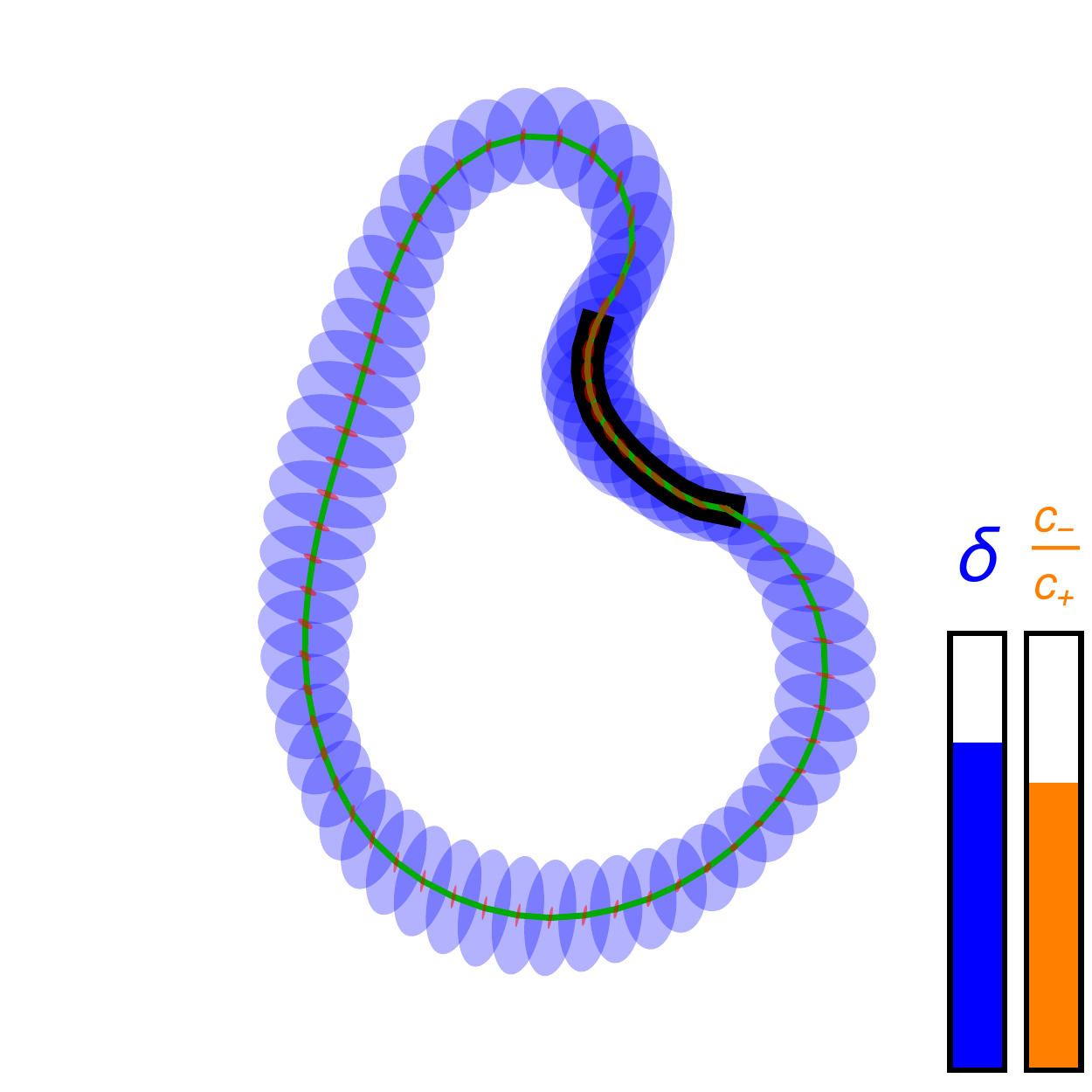}}
&
\raisebox{-.5\height}{\includegraphics[width=\figureSize\linewidth]{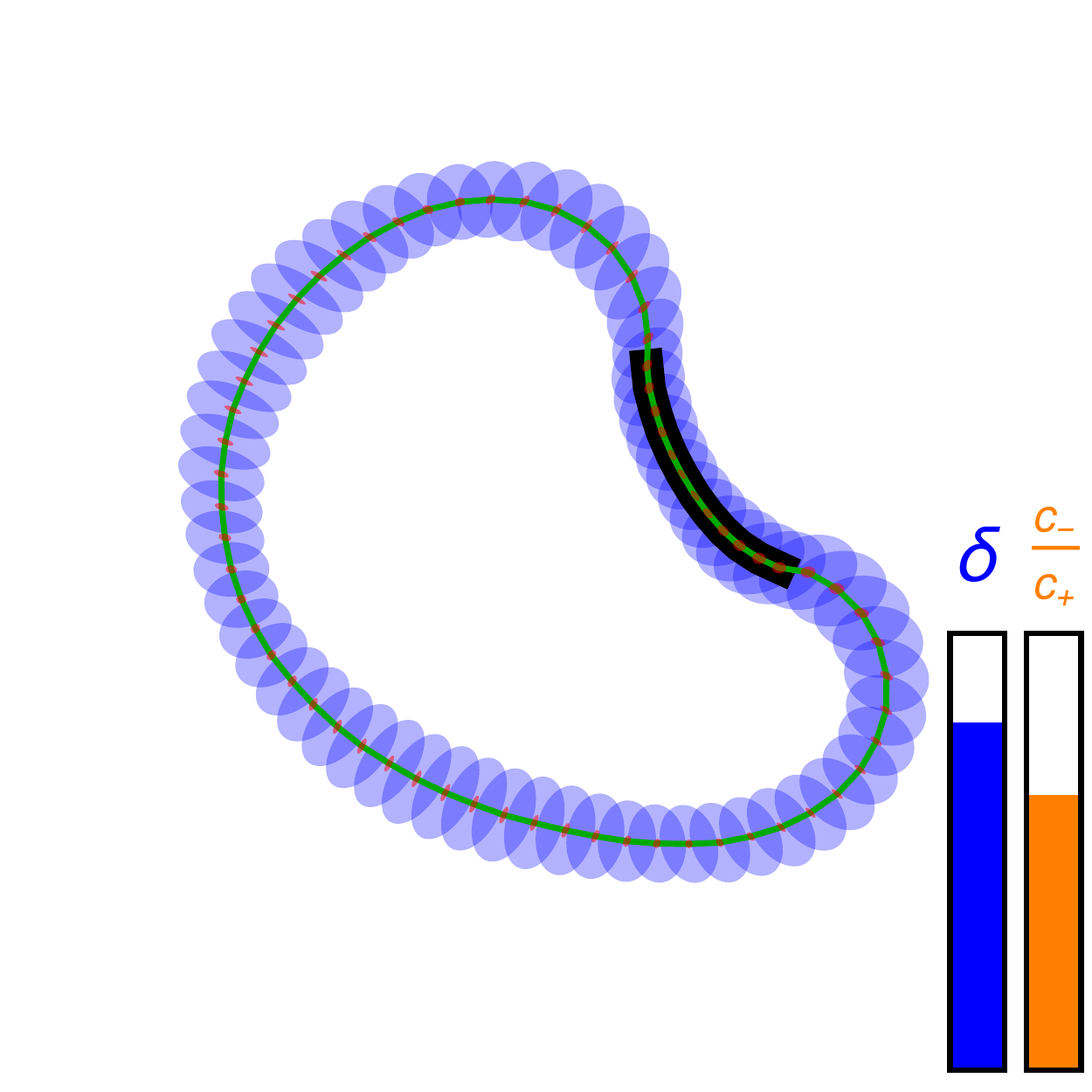}}
\\
\makecell{low\\ contrast} &
\raisebox{-.5\height}{\includegraphics[width=\figureSize \linewidth]{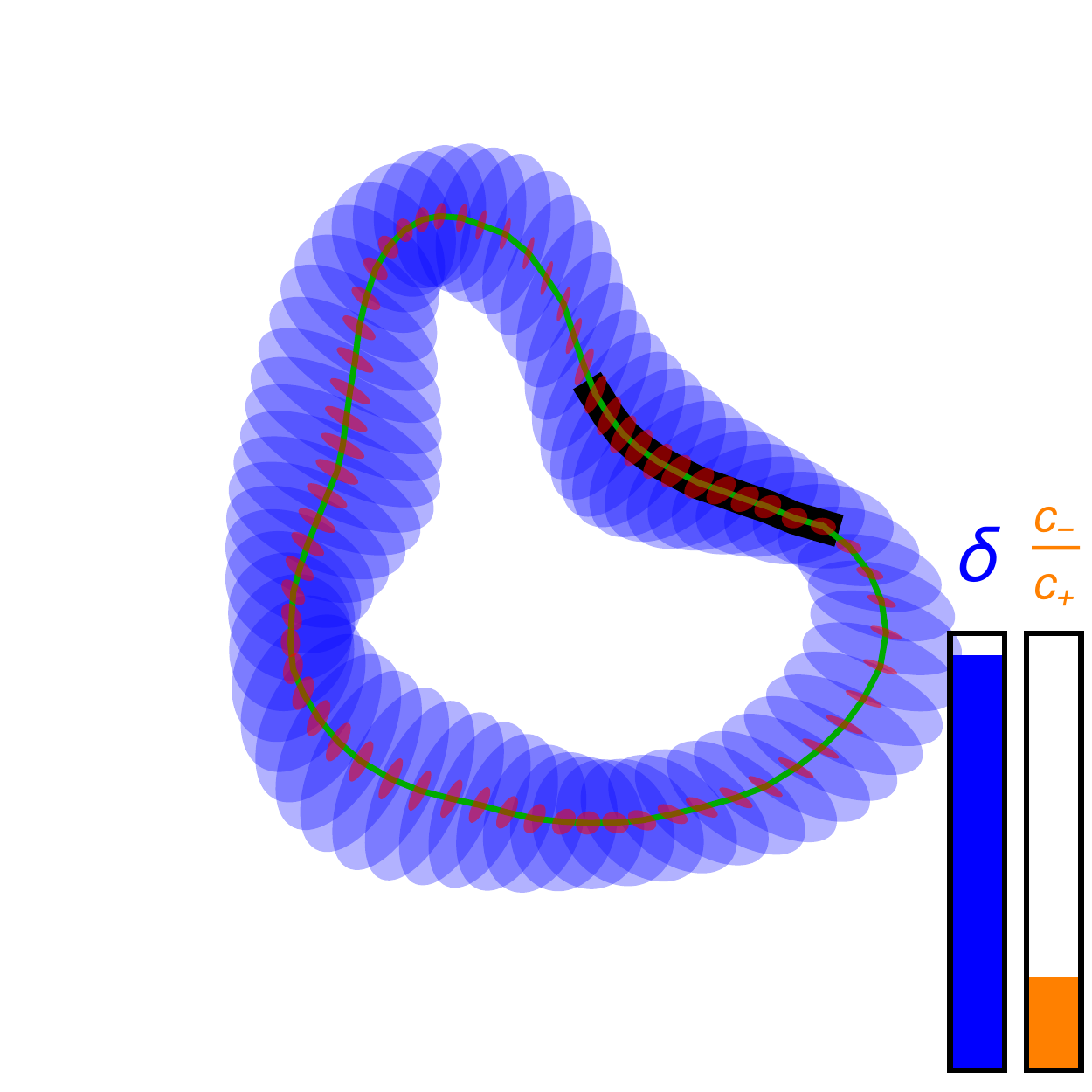}}
&
\raisebox{-.5\height}{\includegraphics[width=\figureSize \linewidth]{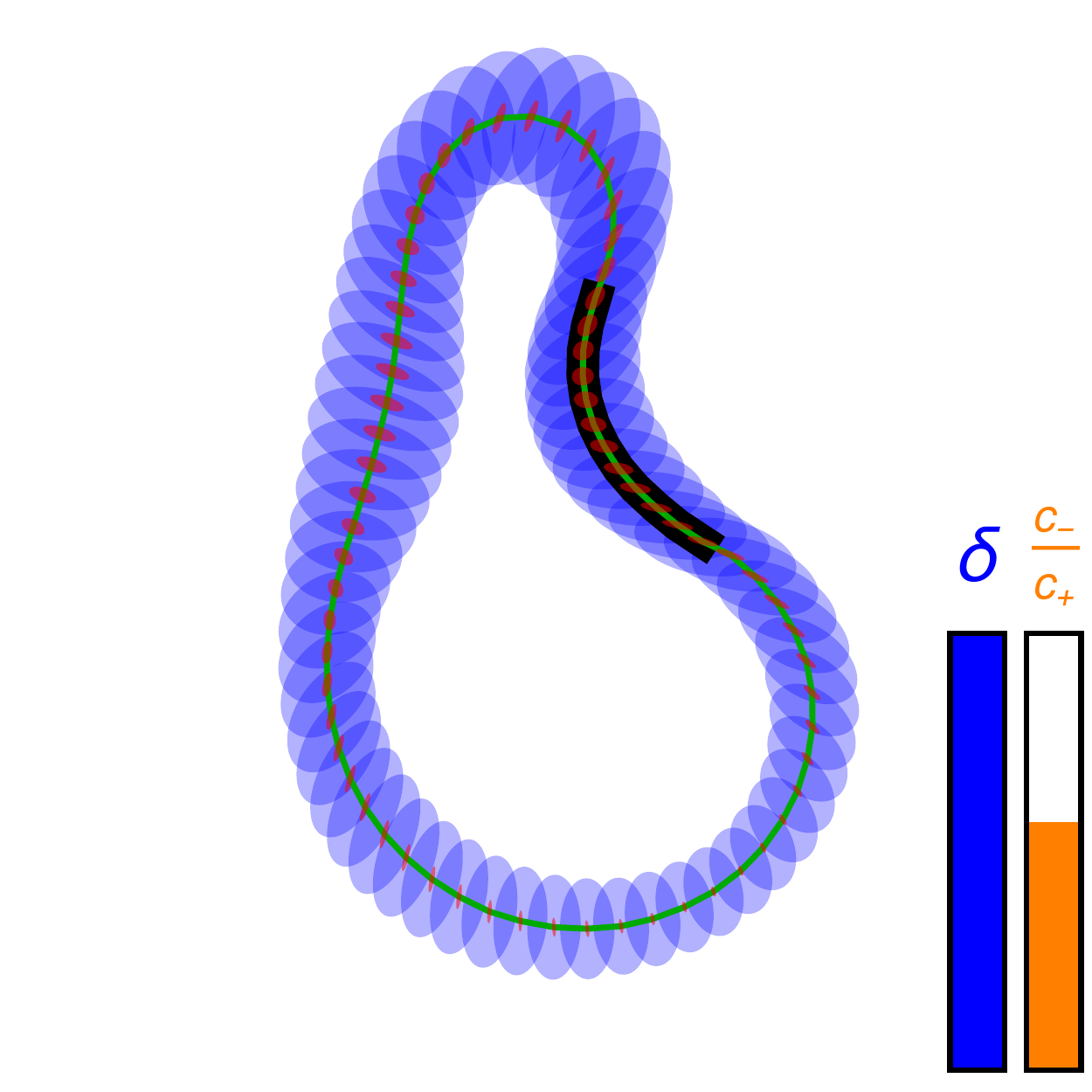}}
 &
\raisebox{-.5\height}{\includegraphics[width=\figureSize \linewidth]{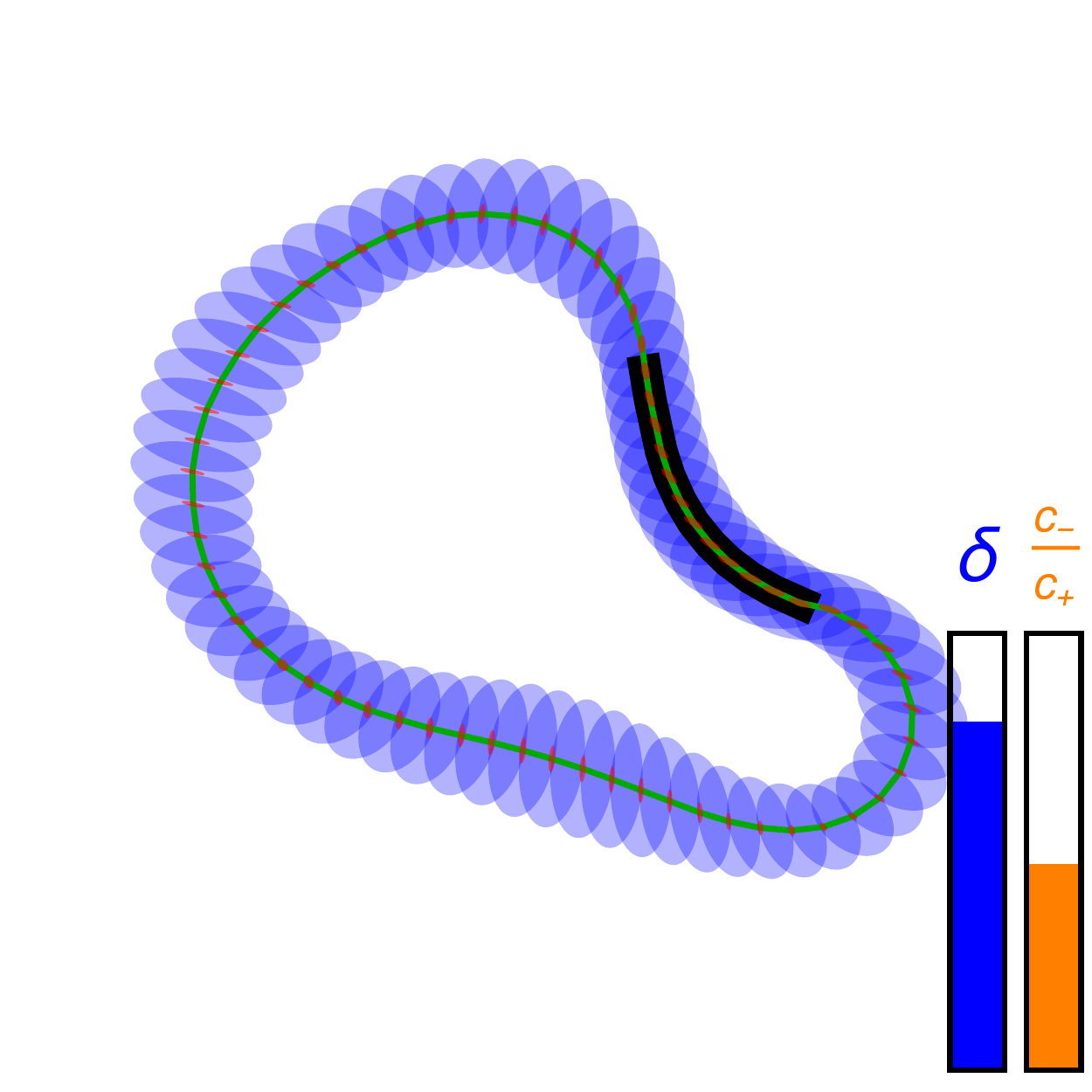}}
  &
\raisebox{-.5\height}{\includegraphics[width=\figureSize\linewidth]{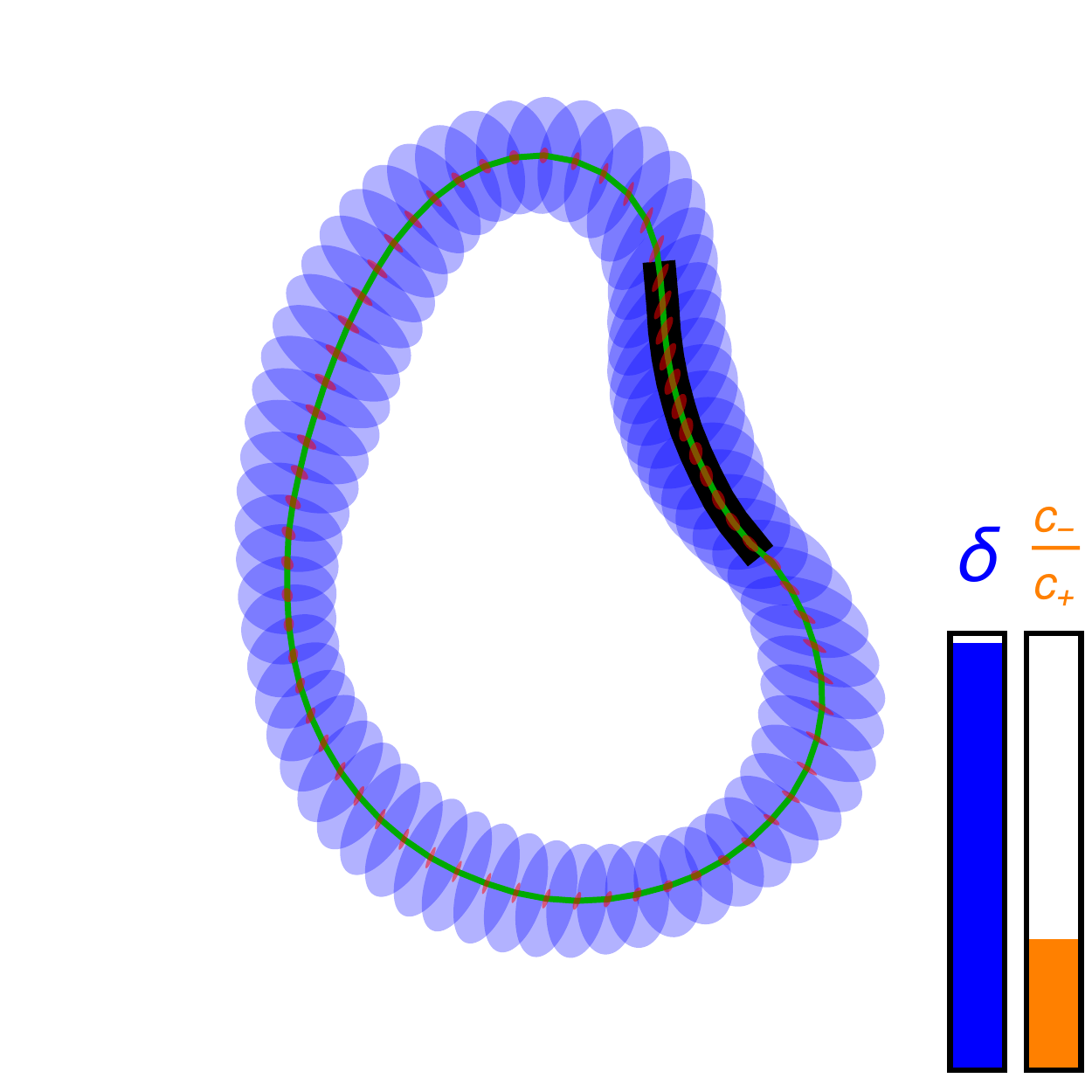}}
\\
\hline 
\end{tabular} 
\caption{(Color online) Average shapes of micelles of various compositions, plotted in the manner of \cref{fig:plotofaverage}. 
The micelle compositions are described in the first paragraph of \cref{sec:results}.
The relative magnitudes of the normalized fluctuation $\delta$ (blue) and curvature ratio $c_-/c_+$ (orange) of each shape are indicated by bars next to the shape. }
\label{tab:resultShapeTable}
\end{table*}

\subsection{Validation of analysis}
\label{subsec:analysisValidation}

In \cref{subsec:consistency}, we introduced two statistics $\chi^2_\nu$ and $\chi^2_{\xi\ell}$, defined in \cref{eq:reducedChiSquared,eq:chiell}, for validating that the variation in the means $\bbbr_\xi$ was consistent with the error $\bbbs_\xi$ in these means.
We stated that a correctly estimated error leads to the statistics being nearly one, while underestimated error lead to large values and overestimated errors lead to small values.
One cause of concern motivating this test is that the simulations may not have been run long enough for the full range of thermal shapes to explore, leading to the estimated mean of a simulation run being strongly biased by the initialization.
This would lead to the means being more different than mere thermal fluctuations would predict, and therefore lead to large $\chi^2_\nu$ and $\chi^2_{\xi\ell}$.
Another cause for concern is that the assumption underpinning \cref{eq:varianceInMean}, namely that the fluctuation mode amplitudes fluctuate independently and are each described by a separate correlation time may be strongly violated to the point that  \cref{eq:varianceInMean} gives an unsatisfactory estimate of the uncertainty in the mean.
A shortcoming of the estimate \cref{eq:varianceInMean} would likely because a systematic dependence of $\chi^2_{\xi\ell}$ on the mode number $\ell$, since we expect the applicability our assumptions to depend on the amplitude or correlation of the mode, both of which vary systematically with $\ell$.

\begin{table} 
\begin{tabular}{|c|cccc|}
\hline 
  & 600 Core & 700 Core & 848 Core & 1000 Core\\
\hline 
\makecell{high\\ contrast}	&5.5	&0.69	&1.8	&---\\[1 em]
\makecell{low\\ contrast} 	&59		&1.2 	&2.3 	&3.2\\
\hline 
\end{tabular} 
\caption{
Reduced chi-squared $\chi^2_\nu$ (defined in \cref{eq:reducedChiSquared}) for each of the averages shown in \cref{tab:resultShapeTable}.
One of the entries is blank since there was only one well-formed result for that micelle composition, in which case the reduced chi-squared statistic is meaningless.
Only one of the simulations runs of the high contrast micelle having $1000$ core beads gave a well-formed average shape, so that all of the other runs were rejected. 
The reduced chi-squared statistic is meaningless in this case, so it is omitted. 
}
\label{tab:reducedChiSquaredTable}
\end{table}
Results for $\chi^2_\nu$ are given in \cref{tab:reducedChiSquaredTable}.
The simulations of low contrast micelles having $600$ core beads had a large $\chi^2_\nu$.
This happened because two of the four simulations fluctuated only modestly about significantly different, but well-defined mean shapes.
The other two of the four simulation runs showed large fluctuations.
This suggests that, for this micelle composition, there are multiple metastable shapes between which the micelle can fluctuate.
This example demonstrates the benefit of doing multiple independent simulation runs and shows the limitations of representing a shape with a single mean and a variance about the mean.

On the other hand, besides the two aforementioned cases, all the $\chi^2_\nu$ are near unity, indicating that the uncertainty in the mean shape from each run is well-estimated.

\begin{figure*}
\subfloat[][]{ %
\centering
\includegraphics[width=\columnwidth]{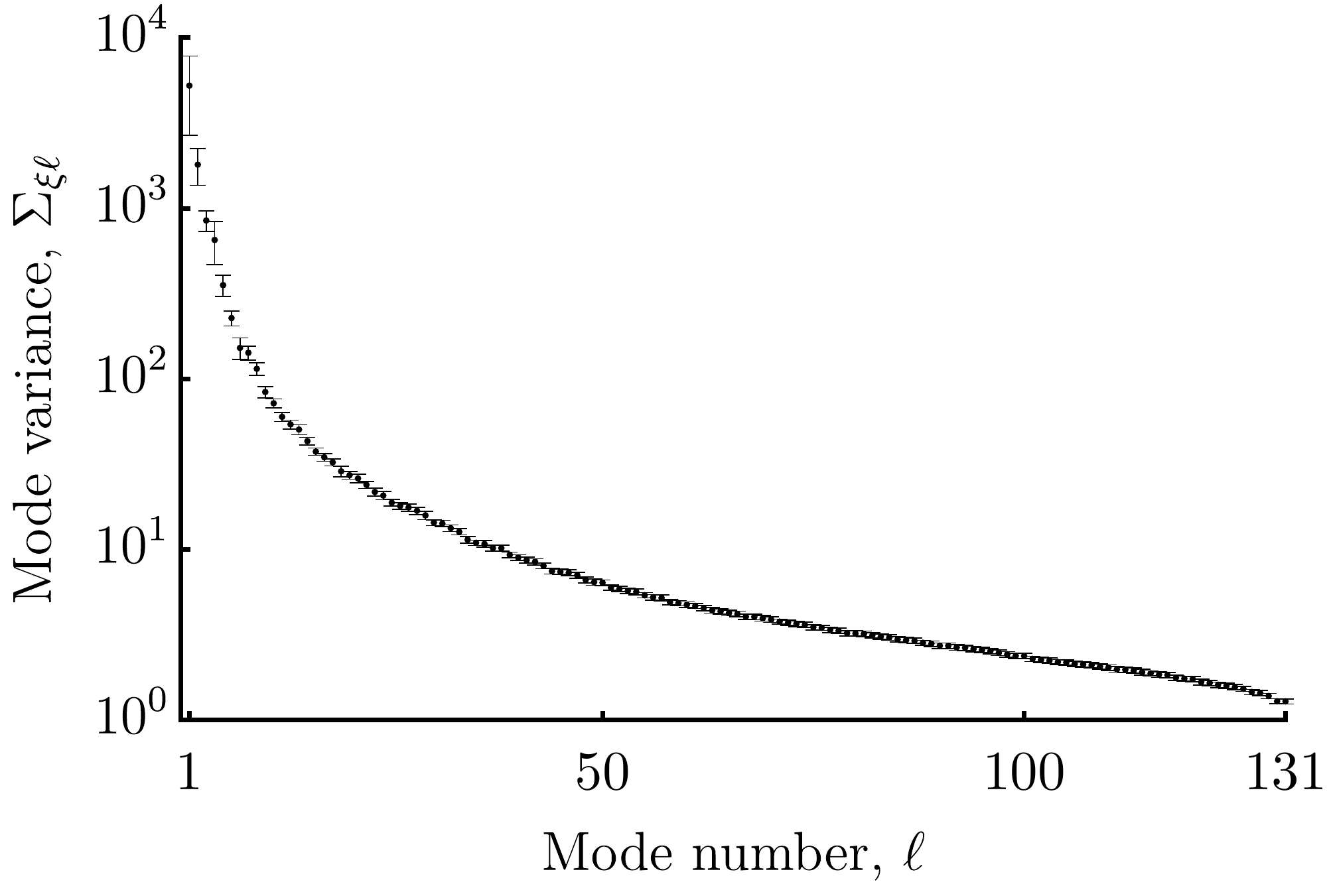}
\label{fig:variancesWithUncertainties}
}
\subfloat[][]{ %
\centering
\includegraphics[width=\columnwidth]{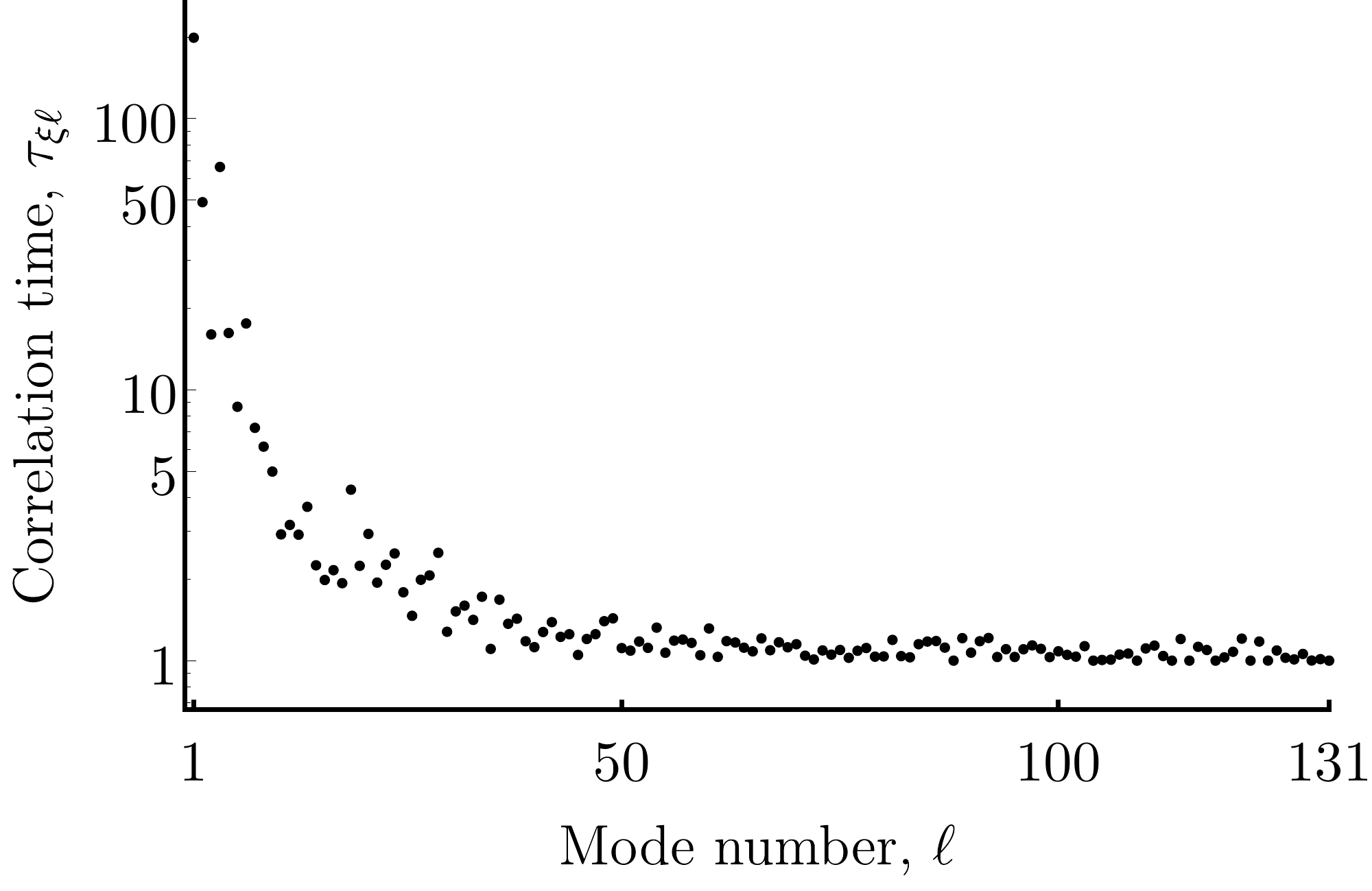}
\label{fig:correlationTimePlot}
}

\subfloat[][]{ %
\centering
\includegraphics[width= \columnwidth]{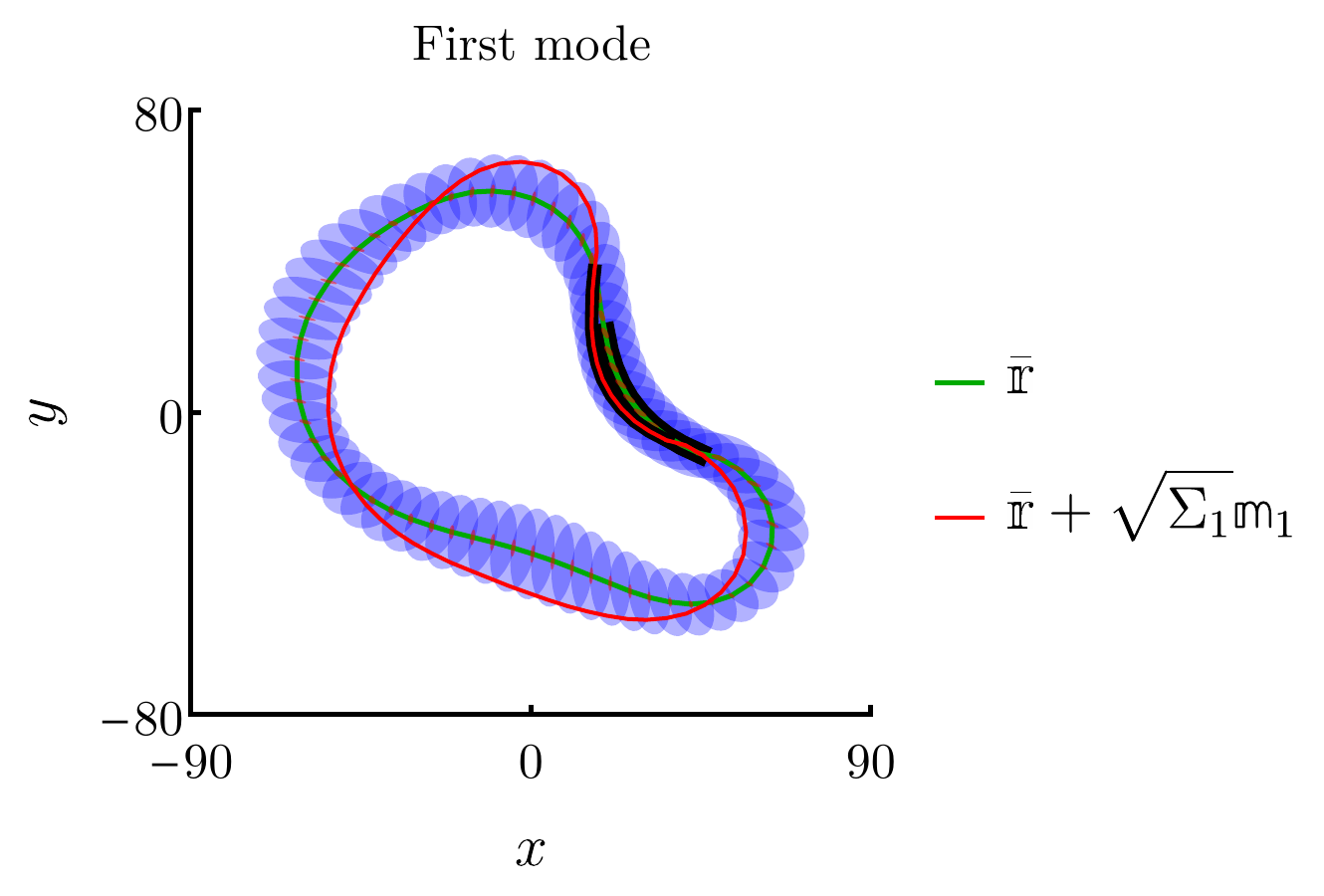}
\label{fig:firstMode}
}
\subfloat[][]{ %
\centering
\includegraphics[width= \columnwidth]{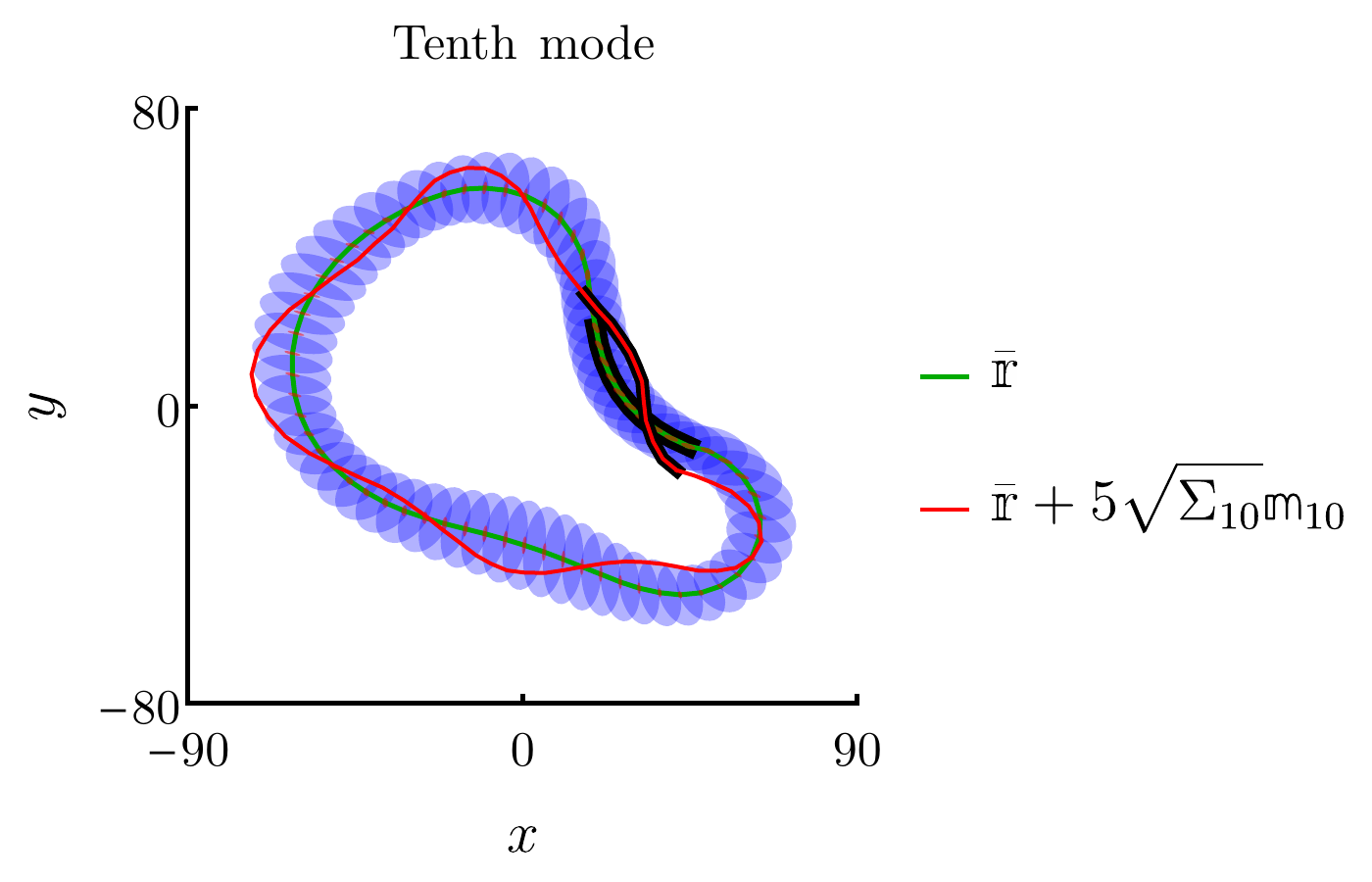}
\label{fig:secondMode}
}

\subfloat[][]{ %
\centering
\includegraphics[width= \columnwidth]{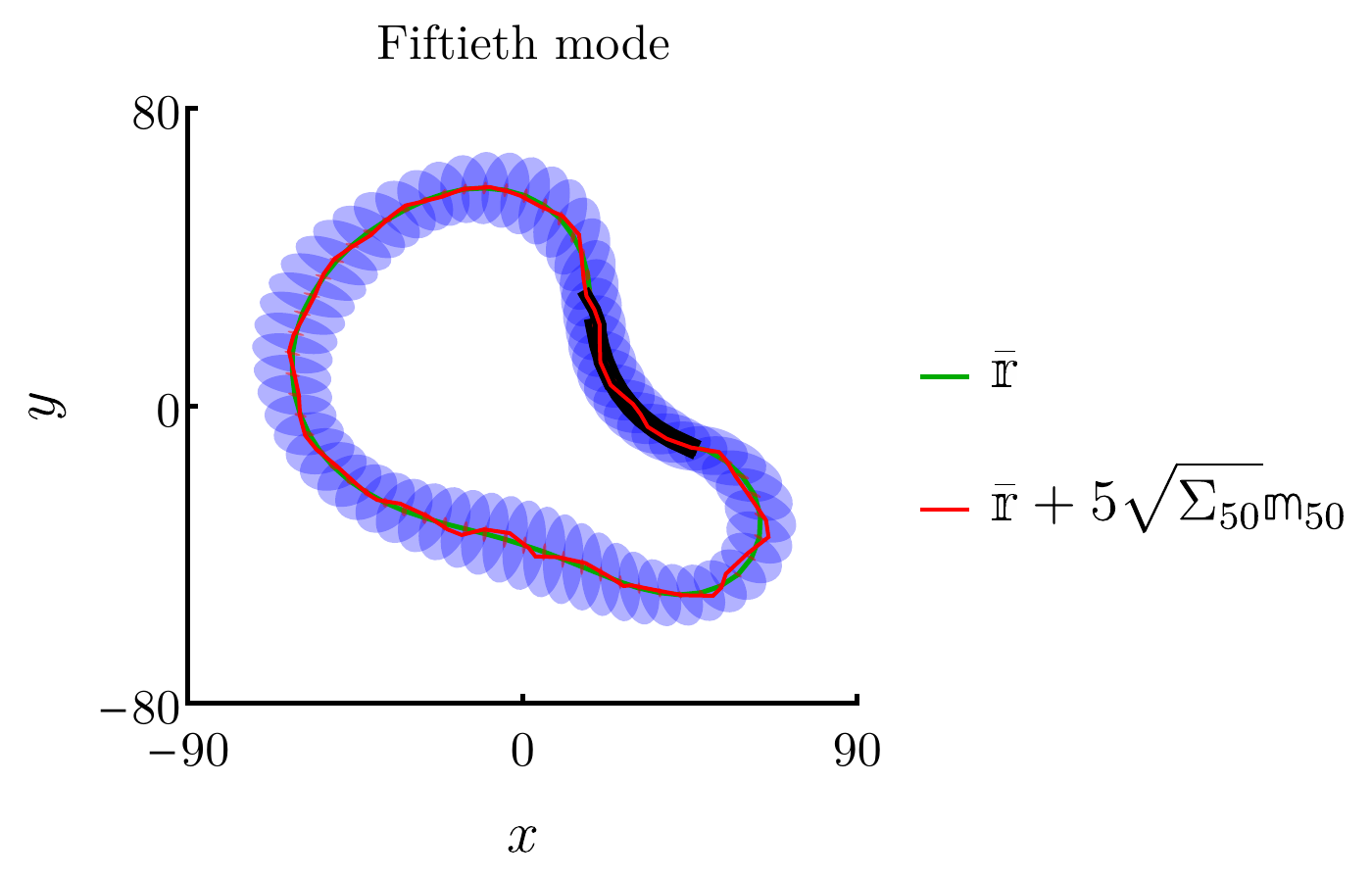}
\label{fig:thirdMode}
}
\caption{(Color online) 
\protect\subref{fig:variancesWithUncertainties} Plot of the mode amplitude variance $\Sigma_{\xi\ell}$ vs mode number $\ell$ and \protect\subref{fig:correlationTimePlot} the correlation time $\tau_{\xi\ell}$ (in units of the sampling interval) from a single simulation run (i.e., single value of $\xi$) for a low contrast micelle containing $848$ core beads.
The three modes corresponding to rigid motions are omitted because their amplitude variance is zero.
The variances range over three orders of magnitude.
\protect\subref{fig:firstMode}, \protect\subref{fig:secondMode}, and \protect\subref{fig:thirdMode} Plots of the $\ell =1$, $\ell=10$ and $\ell = 50$ modes.
The average shapes are plotted as well as a deformation of the shape in the direction of the mode $\mathbb{m}_{\xi\ell}$.
In \protect\subref{fig:firstMode}, the size of this deformation is $\sqrt{\Sigma_{\xi 1}}$, which represents one standard deviation of sampled shape distribution in the direction of $\mathbb{m}_{\xi\ell}$.
In \protect\subref{fig:secondMode} and \protect\subref{fig:thirdMode}, the size of the deformation is increased to five standard deviations for clarity.
}
\label{fig:varianceResultPlots}
\end{figure*}

Next we consider the $\chi^2_{\xi\ell}$ of \cref{subsec:consistency}.
A more detailed diagnostic than the $\chi^2_\nu$, the $\chi^2_{\xi\ell}$ indicate how well the shape error of the $\xi$th simulation run in the direction of the $\ell$th mode is estimated.
To give context to the analysis of the $\chi^2_{\xi\ell}$, we first show a representative plot of the mode amplitude variances $\Sigma_{\xi\ell}$ vs $\ell$, and we show a few modes $\mathbb{m}_{\xi \ell}$ in \cref{fig:varianceResultPlots}.
\begin{figure}
\centering
\includegraphics[width=\linewidth]{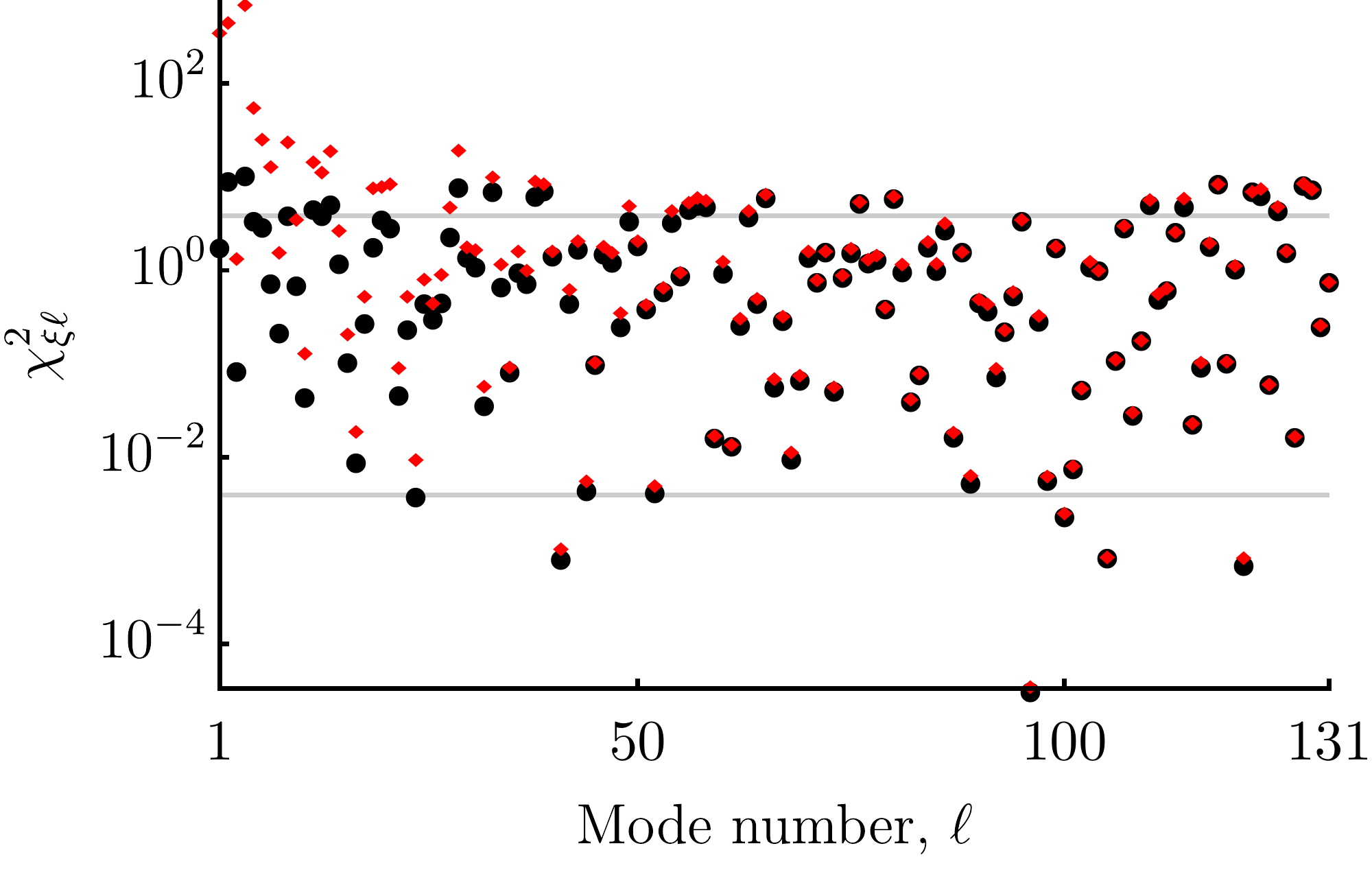}
\caption{(Color online) 
	Plot of $\chi^2_{\xi\ell}$ (defined in \cref{eq:chiell}, black circles) vs $\ell$ for the representative simulation run of \cref{fig:varianceResultPlots} of a low contrast micelle with $848$ core beads.
	As in \cref{fig:varianceResultPlots}, the modes are ordered by their amplitude, so that the first mode ($\ell=1$) is the mode with highest amplitude.
	The horizontal lines bracket the $90\%$ confidence interval for the $\chi^2_{\xi\ell}$ statistic, assuming the mean shape distribution is Gaussian.
	Most of the $\chi^2_{\xi\ell}$ fall in this range, with no apparent systematic dependence on $\ell$.
	For contrast, we present $\chi^2_{\xi\ell}$ with correlations ignored (by substituting $\tau_{\xi\ell} \to 1$ in \cref{eq:varianceInMean}, red diamonds).
	In this case, the $\chi^2_{\xi\ell}$ show a clear dependence on $\ell$, in that $\chi^2_{\xi\ell}$ is much larger than unity for small $\ell$. 
} 
\label{fig:chicorrelationplot}
\end{figure}
We find that the mode amplitude variance $\Sigma_{\xi\ell}$ varies by three orders of magnitude.
With this in mind, we plot in \cref{fig:chicorrelationplot} $\chi^2_{\xi\ell}$ for a representative simulation run of a low contrast micelle with $848$ core beads.
Although the mode variances $\Sigma_{\xi\ell}$ and correlation times $\tau_{\xi\ell}$ vary by several orders of magnitude, the $\chi^2_{\xi\ell}$ remain mostly within their $90\%$ confidence interval with no apparent systematic dependence on $\ell$, giving a positive validation of the assumptions used to calculate $\bbbs_\xi$.
In particular, this validation gives credence to the form of \cref{eq:varianceInMean} used to calculate the error in the mean of a single run, where each mode is assumed to have an independent correlation time estimated by the time series of that mode's amplitude.

If the correlation times were estimated incorrectly, then the $\chi^2_{\xi\ell}$ for small $\ell$ would be significantly different than the $\chi^2_{\xi\ell}$ for large $\ell$.
As an extreme example of an incorrect correlation time estimation, consider the effect of ignoring correlation times completely (i.e., inserting $\tau_{\xi\ell}=1$ into \cref{eq:varianceInMean} as done in \cref{fig:chicorrelationplot}): the error in the mean along the high-amplitude modes is underestimated and so the corresponding $\chi^2_{\xi\ell}$ are much larger than unity.
This analysis was done for a single run using one of our more stable compositions.
Naturally, less well-behaved micelle compositions and runs, such as the low-contrast composition with 600-bead core, are not expected to fare as well under the same analysis.

\begin{figure}
\centering
\includegraphics[width=\linewidth]{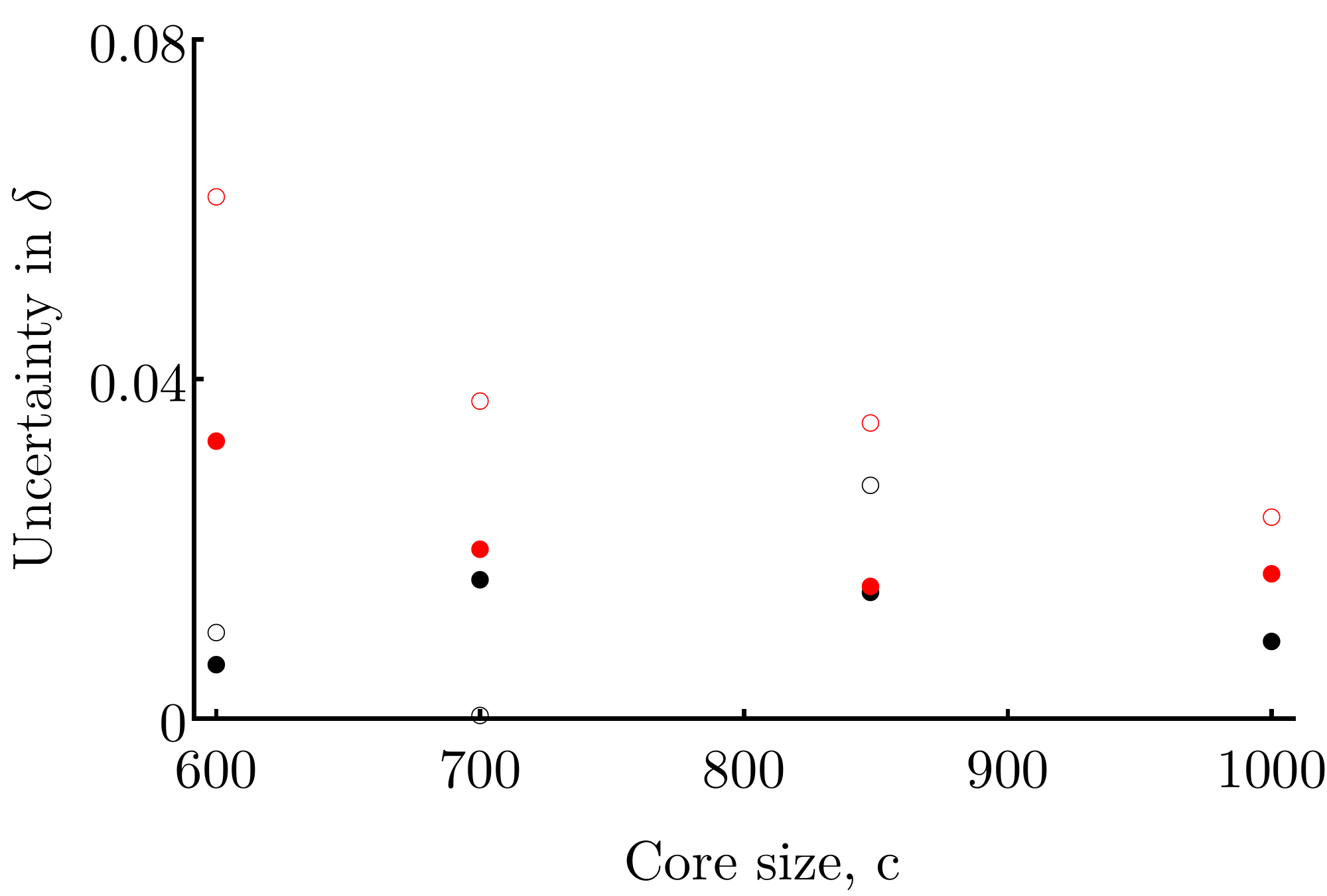}
\caption{(Color online) 
Plot of two estimates of the uncertainty in the normalized fluctuation $\delta$ for the different micelle compositions of \cref{tab:resultShapeTable}.
Black (red) symbols represent high (low) contrast micelles.
Closed symbols represent the first estimate described in \cref{subsec:shapeFeatures}; open symbols represent the second estimate, involving calculation of the standard error of the normalized fluctuation from individual simulation runs.
Since simulations of the high contrast micelle with $1000$ core beads only resulted in one well-formed shape, the second estimate of its normalized fluctuation uncertainty cannot be made.
Besides this case, and the case of the high contrast micelle with $700$ core beads, where the two simulation runs giving well-formed micelles had very similar normalized fluctuations, the estimates agree to within a factor of two.
}
\label{fig:fluctuationComparison}
\end{figure}

Next, in \cref{subsec:shapeFeatures}, we described a way to estimate the uncertainty in the normalized fluctuation $\delta$ by individually estimating the uncertainty in each mode variance $\Sigma_{\ell}$.
Then, to validate this estimate, we proposed performing a comparison to the standard error of the normalized fluctuations individually calculated from each simulation run.
The comparison is shown \cref{fig:fluctuationComparison}.
We expect only rough agreement because the estimate being validated assumed the shape data is drawn from a Gaussian distribution.
In fact the size of the errors are only consistent to about a factor of two.
Therefore, when evaluating the significance of the results in \cref{subsec:shapeFeaturesResult}, it should be remembered that the uncertainties in the normalized fluctuations are determined only to this limited precision.

\begin{figure}
\includegraphics[width=\columnwidth]{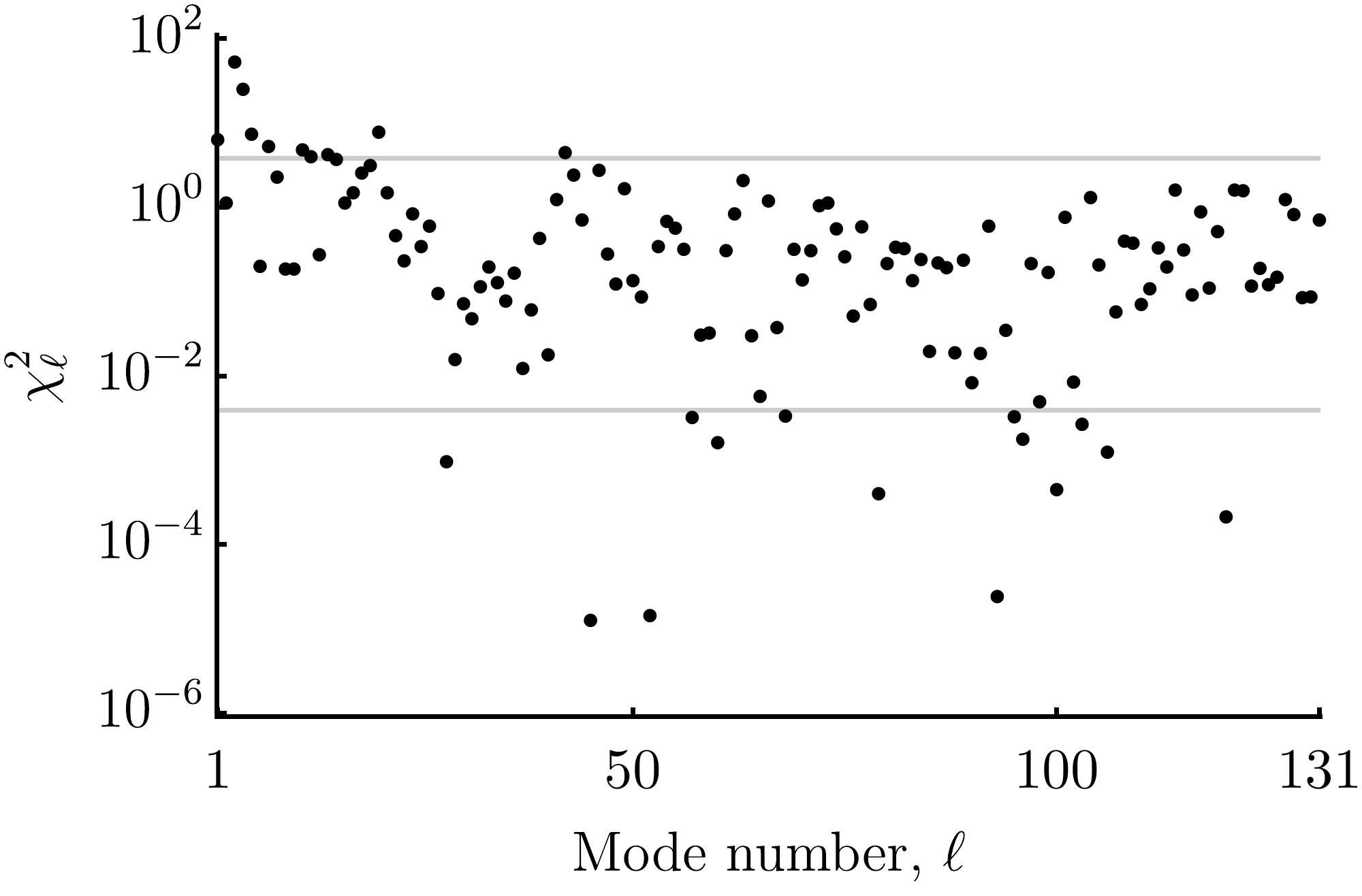}
\caption{
Plot of $\chi^2_\ell$ defined in \cref{eq:combinedChiEll} vs mode number $\ell$.
The two horizontal gray lines demarcate the $90\%$ confidence interval.
}
\label{fig:chiell}
\end{figure}

Finally, we validate the formula \cref{eq:combindAverage} for the combined mean and its uncertainty estimate defined by \cref{eq:varianceInMeanAverageOfRuns}.
This validation is done by comparing the combined mean $\bbbr$ with the combined set of all shape samples, as described in \cref{subsec:validation}.
In \cref{fig:chiell}, we show a plot of the statistic $\chi^2_\ell$, defined in \cref{eq:combinedChiEll}, for the low contrast micelle with $848$ core beads.
The $\chi^2_\ell$ fall within the expected range, and the $\ell$-average of $\chi^2_\ell$ is $1.5$.
The difference between this value and the ideal value of $1$ suggests that the error in the combined mean may be slightly underestimated.
From \cref{fig:chiell}, this underestimation seems to be worst for low $\ell$ modes.

\subsection{Shape features}
\label{subsec:shapeFeaturesResult}

\begin{figure}

\subfloat[][]{ %
\centering
\includegraphics[width=\columnwidth]{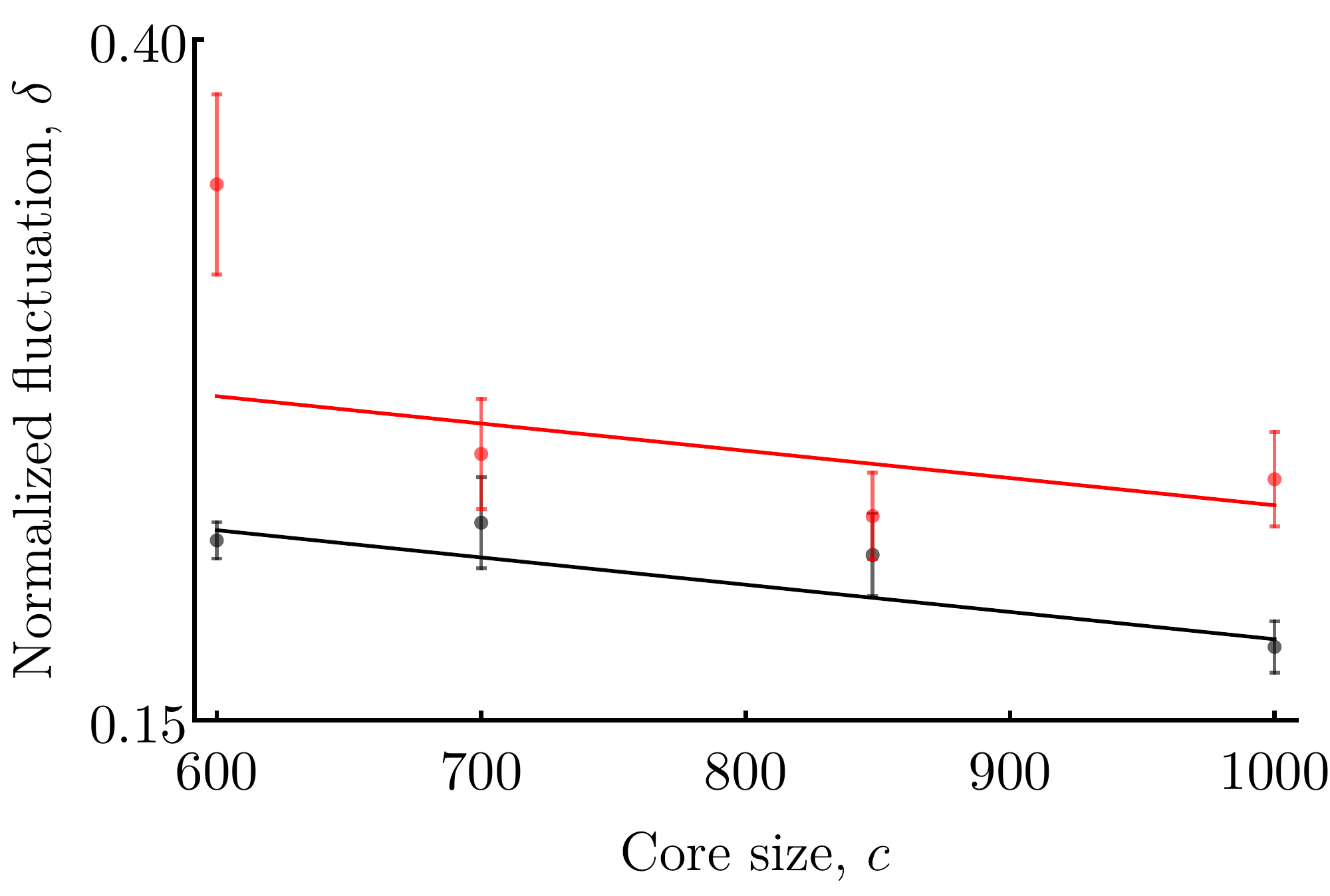}
\label{fig:finegrainedfluctuation}
}

\subfloat[][]{ %
\centering
\includegraphics[width=\columnwidth]{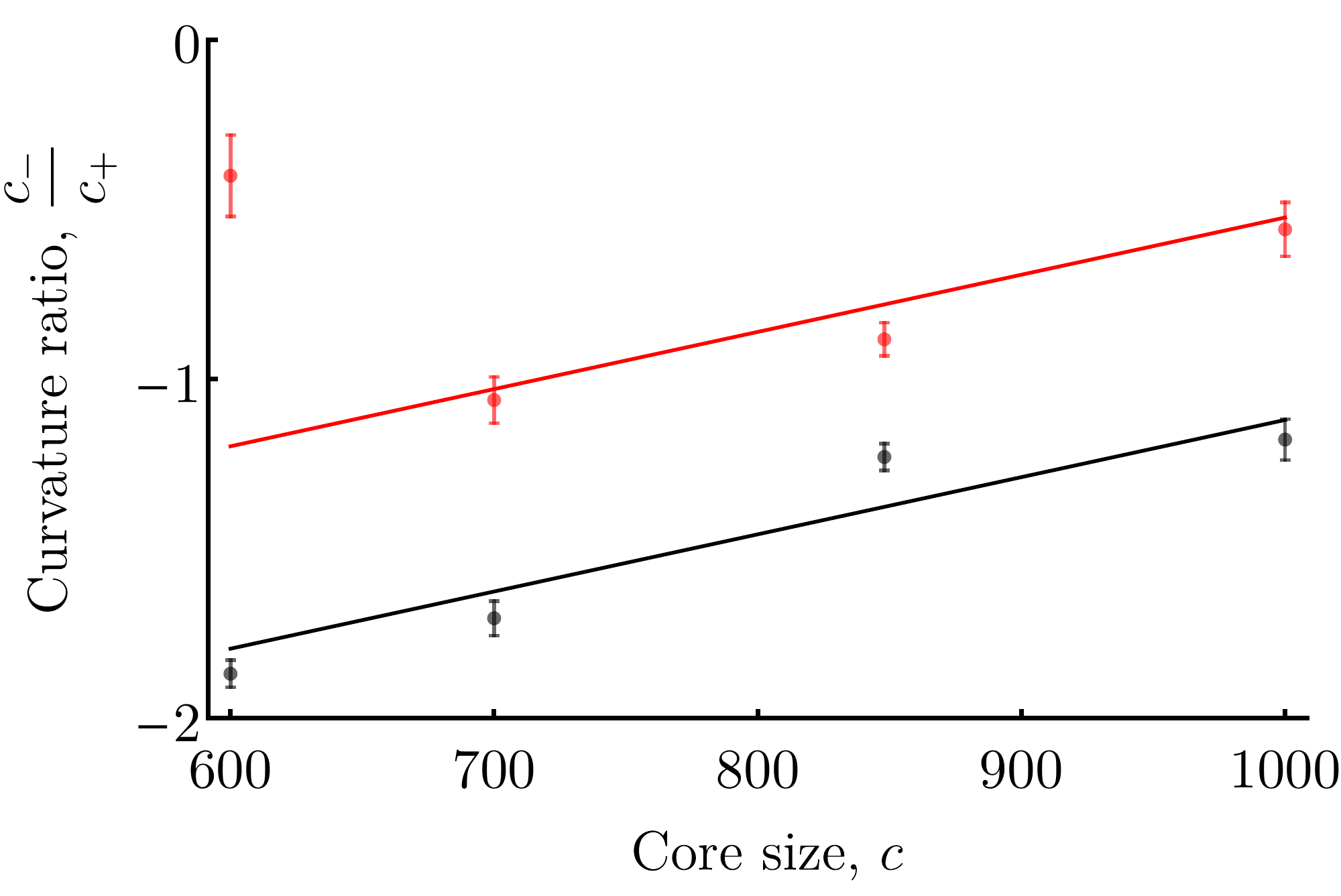}
\label{fig:finegrainedcurvatureratio} 
}

\caption{(Color online) Plots of normalized fluctuation $\delta$ and curvature ratio $c_+/c_-$ vs micelle composition.
The micelle compositions are those of \cref{tab:resultShapeTable}.
In \protect\subref{fig:finegrainedfluctuation}, $\delta$ is plotted against the size of the micelle core. 
Black (red) points represent high (low) contrast micelles.
The data are fit to a model containing only constant and gradient terms: $\delta = \delta_0 + m_c c + m_r r$, with $c$ being the number of core beads and $r$ being the solvophobic-rich diblock asymmetry ratio (defined in \cref{eq:asymmetryRatio}). 
Best fit parameters are found to be $m_c=(-1.0\pm 0.3)\times 10^{-4}$ and $m_r=0.3 7\pm 0.08$. 
The reduced chi-squared for this fit is $1.9$. 
The curvature ratio is treated analogously in \protect\subref{fig:finegrainedcurvatureratio}. 
The best fit parameters for the curvature ratio are given by $m_c=(1.7\pm 0.1)\times 10^{-3}$ and $m_r=4.5 \pm 0.3$. 
The reduced chi-squared for this fit is $14$. 
After omitting the outlier at a core size of 600 and $\nin:\nex = 27:4$, the reduced chi-squared drops to $4.4$.}
\label{fig:resultFitPlots}
\end{figure}
Finally we discuss our findings concerning the shape features discussed in \cref{subsec:shapeFeatures}.
To study quantitatively the dependence of micelle shape on composition already apparent in \cref{tab:resultShapeTable},  we plot in \cref{fig:resultFitPlots} the curvature ratios and normalized fluctuations of these shapes, as well as a fit to a simple linear model.
Several trends are revealed by these plots.
The curvature ratio becomes more negative (meaning that the dimple becomes more pronounced) as the asymmetry ratio becomes more negative (meaning the solvophobic-rich diblocks become even more solvophobic).
This confirms our intuition explained in \cref{subsec:micelleDesign} that the curvature should be positively correlated with the asymmetry ratio.

Another trend is that as the size of the core is increased, the curvature ratio becomes more positive (meaning that the dimple becomes less pronounced).
We propose the following explanation for this behavior.
We note that an increase in the core size increases the volume of the micelle.
This increased volume could be accommodated either by an increase in the perimeter, reducing the density of diblock ``surfactant" on the surface and thereby presumably increasing the micelle surface tension, or by making the micelle more circular, thereby making the curvature ratio more positive.
In practice, we expect both of these happen to some extent, and so increasing the core size would both increase the surface tension and make the curvature ratio more positive.

The normalized fluctuations are less precisely determined, but trends are still apparent.
The data show that the normalized fluctuations decrease as the solvophobic-rich diblocks become more solvophobic.
This effect could also be explained in terms of a competition between the preferred perimeter and preferred curvature.
As the solvophobic-rich diblocks become more solvophobic, the dimple becomes more pronounced, which we propose leads to an increase in perimeter and consequently surface tension.
This increased surface tension would then decrease the amplitude of shape fluctuations.
Also, the data suggest that micelles with more core have lower normalized fluctuations, although the size of this effect is on the same order of the uncertainty.
We have argued in the previous paragraph that increasing the core should increase surface tension.
In addition to reducing the dimple, this increased tension should also reduce fluctuations, explaining the trend.

We note one more feature in the data: the low contrast micelle with $600$ has a significantly more positive curvature ratio than the trend line predicts.
Although we are not sure how to explain this, we suspect this behavior is related to the onset of a transition reported in \cite{leibler89} involving the buckling a two-dimensional vesicle wall upon decreasing the vesicle's interior volume. 
If the energy barrier associated with buckling was high enough to preclude a buckling event from occurring within a simulation, then our observation, previously noted in \cref{subsec:analysisValidation}, of two well-defined, but inconsistent micelle shape averages could be explained.
In any event, this data point is an interesting starting point for further investigation.

Having described our results, we now note that they afford some degree of predictive power.
We have observed a range of curvature ratios with a range extending approximately from $-0.5$ to $-1.8$ exhibiting a mostly regular dependence of this curvature ratio on micelle composition.
Therefore if a micelle with curvature ratio in the observed range is to be constructed, the data provide a way to determine which micelle composition gives the desired curvature ratio.
In this way, we have demonstrated that micelle shape design is possible using our design mechanism.

\section{Discussion}
\label{sec:discussion}
The work presented in this paper is only a first demonstration that our micelle design mechanism can provide for fine control of a micelle shape.
In this section, we describe several directions for future exploration: we describe other aspects of micelle composition to vary; we suggest additional features of the thermal shape distribution to control, we suggest a few alternate schemes for bonding the micelle's constituent diblocks with the idea that they may be more effective at preventing malformed shapes, we propose using a simplified elastic model of the micelle shape energetics, and we discuss how our shape-design strategy might be extended to three dimensions and what challenges may arise.
Finally, we discuss how our results are relevant to the applications mentioned in the introduction.

\subsection{Further variation of micelle composition}
\label{subsec:moreComposition}
In this work, the effect of only two aspects of micelle composition were studied in this paper, and some speculative explanations of the observed behavior were given.
In future work, other aspects of micelle composition may be varied, extending the range of observed micelle shapes and giving further insight into the factors affecting micelle shape.
For example, only the asymmetry of the solvophobic-rich diblocks were studied; the effect of varying the solvophilic-rich diblocks could also be studied.
Another aspect of the micelle composition to address is the length of the diblocks.
In this work, we filled the micelle surface with diblocks of a specific chosen length, and chose the asymmetry of these diblocks to produce the desired curvature.
However, there is freedom in choosing the length of the diblocks: the micelle surface could be filled using a larger number of shorter diblocks, holding fixed the imprinted preferred curvature profile.
A third way of altering the micelle composition is to introduce another species of diblock.
In the micelle shapes presented in \cref{sec:results}, there were three regions of significantly different curvature: the dimple was concave, the surface opposite the dimple was weakly convex, and the surface adjacent the dimple was strongly convex.
Since these micelles only contain two species of diblock, the observed micelle shapes seem to be at odds with our stated design strategy, wherein the diblock's preferred curvature dictates the surface curvature.
We expect that the mismatch between the imprinted curvature and the realized curvature represents a frustration which may affect the dynamics (e.g., fluctuations) of the micelle.
To test this expectation, one could introduce a third species of diblock to better match the realized micelle curvature, and study the resulting micelle shapes.

\subsection{Additional shape features to control}
\label{subsec:moreShapeFeatures}
Just as there are many ways to alter the micelle composition, there are many aspects of the micelle shape to control.
Here we list three extensions to the shape control demonstrated in this work.
First, instead of the curvature ratio of the average shape illustrated in \cref{fig:curvatureRatio}, one could study other quantities characterizing the average shape.
In fact, the difference metric $\din$, defined in \cref{eq:dinDefinition}, provides a way of quantifying the similarity to any chosen target shape.
Second, we chose a static set of interaction parameters in our model; however, applications may require exposing the micelle to varying environments (having variations of e.g., temperature, pH, or salt concentration).
On the one hand, such a varying environment would presumably make it difficult to ensure a fixed micelle shape.
On the other hand, there emerges a challenge of designing a micelle that assumes different designed shapes depending on its changing environment.
Lastly, we have found evidence of a micelle exhibiting two metastable shapes in a single environment.
In this case, there is a breakdown in the representation of the micelle shape as a Gaussian distribution fluctuating about a single mean.
Instead, one could categorize the observed micelle shapes into clusters (each representing a metastable micelle shape), and find the mean of each cluster and the transition rates between the clusters.
Further, each metastable shape presumably may be designed by changing the micelle composition.

\subsection{Avoidance of malformed shapes}
\label{subsec:malformedShapesAvoidance}
\begin{figure}
\begin{center}

\subfloat[][]{ %
\centering
\includegraphics[width=.45\columnwidth]{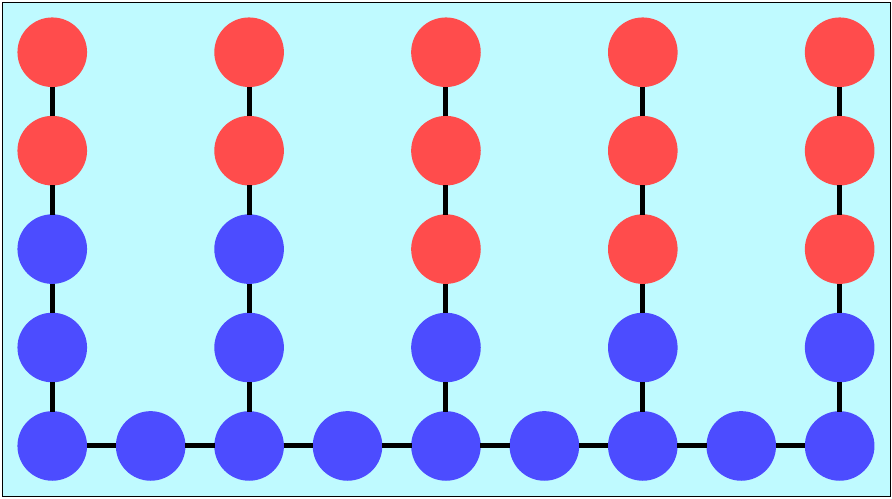}
\label{fig:solvophobicBranch}
} \subfloat[][]{ %
\centering
\includegraphics[width=.45\columnwidth]{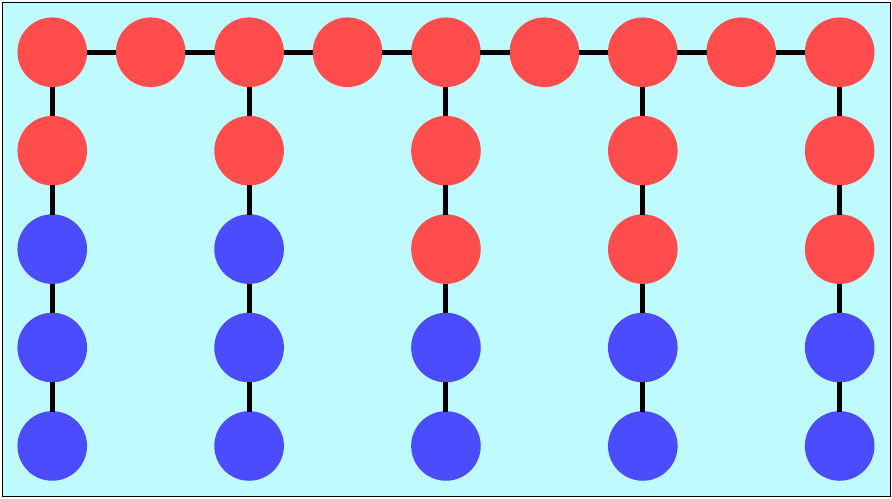}
\label{fig:solvophilicBranch}
}
\end{center}
\subfloat[][]{ %
\centering
\includegraphics[width= .45\columnwidth]{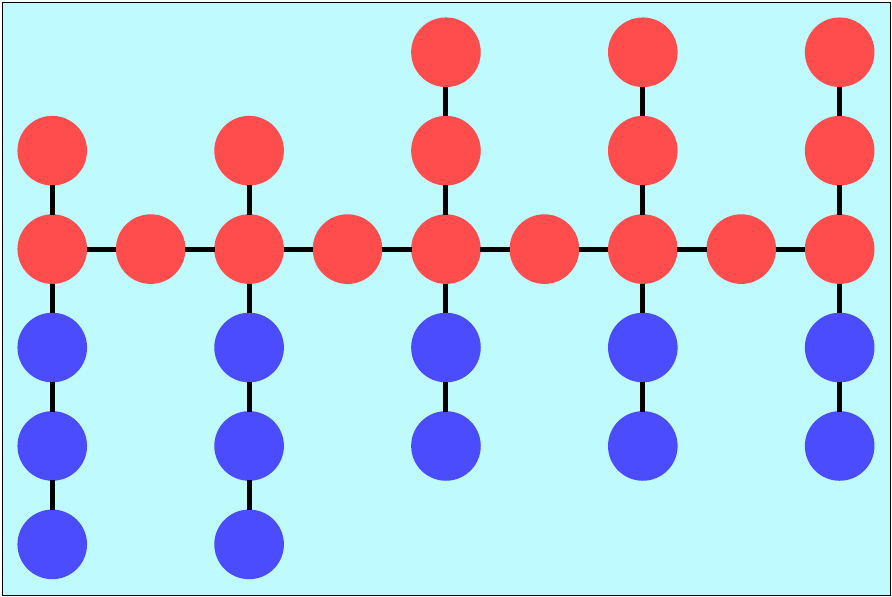}
\label{fig:middleBranch}
}
\caption{(Color online) 
Schematics of alternate bond topologies for single polymer micelles.
In \protect\subref{fig:solvophobicBranch}, diblocks have been attached as side chains to a solvophobic homopolymer.
In \protect\subref{fig:solvophilicBranch}, the diblocks are instead attached on their solvophilic ends to a solvophilic homopolymer.
Lastly, in \protect\subref{fig:solvophilicBranch}, the diblocks are represented as a pair of side chains attached to a solvophilic homopolymer chain.
}
\label{fig:alternateToplogies}
\end{figure}

Another issue to be addressed in future work is the avoidance of malformed shapes.
Although the bond scheme chosen in this work did reduce the occurrence of malformed shapes, it is likely that other bond schemes, while perhaps harder to synthesize, would be even more effective.
One could imagine a solvophobic backbone chain with solvophilic (and perhaps solvophobic) side chains; in fact, these bond schemes have been considered theoretically in \cite{Khokhlov2005}.
Alternatively, the backbone chain could be solvophilic.
Even more intricate possibilities are a solvophobic backbone chain with diblock side chains, or even branched side chains where the degree of branching can be used to control the curvature.
Examples of alternate bond topologies are shown in \cref{fig:alternateToplogies}.
In any case, we imagine that the effect of a diblock's spontaneous curvature on the shape of a well-formed micelle is mostly independent of the bond scheme used to make the micelle well-formed, so that these two problems may be studied independently.

\subsection{Simplified elastic model of micelle shape}
\label{subsec:coarseGrained}
Moving beyond particle-based simulations, further insight could be gained by studying a simplified model containing only the physics believed to be necessary to explain the micelle shape (such as the relationship between a diblock's composition and its preferred curvature), and abstracts away unimportant details (such as the exact nature of the monomer interactions).
We have in mind a model where the degrees of freedom are the positions of the junction points, far fewer in number than the positions of each bead.
These junctions points would be subject to a potential energy function representing the compression energy of the total volume enclosed by the micelle surface and the energies of surface stretching and bending.
However, it is unclear to us if a micelle could faithfully be modeled so simply.
Further, if this model is to be useful in experimental applications, the effective parameters of the simple model must somehow be found from the small-scale monomer interactions. 

\subsection{Extension to three dimensions}
\label{subsec:threeDimensions}
Finally, we address the most readily apparent issue: the extension of our design mechanism to three dimensions.
We expect the basic principle of our design mechanism---that diblock composition can be used to influence preferred surface and therefore shape---should work similarly in three dimensions.
As three dimensional micelles are routinely simulated, we expect that the added computational burden to be minor.

\begin{figure}

\subfloat[][]{ %
\centering
\includegraphics[width=\columnwidth]{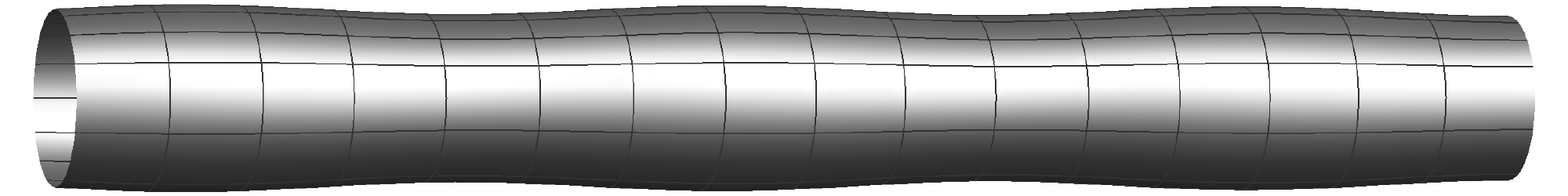}
\label{fig:cylinderlikeUnduloid}
}

\subfloat[][]{ %
\centering
\includegraphics[width=\columnwidth]{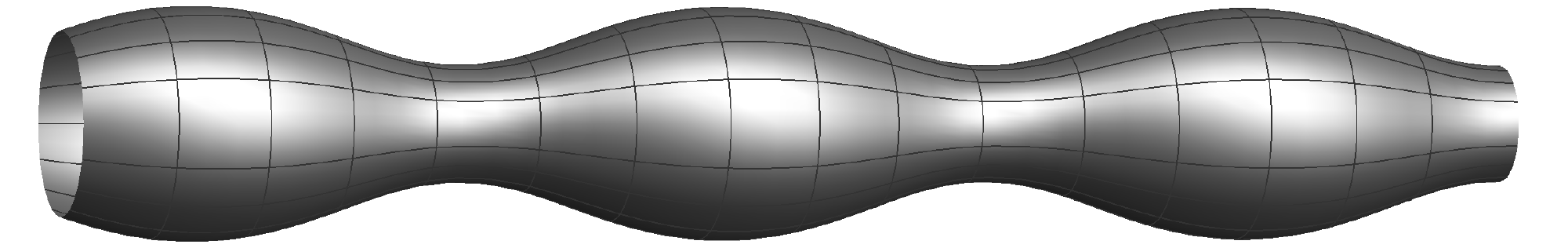}
\label{fig:mediumUnduloid}
}

\subfloat[][]{ %
\centering
\includegraphics[width=\columnwidth]{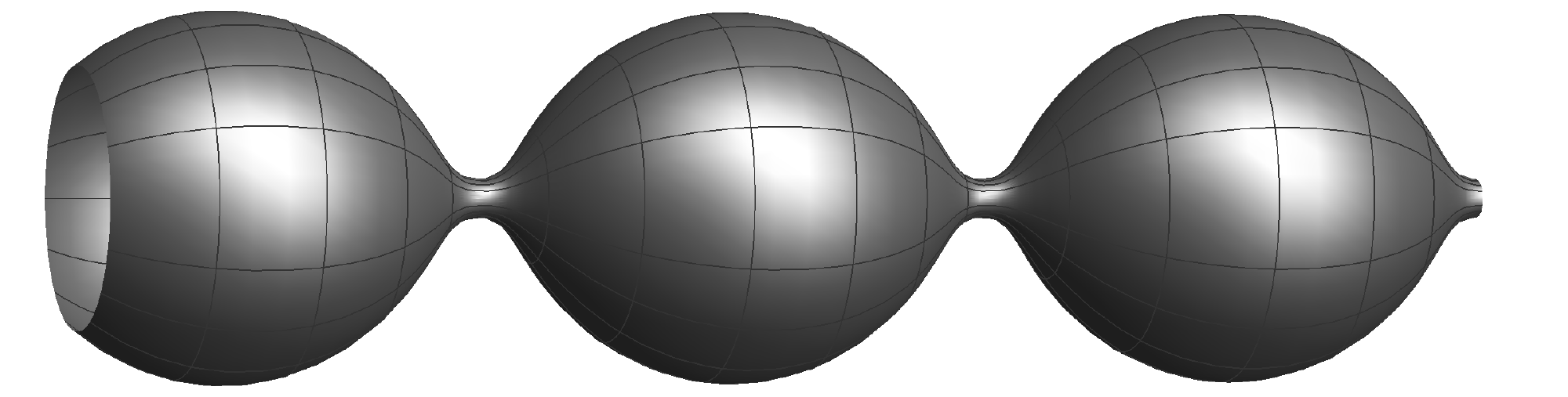}
\label{fig:spherelikeUnduloid}
}

\caption{Plots of three unduloids interpolating between a cylinder and a string of spheres. Each surface shown in \protect\subref{fig:cylinderlikeUnduloid}, \protect\subref{fig:mediumUnduloid}, and \protect\subref{fig:spherelikeUnduloid} has a uniform mean curvature, and furthermore the three curvatures are the same. In fact, these shapes can be continuously deformed into one another with the mean curvature held fixed.} 
\label{fig:unduloids}
\end{figure}

However, there are difficulties that arise in three dimensions.
One difficulty is that to lowest order in the curvature, the local bending energy $U_{\textrm{bend}}$ per unit of the micelle surface area is \cite{helfrich73} given by 
\begin{equation}
\label{eq:helfrich}
U_{\textrm{bend}} = 2k\left(H-c_0\right)^2 + \frac{\bar{k}}{2} K,
\end{equation}
where $H$ and $K$ are the mean and Gaussian curvatures, $c_0$ is the spontaneous mean curvature, and $k$ and $\bar{k}$ are moduli governing the mean and Gaussian curvature, respectively.
We expect that the local spontaneous mean curvature parameter $c_0$ ought to depend on the local diblock composition at the surface.
Even if this is the case, the local diblock composition determines only one preferred curvature, $c_0$, but a micelle shape is described by two principal curvatures at each point.
Therefore it would seem a micelle shape is underspecified by the mean curvature profile induced by diblocks on the micelle surface.
For example, we unduloids~\cite{delaunay}, a family of distinct shapes all having the same mean curvature profile (examples shown in \cref{fig:unduloids}).
The diblocks having mean curvature compatible with the unduloid in \cref{fig:mediumUnduloid}, will likewise have a mean curvature compatible with \cref{fig:spherelikeUnduloid}, so neither one of these shapes can be designed if diblocks affect only the preferred mean curvature.
However, a diblock composition is described by two parameters, a length and an asymmetry ratio, and so it is possible that another parameter in addition to the spontaneous curvature, such as the mean curvature modulus $k$ or Gaussian curvature modulus $\bar{k}$, may be controlled by the diblock composition.
In fact this dependence is described theoretically in \cite{Wang91}.
This may be sufficient to restore full shape-designability.
Alternatively, it is possible that the form of \cref{eq:helfrich} does not accurately describe the highly curved micelles we consider, because higher order terms in the curvature become relevant; in this case, a preferred Gaussian curvature may arise, allowing for more direct control of the surface curvature.
Even if full control of the shape is not possible in three dimensions, it would still be interesting to determine what shape-designability remains.

Another difficulty with extending the shape design mechanism is that a more sophisticated bond scheme is necessary to enforce the relative positioning of the diblocks on the surface of the micelle.
Indeed, three-dimensional simulations amphiphilic linear multiblock copolymers and polymers with side chains have been simulated \cite{vasilevskaya2011}, and while rough shape control has been demonstrated, the junction points do not arrange themselves on the micelle surface in an organized manner, so the fine control we seek does not seem possible.
Instead, the polymer may need to be realized in the form of a branched polymer.
One possibility is to first form a crosslinked polymer network of a roughly spherical shape, and then graft onto the network surface diblocks of the desired compositions.
There remains a question of how to create the crosslinked polymer network with chemically distinct surface regions necessary to graft specific species of polymer to specific regions on the surface.
We hypothesize that such a network could be created either by growing outward from a multifunctional core such as a silsesquioxane \cite{cordes2010} or by growing inward from an external, rigid scaffold such as a protein cage \cite{abedin09}.
In either case, after the network is formed, the original core or scaffold could be disassembled, leaving only the flexible polymer network.

\subsection{Relevance to applications}
\label{subsec:applications}
We now discuss how our results could be used to address the applications discussed in the introduction.
One application was to use shape-designed micelles as drug carriers.
It has been found that the carrier shape can affect how much drug can be loaded into the micelle \cite{Cai2007}.
Also, it has been found that nanoparticle shape can affect where in the body (e.g., in which organ) the particles accumulate \cite{Decuzzi10,Devarajan09,Muro08,Patil08,Gillies04}.
Since we have found it possible to precisely control at least one feature of the micelle shape (presumably more shape features could be controlled with additional effort), one could expect to fully optimize the micelle shape to target a specific organ.
Additionally, it was found that carrier shape flexibility affects how quickly the drug is cleared from the body \cite{Chen09,Fox09,Gillies04,Lee05}.
Our results have demonstrated that this flexibility can be controlled precisely, potentially allowing precise control of drug clearance.

Another application was the lock and key mechanism.
In \cite{Sacanna2010Lock}, the assembly of concave objects, similar to the ones designed in this work, were studied.
It was found that in the presence of depletants, ensembles of these object could be made to aggregate.
It was further found that the size of the concave feature affects the aggregation: the concave curvature of the dimple must match the convex curvature of the object to which it will bind.
Precise control of the dimple size thereby allows precise control of how these objects aggregate.
Additionally, one could imagine that a micelle with large shape fluctuations could deform to fit a wide range of curvature, thus acting as a ``master key".
This last possibility highlights the benefits of using diblock copolymer micelles as opposed to more rigid shapes such as proteins.

\section{Conclusion}
\label{sec:conclusion}
Molecular dynamics simulations were used to study the fluctuating shape of a polymeric micelle at finite temperature in two dimensions.
The micelle was constructed from a single, linear, multiblock copolymer.
A globular state with the multiblock's junction points sequentially ordered around the micelle perimeter is often maintained during the course of the simulation when such a state is used as the initial configuration.
We demonstrated the effectiveness of a strategy where the multiblock is viewed as a collection of diblock copolymers joined end to end, and the asymmetry of these diblocks is selected to dictate the micelle surface curvature.
Specifically, we found that positioning solvophobic-rich diblocks preferring concave curvature on the micelle surface caused the formation of a concave dimple in the surface region occupied by these diblocks.
Further, the strength of the dimple is controlled by both the asymmetry of these diblocks, and the size of a homopolymer core chain located in the micelle interior.
In addition to the strength of the dimple, the asymmetry of the solvophobic-rich diblocks and the size of the core chain affected the amount of fluctuations in the micelle shape.
In future work, the micelle shape design strategy could be studied in three dimensions where it is as yet unclear how precisely polymeric micelles shapes could be controlled.

\section{Acknowledgements}

The authors thank Juan de Pablo, Greg Voth, Kurt Kremer, and Jeff Vieregg for enlightening conversations regarding the context of this work within the literature and for suggesting modifications to our design scheme to ease experimental realization.
We also thank Jian Qin, Steve Tse, James Dama, and Glen Hocky for their guidance in carrying out the simulations presented in this work.
This work was completed in part with resources provided by the University of Chicago Research Computing Center.
This work was principally supported by the University of Chicago Materials Research Science and Engineering Center, which is funded by the National Science Foundation under award number DMR-1420709.

\appendix*

\section{Equivalence of two definitions of mean shape}
\label{sec:appendix}
To justify our definition \cref{eq:averageDefinition} of $\bbbr$, we show that it is equivalent to an alternative notion of average.
The alternative definition involves first rotating the shapes so that the summed pairwise square differences are minimized, and then simply taking an arithmetic average.
In symbols, we define the minimizing rotation angles $\ct_\alpha$ by
\begin{equation}
\label{eq:brCheck}
(\ct_1,\dots,\ct_{N_s})= \argmin_{(\theta_1,\dots,\theta_{N_s})} \sum_{\alpha,\beta=1}^{N_s} \dex(\bR_{\theta_\alpha} \bbr_\alpha ,\bR_{\theta_\beta} \bbr_\beta),
\end{equation}
and then we define the arithmetic average $\tilde{\bbbr}$ of the shapes by
\begin{equation}
\label{eq:arithmeticAverageShape}
\tilde{\bbbr} = \frac{1}{N_s}\sum_{\alpha=1}^{N_s}\bR_{\ct_\alpha} \bbr_\alpha.
\end{equation}
To transform \cref{eq:arithmeticAverageShape} into a form more similar to \cref{eq:averageDefinition}, we use a standard identity relating the expected square differences between two independent samples to expected square difference of a single sample to the mean\footnote{
Recall if $X$ and $Y$ are two independent, identically distributed random variables, then (using $\langle\dots\rangle$ to denote expected value) $\langle(X-Y)^2\rangle = \langle X^2 -2 X Y +Y^2 \rangle= 2\left(\langle X^2\rangle-\langle X\rangle^2\right) = 2\langle \left(X-\langle X\rangle\right)^2\rangle$. This identity applies in our case because $\dex(X,Y)$ is of the form $(X-Y)^2$.
}:
\begin{multline}
\label{eq:sumSquareErrorTransform}
\frac{1}{N_s^2} \sum_{\alpha,\beta=1}^{N_s} \dex(\bR_{\theta_\alpha} \bbr_\alpha ,\bR_{\theta_\beta} \bbr_\beta) = \\ 2 \frac{1}{N_s} \sum_{\alpha=1}^{N_s} \dex(\bR_{\theta_\alpha} \bbr_\alpha ,\frac{1}{N_s} \sum_{\beta=1}^{N_s} \bR_{\theta_\beta} \bbr_\beta).
\end{multline}
In fact we can proceed further by recognizing that the arithmetic mean minimizes the sum of square differences.
Using this fact to transform the right hand side of \cref{eq:sumSquareErrorTransform}, we obtain
\begin{multline}
\label{eq:sumSquareErrorTransformedAgain}
\frac{1}{N_s^2} \sum_{\alpha,\beta=1}^{N_s} \dex(\bR_{\theta_\alpha} \bbr_\alpha ,\bR_{\theta_\beta} \bbr_\beta) = \\ 2 \frac{1}{N_s} \min_{\bba} \sum_{\alpha=1}^{N_s} \dex(\bR_{\theta_\alpha} \bbr_\alpha ,\bba).
\end{multline}

Using this identity (and ignoring an unimportant multiplicative factor of $2 N_s$), we may transform \cref{eq:brCheck} into
\begin{multline}
\label{eq:brCheckTransformed}
(\ct_1,\dots,\ct_{N_s})=\\ \argmin_{(\theta_1,\dots,\theta_{N_s})} \min_{\bba} \sum_{\alpha=1}^{N_s} \dex(\bR_{\theta_\alpha} \bbr_\alpha ,\bba).
\end{multline}
From this equation, we see that the $\ct_\alpha$ result from performing a double minimization of $\sum_{\alpha=1}^{N_s} \dex(\bR_{\theta_\alpha} \bbr_\alpha ,\bba)$ with respect to both the $\theta_\alpha$ and $\bba$.
Now we have already stated above \cref{eq:sumSquareErrorTransformedAgain} that the minimizing $\bba$ must be the arithmetic average given by \cref{eq:arithmeticAverageShape}, so that we may simply write
\begin{equation}
\label{eq:arithmeticAverageArgmin}
\begin{aligned}
\tilde{\bbbr} &=\argmin_{\bba}\min_{(\theta_1,\dots,\theta_{N_s})}  \sum_{\alpha=1}^{N_s} \dex(\bR_{\theta_\alpha} \bbr_\alpha ,\bba)\\
&=\argmin_{\bba} \sum_{\alpha=1}^{N_s} \min_{\theta_\alpha} \dex(\bR_{\theta_\alpha} \bbr_\alpha ,\bba)\\
&=\argmin_{\bba} \sum_{\alpha=1}^{N_s} \din(\bbr_\alpha ,\bba)\\
&=\bbbr,
\end{aligned}
\end{equation}
where the second line is obtained by noting that each term in this sum of square differences depends on only one $\bR_{\theta_\alpha}$, and the third and fourth lines are obtained by applying the definitions \cref{eq:dinDefinition} and \cref{eq:averageDefinition} respectively. 
We conclude that the two notions of average $\bbbr$ and $\tilde{\bbbr}$ are indeed the same.

\bibliography{references/abedin09,%
references/bergstrom2008,%
references/brandt2011,%
references/binder11,%
references/Chen09,%
references/cordes2010,%
references/crcHandbook,%
references/dataReduction,%
references/Decuzzi10,%
references/delaunay,%
references/detcheverry2009CoarseGrained,%
references/Devarajan09,%
references/Dimitrakopoulos04,%
references/Doi1986ChainLength,%
references/Doi1986KuhnLength,%
references/drugDelivery14,%
references/drugDeliveryShape2011,%
references/fisher,%
references/Fox09,%
references/geyer2011,%
references/Gillies04,%
references/Goetz1999,%
references/gompper2002,%
references/golas2009Synthesis,%
references/gregory2012,%
references/guo2010,%
references/harmandaris2006,%
references/helfrich73,%
references/Hsieh06,%
references/Hu2012,%
references/hu2015,%
references/Israelachvili,%
references/jelonek2015,%
references/khokhlov2005,%
references/laradji2000,%
references/Lee05,%
references/leibler,%
references/lewandowski2009,%
references/lipowsky2014,%
references/lockAndKeyColloids,%
references/lockAndKeyPicture,%
references/Lof2009,%
references/lammps,%
references/ljunggren89,%
references/rikken2016,%
references/Zhao2010,%
references/loverde2010,%
references/Matyjaszewski,%
references/micelleDrugDeliverykataoka,%
references/mixedMicelle2011,%
references/Muro08,%
references/Patil08,%
references/pdmsHandbookOfPolymers,%
references/physPropsOfPols,%
references/poorgholami-bejarpasi2010,%
references/rao73,%
references/rekvig2004,%
references/Rozycki2015,%
references/sheng2013,%
references/srinivas2004,%
references/Tarazona2013,%
references/velinova2011,%
references/venable2015,%
references/cai2007,%
references/watson2011,%
references/vasilevskaya2011,%
references/Wang91,%
references/Zhang98} 
\end{document}